\documentclass[12pt, hyper]{article}
\usepackage{amsmath,amsopn,amssymb,amsfonts}
\numberwithin{equation}{section}
\usepackage{amsthm}
\usepackage{hyperref}
\usepackage[pdftex]{graphicx}
\usepackage{epsfig}
\usepackage{cleveref}
\usepackage{subfigure}
\usepackage{mathrsfs}
\DeclareMathAlphabet{\mathpzc}{OT1}{pzc}{m}{it}
\usepackage{verbatim}
\usepackage{multirow}
\usepackage{tikz}
\usepackage{pgfplots}
\usepackage{makecell}
\usepackage{arydshln}
\usepackage{enumitem}
\setlist{leftmargin=2mm}
\usepackage{cite}
\usepackage{amsthm}
\usepackage[margin=1cm]{caption}
\usepackage[font={small}]{caption}

\crefname{equation}{}{}

\def\IZ{\mathbb{Z}}

\def\CD{{\cal D}}

\def\CH{{\cal H}}

\def\CO{{\cal O}}

\def\CV{{\cal V}}
\def\CW{{\cal W}}
\def\CZ{{\cal Z}}

\def\fg{\mathfrak{g}}

\def\a{\alpha}\def\b{\beta}
\def\d{\delta}

\def\l{\lambda}

\def\r{\rho}
\def\t{\tau}

\def\D{\Delta}

\def\half{\frac{1}{2}}

\def\tr{{\rm Tr}}

\newcommand{\ket}[1]{|{#1}\rangle}

\usepackage{cleveref}

\begin{document}

\begin{titlepage}
\vfill
\begin{flushright}
{\tt\normalsize KIAS-P17036}\\
\end{flushright}
\vfill
\begin{center}
{\Large\bf Modular Constraints on

Conformal Field Theories with Currents}
\vfill

Jin-Beom Bae, Sungjay Lee and Jaewon Song

\vskip 5mm
{\it Korea Institute for Advanced Study \\
85 Hoegiro, Dongdaemun-Gu, Seoul 02455, Korea}

\end{center}
\vfill

\begin{abstract}
\noindent
We study constraints coming from the modular invariance of the partition function of two-dimensional conformal field theories. We constrain the spectrum of CFTs in the presence of holomorphic and anti-holomorphic currents using the semi-definite programming.
In particular, we find the bounds on the twist gap for the non-current primaries depend dramatically on the presence of holomorphic currents, showing numerous kinks and peaks.
Various rational CFTs are realized at the numerical boundary of the twist gap, saturating the upper limits on the degeneracies.
Such theories include Wess-Zumino-Witten models for the Deligne's exceptional series, the Monster CFT and the Baby Monster CFT.
We also study modular constraints imposed by $\CW$-algebras of various type and observe that the bounds on the gap depends
on the choice of $\CW$-algebra in the small central charge region.

\end{abstract}

\vfill
\end{titlepage}

\parskip 0.1 cm
\tableofcontents
\renewcommand{\thefootnote}{\#\arabic{footnote}}
\setcounter{footnote}{0}

\parskip 0.2 cm

\section{Introduction} \label{sec:intro}

Conformal field theories (CFT) are highly constrained by underlying symmetry.
For the case of two-dimensional CFTs, the modular invariance and
crossing symmetry put strong constraints on the spectrum and operator product expansion (OPE) coefficients.
Especially, the infinite dimensional Virasoro symmetry makes it possible
to have CFTs with finite number of primary operators. Those CFTs are called as rational CFTs \cite{Belavin:1984vu, Moore:1988qv}
and it has been known that they can be completely solved using the crossing symmetry constraints when $c<1$.

The full classification of the unitary two-dimensional CFTs is however still out of reach.
For the CFTs with $c>1$ (without any extended chiral algebra) there are infinite number of primary operators.
For such theories, modular invariance and crossing symmetry were not enough to completely solve the theory.
Nevertheless, in recent years, applying the universal constraints originated from the conformal symmetry and unitarity, a.k.a
conformal bootstrap has produced many fruitful results for the higher-dimensional CFTs \cite{Rattazzi:2008pe, Poland:2011ey}
and two-dimensional CFTs with $c>1$\cite{Chang:2015qfa,Chang:2016ftb,Lin:2015wcg,Lin:2016gcl}

In the current paper, we further study the consequences of modular invariance for the two-dimensional CFTs with $c>1$,
along the line of \cite{Hellerman:2009bu, Friedan:2013cba, Hartman:2014oaa, Collier:2016cls}. See also \cite{Hellerman:2010qd, Keller:2012mr, Qualls:2013eha,Qualls:2014oea,Qualls:2015bta}.\footnote{Very recently, constraints on 2d CFT from the genus two partition function have been studied \cite{Cardy:2017qhl,Cho:2017fzo,Keller:2017iql}.} The previous works mostly
focused on the case without extended chiral algebra. Instead we focus on the case of CFTs with holomorphic/anti-holomorphic
(higher-spin) currents and also with $\CW$-algebra symmetry.
We first assume that there are conserved currents of conformal weights $(h, \bar{h})=(j, 0)$ and $(h, \bar{h})=(0, j)$.
Under this assumption, we investigate the gap on the twist $\Delta_t \equiv \Delta - |j| = 2 \textrm{min}(h, \bar{h})$ for
the non-current operators.
We observe rather dramatic consequences for the upper bounds on the twist gap for the non-current operators.
Also, we explore modular constraints for the CFTs with $\CW$-algebra symmetry.
We focus on the $\CW$-algebra associated with a simple Lie algebra $\fg$, which we denote as $\CW(\fg)$.
We find that the numerical upper bounds depend on the choice of $\CW(\fg)$-algebra when the central charge is small,
namely $c \lesssim \textrm{rank}(\fg)$.
For the case of $\CW(A_2)$-algebra, this problem has been recently discussed in \cite{Afkhami-Jeddi:2017idc,Apolo:2017xip}.


The torus partition function of an arbitrary $c >1$ CFT admits a character decomposition of the form
\begin{align}
\begin{split}
Z(\tau, \bar{\tau}) &=  \chi_{0}(\tau) \bar{\chi}_{0}(\bar{\tau}) + \sum_{h,\bar{h}} d_{h,\bar{h}}
\chi_h(\tau) \bar{\chi}_{\bar{h}}(\bar{\tau}) \\
 &\quad + \sum_{j=1}^{\infty} \left[ d_j \chi_j(\tau) \bar{\chi}_0(\bar{\tau})
 + \tilde{d}_j \chi_0(\tau) \bar{\chi}_j (\bar{\tau})\right] \ ,
\label{partition function}
\end{split}
\end{align}
where $\chi_h (\tau)$ denotes the Virasoro character for the primary operator of weight $h$.
Here the degeneracies $d_j, \tilde{d}_j, d_{h,\bar{h}}$ have to be non-negative integers.
The invariance of (\ref{partition function}) under $T$-transformation forces the given CFT to have
states of integer spin $j = |h-\bar{h}|$ while the invariance under $S$-transformation implies
\begin{align}
Z(\tau, \bar{\tau}) = Z(-\frac{1}{\tau}, -\frac{1}{\bar{\tau}})\ .
\label{S constraint}
\end{align}
Combining (\ref{partition function}) and (\ref{S constraint}), we obtain a constraint
\begin{align}
\mathcal{Z}_{0}(\tau, \bar{\tau}) + \sum_{j=1}^{\infty} \Big[ d_j \mathcal{Z}_{j}(\tau, \bar{\tau})
+  \tilde{d}_{j}  \mathcal{Z}_{\bar{j}}(\tau, \bar{\tau}) \Big]
+ \sum_{h,\bar{h}} d_{h,\bar{h}} \mathcal{Z}_{h,\bar{h}}(\tau, \bar{\tau}) = 0,
\label{MBE}
\end{align}
where the function $\mathcal{Z}_{\lambda}(\tau, \bar{\tau})$ is
defined as ${\chi}_{\lambda}(\tau) {\bar{\chi}}_{\lambda}(\bar{\tau})
- {\chi_{\lambda}}(-\frac{1}{\tau}) {\bar{\chi}}_{\lambda}(-\frac{1}{\bar{\tau}})$.
We call \eqref{MBE} as the modular bootstrap equation.

We utilize the semi-definite programming with the help of powerful numerical package
{\tt SDPB}\cite{Simmons-Duffin:2015qma} to examine the modular bootstrap equation
(\ref{MBE}). The semi-definite programming method has been first employed in \cite{Friedan:2013cba} and then further refined in \cite{Collier:2016cls}, leading to universal constraints on the CFT spectrum.
For example, the upper bound on the spin-independent gap $\Delta_{\textrm{gap}}$ in the operator dimension is given as
\begin{align}
 \frac{1}{12} \le \lim_{c\to \infty} \frac{\Delta_{\textrm{gap}}}{c} <  \frac{1}{9} \ .
\end{align}
A similar bound in the presence of a $U(1)$ global symmetry has been found in \cite{Benjamin:2016fhe}.
In \cite{Collier:2016cls}, it was also observed that
the level-one ${\fg} = A_1, A_2, G_2, D_4, E_8$ Wess-Zumino-Witten (WZW) models
can be realized on the upper limit of the dimension gap of scalar primaries
at the corresponding values of central charge.
\begin{figure}[t]
\begin{center}
\includegraphics[width=.72\textwidth]{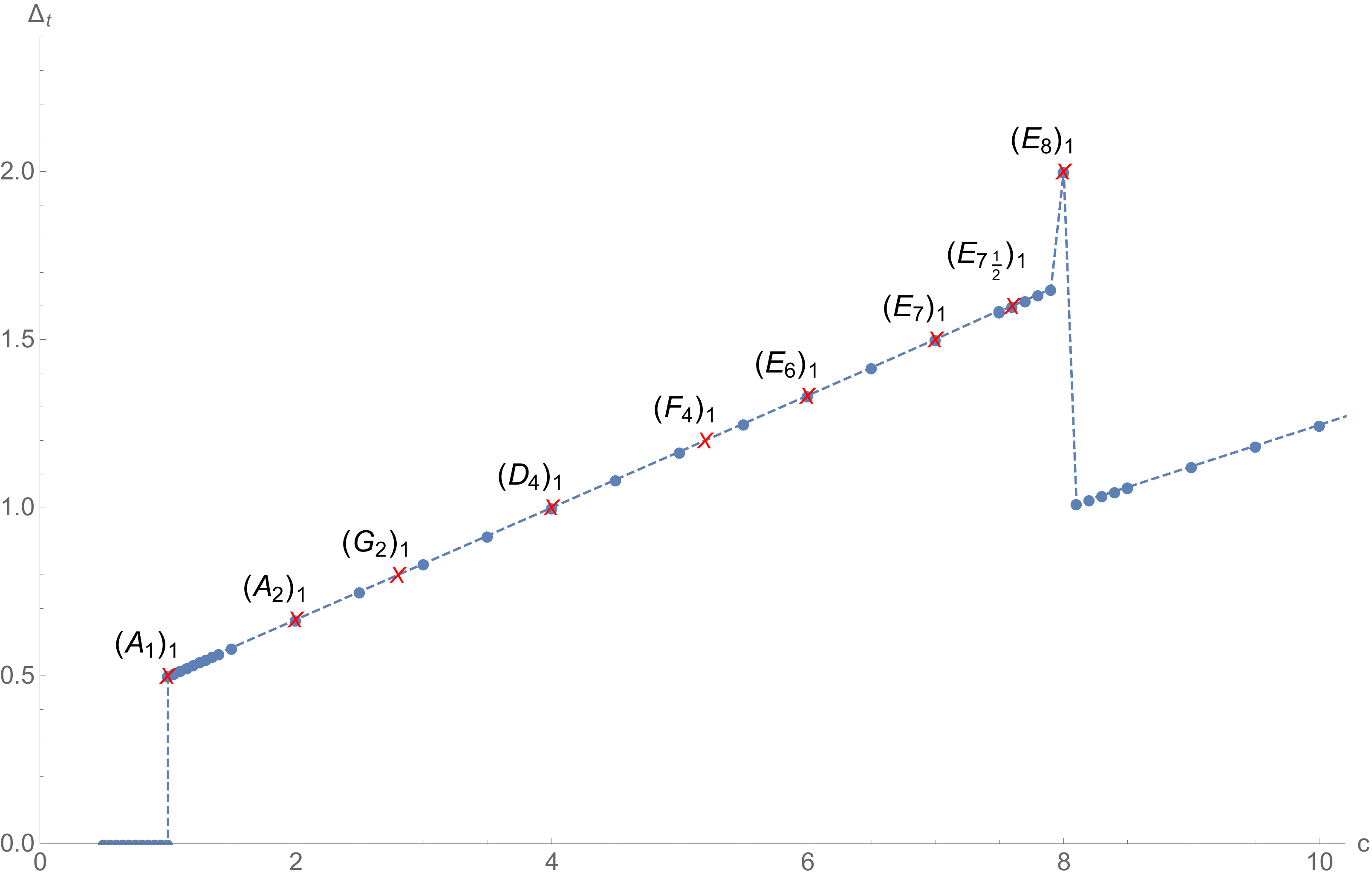}
\end{center}
\caption{Numerical bound on the twist gap $\Delta_t \equiv \Delta - |j|$ for the non-current operators in the presence of holomorphic currents.}
\label{fig:TwistgapW2}
\end{figure}
We further refine this analysis by either relaxing or imposing extra constraints on the spectrum.
Let us highlight some of our results. We first assume the existence of the holomorphic currents,
and investigate the bounds on the twist gap ($\Delta \ge \Delta_t + |j|$) for the non-current primaries
as a function of central charge $c$.
The resulting numerical upper bound exhibits rather distinctive behavior
compared to that of the upper bounds studied in \cite{Collier:2016cls}.
It is natural to ask if any known theories are realized on the numerical boundary of the twist gap.

We find that the numerical boundary realizes the level-one WZW models with $\fg=A_1, A_2, G_2, D_4, F_4, E_6, E_7, E_8$.
This set of $\fg$ agrees with the so-called Deligne's exceptional series.
They are the simplest examples of rational conformal field theories (RCFT) with extended chiral algebras
whose characters are given by the solutions of degree-two modular differential equation\cite{Mathur:1988na, Hampapura:2015cea}
with
\begin{align}
 c=1,2, \frac{14}{5}, 4, \frac{26}{5}, 6, 7,  8\ .
\end{align}
Moreover, our numerical analysis also supports a conjecture that there can exist a
two-channel RCFT with $c=\frac{38}{5}$ that has the extended chiral algebra
$\widehat{\fg}_{k=1}= (\widehat{E}_{7\frac12})_{k=1}$. We find the modular invariant partition function
consistent with this conjecture.
\begin{figure}[t]
\begin{center}
\includegraphics[width=.84\textwidth]{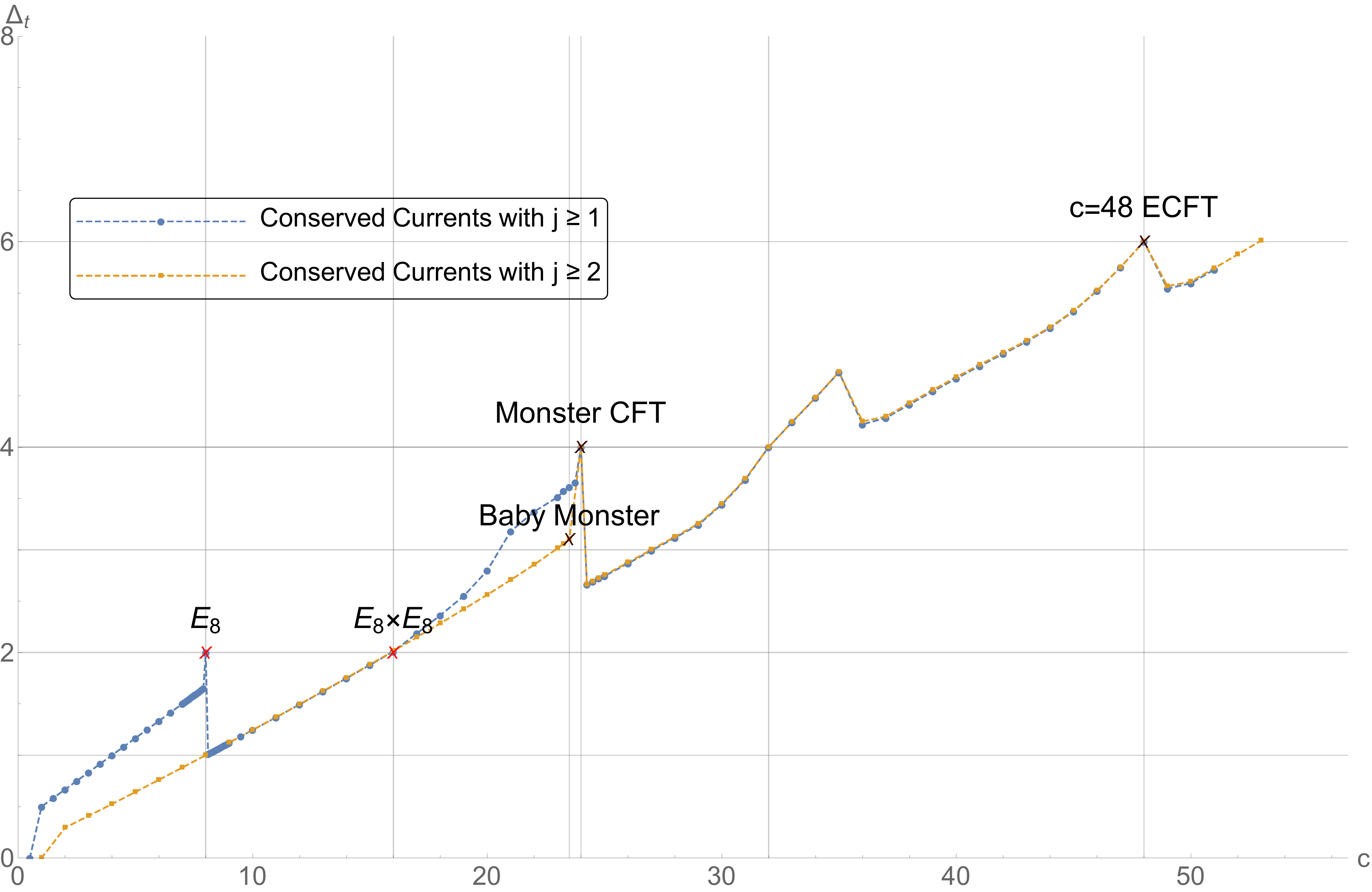}
\end{center}
\caption{Numerical bound for the twist gap, in the presence of the conserved currents of spin $1 \le j \le j_{\textrm{max}}$ and the conserved currents of spin $2 \le j \le j_{\textrm{max}}$.}
\label{W2_Fig4_largeC2}
\end{figure}
Figure \ref{W2_Fig4_largeC2} shows the upper bound on the twist gap in the the large central charge region.
We find numerous kinks and peaks on the numerical boundary. Among them,
the three points at $c=8$, $c=24$ and $c=48$ can be identified
with the $(\widehat{E_8})_1$ WZW, Monster CFT \cite{frenkel1984natural} and $c=48$ extremal CFT, respectively.
We also find that the $(\widehat{E_8})_1 \times (\widehat{E_8})_1$ WZW model can be
placed at the boundary point at $c=16$. On the other hand, one can
obtain different numerical bounds on the twist gap depicted in Figure \ref{W2_Fig4_largeC2_CC2}
when the currents are restricted to have spins of $j\ge2$.
Interestingly, we are able to shows that some RCFTs with finite group symmetry of
large degree can sit on the numerical upper bound. For instance,
the baby Monster CFT\cite{Hoehn:2007aa} is realized at the boundary point
of $c=\frac{47}{2}$.
\begin{figure}[t]
\begin{center}
\includegraphics[width=.84\textwidth]{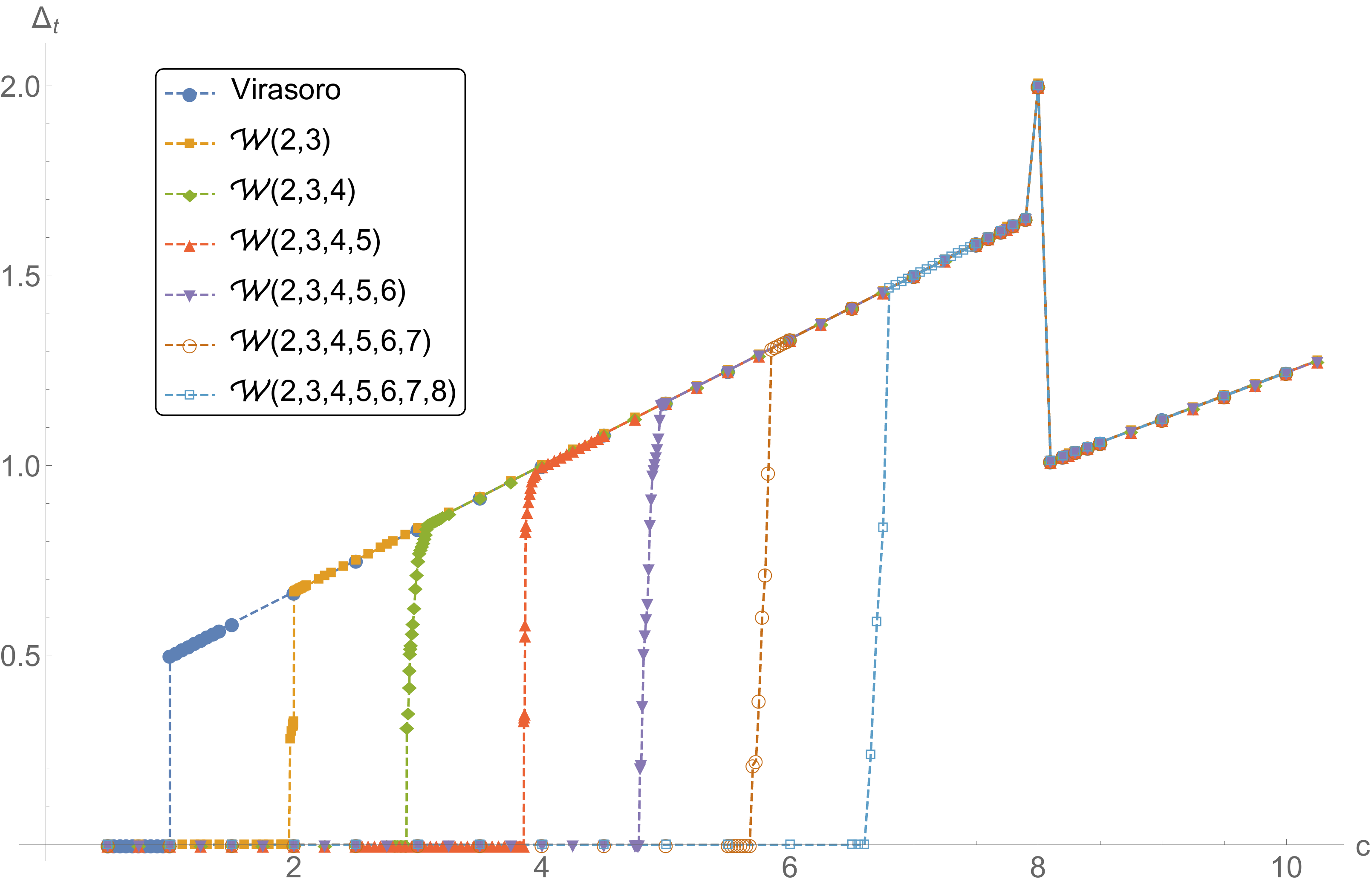}
\end{center}
\caption{Numerical bounds on the twist gap with various $\mathcal{W}$-algebras. We also assume the presence of the conserved currents of $j \ge 1$ in the spectrum.}
\label{fig:IntroWtwistgap}
\end{figure}
Finally, we impose the $\CW$-algebra symmetry to our analysis, which can be applied to constraining higher-spin gravity in AdS$_3$ \cite{Henneaux:2010xg}. This can be done by using the character for the $\mathcal{W}$-algebra instead of the Virasoro character in the equation \eqref{partition function}. We assume that there is no degenerate state besides the vacuum.
When the central charge is larger than the rank $r=\textrm{rank}(\fg)$, the numerics do not shows a significant difference with the Virasoro case.
This might be due to the fact that we are not using the full character of the $\CW$-algebra, but rather an `unrefined' one for simplicity.
However, we find that the numerical bound shows a sharp cliff at $c \sim r$, as presented in Figure \ref{fig:IntroWtwistgap}. It means there is no modular invariant partition function for $c < r$, which can be considered as a unitary bound of $\mathcal{W}_k$ algebra. Additionally, we find that the rank-three $\mathcal{W}(A_3)$ character realize the level-one $A_3$ WZW model at the numerical boundary, but not for the other choice of $\CW(\fg)$.

This paper is organized as follows. In section \ref{sec:Prelim}, we review the solutions to the degree-two and degree-three modular differential equation (MDE).
It is known that the solutions to the MDE can be identified with the vacuum and primary characters of a two-character or three-character RCFT.
We also summarize the basic aspect of the `unrefined' character of $\CW$-algebra.
In section \ref{sec:Virasoro}, we present numerical upper bounds on the scalar gap, overall gap and twist gap
for the parity preserving CFTs with or without conserved currents.
We identify 16 RCFTs that are located on the numerical boundary.
In section \ref{sec:Spec}, we provide the evidences for the identification of the 16 special points
on the numerical boundary with the known and conjectured RCFTs.
We find the modular invariant partition function of various RCFTs by assuming
every extremal spectrum saturate the bounds on the maximal degeneracies.
In section \ref{sec:Walgebra}, we repeat the numerical analysis with the character for the $\CW$-algebra.

\section{Preliminaries}                        
\label{sec:Prelim}                               

\subsection{Modular Differential Equation}

A conformal field theory with a finite number of primary operators (maybe under an
extended chiral algebra) is called a rational conformal field theory (RCFT).
From this definition, the partition function of a RCFT
can be expressed as
\begin{align}
  Z(\tau, \bar \tau) = \sum_{i,j=1}^{n} M_{ij} f_i(\tau) \bar f_j(\bar \t),
\end{align}
where $f_i(\tau)$ denote the characters of the (extended) chiral algebra including
the Virasoro algebra.

There has been an attempt to classify RCFTs using the
modular differential equation\cite{Mathur:1988na}. The main idea is to regard $n$ characters
$f_i(\tau)$ of a given RCFT as independent solutions to a modular-invariant differential
equation. One can use the covariant derivative on a modular form of weight $r$,
\begin{align}
  \CD_\tau= \partial_\t - \frac16 i\pi r E_2(\t)
\end{align}
with the second Eisenstein series $E_2(\t)$, to express the most general $n$-th order
holomorphic differential equation as
\begin{align}
  \left[ \CD_\t^n + \sum_{k=0}^{n-1} \phi_{2(n-k)}(\t) \CD_\t^k \right] f(\t) = 0 \ .
  \label{MDE01}
\end{align}
Here $\phi_{2k}(\t)$ denotes a modular form of weight $2k$ that becomes singular
at the zeros of the Wr\"onskian $W=\det W^i_{\ j}$ with $W^i_{\ j} = \CD^i_\t f_j(\t)$ and $i,j=1,\ldots, n$. It is shown in \cite{Mathur:1988na} that
the total number of zeroes of the Wr\"onskian $W$ is given by
\begin{align}
  - \sum_{i=1}^{n-1} h_i + \frac{c}{24} + \frac{n(n-1)}{12} = \frac \ell6\ ,
\end{align}
where $h_i$ denote the conformal weights of non-vacuum primary operators in a given RCFT
and $\ell$ is either $0$ or an integer greater than or equal to two.

In the present work, we are mostly interested in the modular differential equation
with $\ell=0$ where $\phi_{2k}(\t)$ of (\ref{MDE01}) can be expressed as
a sum of monomials of $E_4(\t)^a E_6(\t)^b$ with $4a+6b=2k$.
The Eisenstein series are normalized such that they can be expanded as
\begin{align}
\begin{split}
  E_4(\t) & = 1 +240 q + 2160 q^2 + 6720 q^3 + 17520 q^4   + \mathcal{O}(q^5) \ , \qquad \\
  E_6(\tau) & = 1 -504 q - 16632 q^2 - 122976 q^3 - 532728 q^4 + \mathcal{O}(q^5) \ ,
\end{split}
\end{align}
in the limit $q=e^{2\pi i\t} \to 0$.
We further demand the modular differential equation to have a solution of the form
\begin{align}
  f_\text{vac} = q^{-\frac{c}{24}} \left[ \prod_{m=2}^\infty \frac{1}{1-q^m} + \CO(q^{n-1})\right] \ ,
  \label{vac_ch_ansatz}
\end{align}
which will be identified as the vacuum character of a corresponding RCFT with the central charge $c$.
We can then determine the modular differential equations unambiguously up to fifth order. 
For instance, the modular differential equation of order two is given by
\begin{align}
  \left[ \left(q \frac{d}{dq} \right)^2 - \frac16 E_2(\t) \left(q \frac{d}{dq} \right)
  - \frac{c(c+4)}{576}  E_4(\t) \right] f(\t) = 0 \ .
  \label{MDE02}
\end{align}
And the modular differential equation of order three is given by
\begin{align}   \label{3rd MDE}
  \left[ \big(q \frac{d}{dq} \big)^3 - \frac12 E_2(\t) \big(q \frac{d}{dq} \big)^2
  + \Big\{ \frac{1}{24} E_2(\t)^2 - \a E_4(\t) \Big\}
  \big(q \frac{d}{dq} \big) + \b E_6(\t) \right] f(\t) = 0 \
\end{align}
with
\begin{align}
  \a & = \frac{152 + 80 c + 7 c^2}{5952}, \ \qquad
  \b  = \frac{c(144 + 66 c + 5c^2) }{214272}\ .
\end{align}

\subsection{Deligne's Exceptional Series and Monsters}

\paragraph{Deligne's exceptional groups and WZW models}
The ``vacuum'' character of (\ref{MDE02}) takes the following form
\begin{align}
  f_\text{vac}(\tau) = q^{-c/24} \left[ 1 + \frac{c(5c+22)}{10 -c } q + \cdots \right]\ ,
  \label{vac_ch}
\end{align}
which implies that the corresponding RCFT with central charge $c$ should contain
$\frac{c(5c+22)}{10-c }$ spin-one conserved currents. In other words,
the chiral algebra of the RCFT we are looking for could be generated by Virasoro
and also Kac-Moody algebras as long as $\frac{c(5c+22)}{10 -c}$ becomes a positive integer. When $c=-22/5$, there is no conserved-current. This gives the simplest (non-unitary) minimal model, the Yang-Lee model.

One can show that the Wess-Zumino-Witten (WZW) models for
$A_1$, $A_2$, $G_2$, $D_4$, $F_4$, $E_6$, $E_7$ and $E_8$ with level one
satisfy the above conditions; In general, the central charge of the $\widehat{\mathfrak{g}}_k$ WZW model with level $k$
is given by
\begin{align} \label{WZWc}
  c(\widehat{\mathfrak{g}}_k) = \frac{k \dim \mathfrak{g}}{k+h^\vee}\ ,
\end{align}
where $h^\vee$ denotes the dual Coxeter number of $\mathfrak{g}$.
It is straightforward to show that
\begin{align}
  \dim \mathfrak{g} = \frac{c(\widehat{\mathfrak{g}}_{1}) \left(5c(\widehat{\mathfrak{g}}_{1}) + 22\right)}
  {10 -c(\widehat{\mathfrak{g}}_{1})}\ ,
\end{align}
when $\mathfrak{g}=$ $A_1$, $A_2$, $G_2$, $D_4$, $F_4$, $E_6$, $E_7$ and $E_8$.
These groups are often referred as the Deligne's exceptional series.

Indeed, $f_\text{vac}(\t)$ of (\ref{vac_ch}) at $c=c(\widehat{\mathfrak{g}}_{1})$ is the vacuum character
of the WZW model for Deligne's exceptional groups with level one. Moreover, one can identify the other
solution to (\ref{MDE02}) with $c=c(\widehat{\mathfrak{g}}_{1})$\footnote{Two independent solutions $f_\text{vac}(\t)$ and $f_{h}(\tau)$ can be expressed as hypergeometric series\cite{Mathur:1988gt}.},
\begin{align}
  f_{h}(\tau) = a_0 q^{h - \frac{c}{24}}  \left[ 1 +
  \frac{(5c-2)(c+4)}{c+14} q
  + \cdots \right] \text{ with } h = \frac{2+ c}{12}\ ,
  \label{primary_ch}
\end{align}
as the characters associated with the primary operators of the corresponding theory;
Here $a_0$ is a constant which is not determined by the modular differential equation.
The $\widehat{\mathfrak{g}}_1$ WZW model with level one has primary operators in the
dominant highest-weight representations of the affine algebra $\widehat{\mathfrak{g}}$, i.e.,
\begin{align}
  k = 1 \geq ( \l , \theta)_{\widehat{\mathfrak{g}}}\ ,
\end{align}
where $\theta$ is the highest root of the Lie algebra $\mathfrak{g}$ (See Table \ref{LG data 1}).
Their conformal weights are
\begin{align} \label{WZWh}
  h_\l(\mathfrak{g}_k) = \frac{(\l, \l +2 \r)}{2(k+h^\vee)}\ ,
\end{align}
where $\r$ denote the Weyl vector.  The conformal weight $h_\lambda $ coincides with $\frac{1}{12}\big(2+c(\widehat{\mathfrak{g}}_{1})\big)$
for the $\widehat{\fg}_1$ in the Deligne's exceptional series except for $E_8$.
\begin{table}[t]
\centering
{%
\begin{tabular}{c |c | c | c | c| c }
  \hline
   \rule{0in}{2.5ex} $\mathfrak{g}$ & $h^\vee$ & $\dim \mathfrak{g}$ & $c(\widehat{\mathfrak{g}}_{1})$ &
  dominant highest-weight rep.
  & $h_\l(\widehat{\mathfrak{g}}_{1})$
   \\ [0.7ex] \hline
  \rule{0in}{3ex}$A_1$ & $2$ & $3$ & $1$ & $[0;1]$ & $\frac14$ \\ [0.3ex]
  \rule{0in}{1.5ex}$A_2$ & $3$ & $8$ & $2$ & $[0;1,0]$, $[0;0,1]$  & $\frac13$ \\ [0.3ex]
  \rule{0in}{1.5ex}$G_2$ & $4$ & $14$ & $\frac{14}{5}$ & $[0;1,0]$ & $\frac25$ \\ [0.3ex]
  \rule{0in}{1.5ex}$D_4$ & $6$ & $28$ & $4$ & $[0;1,0,0,0]$, $[0;0,0,1,0]$, $[0;0,0,0,1]$ & $\frac12$ \\ [0.3ex]
  \rule{0in}{1.5ex}$F_4$ & $9$ & $52$ & $\frac{26}{5}$ & $[0;0,0,0,1]$ & $\frac35$ \\ [0.3ex]
  \rule{0in}{1.5ex}$E_6$ & $12$ & $78$ & $6$ & $[0;1,0,0,0,0,0]$, $[0;0,0,0,0,1,0]$ & $\frac23$ \\ [0.3ex]
  \rule{0in}{1.5ex}$E_7$ & $18$ & $133$ & $7$ & $[0;0,0,0,0,0,1,0]$ & $\frac34$ \\ [0.3ex]
  \rule{0in}{1.5ex}$E_8$ & $30$ & $248$ & $8$ & none &
\end{tabular}%
\caption{Dual Coxeter numbers, dimensions, central charges $c(\mathfrak{g}_{k=1})$,
dominant affine weights of $\widehat{\mathfrak{g}}_{k=1}$ other than the basic weight $[1;0,0,..,0]$,
and conformal weights of corresponding primary operators.}
\label{LG data 1}
}
\end{table}
One can also show that, when $a_0=\dim\l$, all the coefficients of the $q$-expansion
of $f_h(\t)$ (\ref{primary_ch}) with $c=c(\widehat{\mathfrak{g}}_{1})$ become positive integers
that agree with those of the characters associated with the primary operators of weight
$h=h_\l(\widehat{\mathfrak{g}}_{1})$. For example, for $c=4$ (when we turn off the chemical potentials for the flavor charges),
\begin{align}
\begin{split}
  f_{h=1/2}(\t) &= \chi_{[0;1,0,0,0]}(\t) = \chi_{[0;0,0,1,0]}(\t) = \chi_{[0;0,0,0,1]}(\t)  \\
                &= 8 + 64 q + 288 q^2 + 1024 q^3 + 3152 q^4 + 8704 q^5 + \cdots
\end{split}
\end{align}
where $\chi_{[0;\l]}$ are the characters associated with
the primary operators with the highest-weights $\l$
that are present in the $\widehat{so}(8)_1$ WZW model.
The three representations are mapped to each other via triality of $so(8)$.

For later convenience, let us also discuss the modular-invariant partition
function of the WZW models with level one.
For the Deligne exceptional groups,
it is shown in \cite{Gannon:1992nq} that the torus partition functions (in the limit of zero chemical potential for the global currents) are
\begin{align}
  Z_{\widehat{\mathfrak{g}}_{1}}(\t,\bar \t) = f_\text{vac}(\t) \bar{f}_\text{vac}(\bar \t) +
  N\left( \widehat{\mathfrak{g}}_{1} \right) f_{h}(\t) \bar{f}_h( \bar \t)\ ,
  \label{modular inv ptn}
\end{align}
where $N\left(\widehat{\mathfrak{g}}_{1}\right)$ denotes the number of dominant highest-weight representations of $\widehat{\mathfrak{g}}_{1}$
other than the basic representation. For instance, the partition function of $\widehat{so}(8)_1$ WZW model can be expressed as
\begin{align}
  Z_{\widehat{so}(8)_1}(\t,\bar \t) = f_\text{vac}(\t) {\bar f}_\text{vac}(\bar \t) +
  3 f_{h=1/2}(\t) \bar f_{\bar h=1/2}(\bar \t)
\end{align}
with $c=4$.

\paragraph{Monster and its cousins}

The modular differential equation of order three (\ref{3rd MDE}) with the ansatz (\ref{vac_ch_ansatz})
has been studied relatively recently in order to explore unitary RCFTs
that has no Kac-Moody symmetry (but may have some extended chiral algebras) \cite{Tuite:2008pt,Hampapura:2016mmz}.
Let us first discuss the possible values of the central charge
for which RCFTs without the Kac-Moody symmetry may exist.
The ``vacuum character'' of (\ref{3rd MDE}) is given by
\begin{align}
\begin{split}
   f_\text{vac}(\t) & = q^{-\frac{c}{24}} \bigg[ 1 +
   \frac{c\left(70 c^2+955 c+2388\right)}{2 ((c-55) c+748)} q^2
  \\ & +
   \frac{5530 c^5+114000 c^4+919648 c^3+3949824 c^2+5656576 c}{3 (c^2-86c+1864) (c^2-55c+748)}  q^3
   +\cdots \bigg]\ .
   \label{vac_ch_3MDE}
\end{split}
\end{align}
It is not difficult to show that the requirement to have positive integer coefficients
of the $q$-expansion of $f_\text{vac}(\t)$ (up to the $\mathcal{O}(q^{400})$)
can restrict the central charge $c$ to nine possible values\cite{Tuite:2008pt,Hampapura:2016mmz},
\begin{align}
  c = -\frac{44}{5}, 8 , 16, \frac{47}{2}, 24, 32, \frac{164}{5}, \frac{236}{7}, 40\ .
\end{align}

One can also identify the other two solutions to (\ref{3rd MDE}) as two different
characters associated with primary operators of conformal weights $h=h_\pm(c)$,
\begin{align}
  f_{h_\pm}(\t) = q^{h_\pm - \frac{c}{24}} \Big[ a_0 + \CO(q) \Big]
  \label{pri_ch_3MDE}
\end{align}
with
\begin{align}
  h_\pm(c) = \frac{c+4}{16} \pm \frac{\sqrt{(24-c)c + 368}}{16\sqrt{31}}\ ,
  \label{hpm}
\end{align}
in the corresponding would-be RCFT with central charge $c$.
It is clear from (\ref{hpm}) that the above two ``primary operators'' of weights $h_\pm(c)$
would not exist for $c\geq12 + 16\sqrt2$.
This suggests that the RCFT of our interest with $c=40$ can correspond to the single-character RCFT. 

The vacuum character of a single-character theory has to be of the form\cite{Mathur:1988na}
\begin{align}
  j(\t)^\a \big(j(\t)-1728\big)^\b P_k\big( j(\t)\big)\ ,
\end{align}
where $\a=0,1/3,2/3$ and $\b=0,1/2$ while $P_k(j(\t))$ denotes a polynomial of order $k$
in the $j$-function with integer coefficients\footnote{
Here the $j$-function is normalized as follows
\begin{align}
  j(\t) = \frac{\left(12 E_4(\t)\right)^3}{E_4(\t)^3 - E_6(\t)^2}\ .
  \label{j inv}
\end{align}
}.
One can indeed show that $f_\text{vac}(\t)$ (\ref{vac_ch_3MDE}) for $c=40$ can be expressed
in terms of the $j$-function,
\begin{align}
\begin{split}
  f_\text{vac}(\t) & = q^{-\frac{40}{24}} \left[ 1 + 20620q^2 + 86666240q^3 + \cdots \right]
  \\ & = j^{\frac{2}{3}}(\t) \Big[ j(\t) - 1240 \Big]\ .
\end{split}
\end{align}

In summary, it is plausible from (\ref{3rd MDE}) with (\ref{vac_ch_ansatz}) that
unitary RCFTs without Kac-Moody algebra can exist for $c=8$, $16$, $47/2$, $24$, $32$, $164/5$, $236/7$ and $40$. (See Table \ref{LG data 2})
Some of their vacuum characters (\ref{vac_ch_3MDE}) can be identified as
the Monster module ($c=24$) \cite{Frenkel:1988xz} and the H\"ohn Baby Monster module ($c=47/2$)\cite{Hoehn:2007aa}.
Furthermore, the vacuum characters for $c=8,16$ can be realized as a certain fixed-point free lattice
for the rank $8$ even lattice and the rank $16$ Barnes-Wall even lattice whose automorphism
groups are related to $O_{10}^+(2).2$ and $O_{10}^+(2).2^{16}$ respectively.
The extended chiral algebras for $c=32,\frac{164}{5},\frac{236}{7},40$ are however not much known and need further investigation.
Moreover, beyond the Monster CFT,
some of their modular-invariant partition functions have been poorly understood.
\begin{table}[t]
\centering
{%
\begin{tabular}{c |c | c | c }
  \hline
  \rule{0in}{2.5ex}$c$ & \multicolumn{1}{c|}{$f_\text{vac}(\t)$} & \multicolumn{1}{c|}{$f_{h_\pm}(\t)$} & automorphism group
  \\ [1ex]\hline
  \rule{0in}{5ex}$8$ & $q^{-\frac{8}{24}} \left[ 1 + 156 q^2 +\cdots \right]$ &
  $\left\{\begin{tabular}{@{\ }l@{}}
    $q^{\frac12 -\frac{8}{24}} a_0 \left[ 1 + 36q + \cdots \right]$ \\ 
    $q^{1 -\frac{8}{24}} b_0 \left[ 1 + 16q + \cdots \right]$
  \end{tabular}\right.$ & $O_{10}^+(2).2$
  \\ [2.0ex]
  \rule{0in}{5ex}$16$ & $q^{-\frac{16}{24}} \left[ 1 + 2296 q^2 +\cdots \right]$ &
  $\left\{\begin{tabular}{@{\ }l@{}}
  $q^{1 -\frac{16}{24}} a_0 \left[ 1 + 136q + \cdots \right]$ \\
  $q^{\frac32 -\frac{16}{24}} b_0 \left[ 1 + 52q + \cdots \right]$ 
  \end{tabular}\right.$ & $O_{10}^+(2).2^{16}$
  \\ [0.5ex]
  \rule{0in}{5ex}$\frac{47}{2}$ & $q^{-\frac{47}{48}} \left[ 1 + 96256 q^2 +\cdots \right]$ &
  $\left\{\begin{tabular}{@{\ }l@{}}
  $q^{\frac32 -\frac{47}{48}} a_0 \left[ 3 + 785 q + \cdots \right]$ \\
  $q^{\frac{31}{16} -\frac{47}{48}} b_0 \left[ 47 + 5177 q + \cdots \right]$
  \end{tabular}\right.$ & Baby Monster
  \\ [0.5ex]
  \rule{0in}{2.5ex}$24$ & $q^{-\frac{24}{24}} \left[ 1 + 196884 q^2 +\cdots \right]$ &
  \multicolumn{1}{c|}{none} & Monster
  \\ [0.5ex]
  \rule{0in}{2.5ex}$32$ & $q^{-\frac{32}{24}} \left[ 1 + 139504 q^2 +\cdots \right]$ &
  \multicolumn{1}{c|}{none} & unknown
  \\ [0.5ex]
  \rule{0in}{5ex}$\frac{164}{5}$ & $q^{-\frac{41}{30}} \left[ 1 + 90118q^2 + \cdots \right] $ &  
  $\left\{\begin{tabular}{@{\ }l@{}}
  $q^{\frac{11}{5}-\frac{41}{30}} [248 + 90365 q + \cdots]$ \\
  $q^{\frac{12}{5}-\frac{41}{30}} [484 + 120032q + \cdots]$
  \end{tabular}\right.$
   & unknown
  \\ [0.5ex]
  \rule{0in}{5ex}$\frac{236}{7}$ & $q^{-\frac{59}{42}} \left[ 1 + 63366q^2 + \cdots \right]$ & 
  $\left\{\begin{tabular}{@{\ }l@{}}
  $q^{\frac{16}{7}-\frac{59}{42}} [391 + 140896 q + \cdots]$ \\
  $q^{\frac{17}{7}-\frac{59}{42}} [7192 + 1971507q + \cdots]$
  \end{tabular}\right.$
   & unknown
  \\ [0.5ex]
  \rule{0in}{2.5ex}$40$ & $q^{-\frac{40}{24}} \left[ 1 + 20620 q^2 +\cdots \right]$ &
  \multicolumn{1}{c|}{none} & unknown
\end{tabular}%
\caption{Central charge, characters with unfixed integers $a_0$ and $b_0$, and automorphism group of eight possible rational conformal field theories.}
\label{LG data 2}
}
\end{table}

We will discuss in the next section that the above RCFTs, both of the WZW models with level one
for Deligne's exceptional series and the Monster CFT and its cousins, are realized
on the numerical bounds of the twist gap.

\subsection{Character of $\mathcal{W}$-algebra}

$\CW$-algebra is an extension of the Virasoro algebra, augmented with generators with higher spins $s \ge 2$. In section \ref{sec:Walgebra}, we will investigate consequences of the $\CW$-algebra symmetry in the modular invariant partition function. We label the $\CW$-algebras by the spin (or dimension) of the generators as $\CW(d_1, d_2, \ldots, d_r)$, where we call $r$ as the rank of the algebra. For the $\CW(\fg)$-algebra associated to a Lie algebra $\fg$, the spin of the generators agree with the degrees of the Casmirs of $\fg$. In this section, we briefly review necessary aspects of the $\CW$-algebra for our numerical bootstrap program. See \cite{Bouwknegt:1992wg} for more details.

\paragraph{$\CW(2, 3)$-algebra}
The first example of $\CW$ algebra is the $\CW(A_2) = \mathcal{W}(2,3)$ \cite{Zamolodchikov:1985wn}. It has two generators, one comes from the stress tensor $T(z)$, and the spin-$3$ generator $W(z)$, which upon mode expansion
\begin{align}
 T(z) = \sum_{n \in \IZ} L_n z^{-n-2} \ , \qquad W(z) = \sum_{n \in \IZ} W_n z^{-n-3} \ .
\end{align}
The commutation relations between generators are given by
\begin{align}
\begin{split}
[L_m, L_n] &= (m-n) L_{m+n} + \frac{c}{12} m(m^2-1)\delta_{m+n,0}  \\
[L_m, W_n] &= (2m-n) W_{m+n}  \\
[{W}_m, {W}_n] &= (m-n) \left[ \frac{1}{15}(m+n+3)(m+n+2)-\frac{1}{6}(m+2)(n+2)\right] L_{m+n}  \\
           &\quad + \frac{16}{22+5c} (m-n) \Lambda_{m+n} + \frac{c}{360} m (m^2-1)(m^2-4)\delta_{m+n,0}
\end{split}
\end{align}
where the operator $\Lambda_n$ is
\begin{align}
\Lambda_n = \sum_{p \le -2} L_p L_{n-p} + \sum_{p \ge -1} L_{n-p} L_p - \frac{3}{10}(n+2)(n+3) L_n.
\end{align}
The highest weight states $\ket{h, w; c}$ are labeled by $L_0$ eigenvalue $h$ and $\mathcal{W}_0$ eigenvalue $w$ and also by the central charge $c$. The Verma module is generated by acting negative modes $L_{-m}$ and $\mathcal{W}_{-m}$ to the highest weight state. When the central charge $c$ and the highest weights $(h, w)$ satisfy some relation, the Verma module might carry null states that we need to mod out to form a faithful representation. For a generic value of $c, h, w$, the Verma module does not have a null state, which is the scenario we are mostly interested in.

One can define the character for a given representation $\CV_{h, w}^c$ as
\begin{align}
 \chi_{(h, w; c)}(\tau; p) = \tr_{\CV_{h, w}^c} \left( q^{L_0 - \frac{c}{24}} p^{W_0} \right) \ .
\end{align}
This quantity turns out to be rather challenging to compute, even for the case of the Verma module. This has to do with the fact that one need to simultaneously diagonalize vectors in $\CV_{h, w}$ with respect to $L_0$ and $W_0$. See \cite{Iles:2013jha,Iles:2014gra} for a recent development on this issue. Due to its computational difficulty, we will focus on the `unrefined' character, which sets $p=1$. The unrefined character turns out to be very simple, just given by
\begin{align}
 \chi_{(h, w;c)} (\t) = \tr_{\CV^c_{h, w}} q^{L_0 - \frac{c}{24}} = q^{h-\frac{c}{24}} \prod_{n \ge 1} \frac{1}{(1-q^n)^2} = \frac{q^{h-\frac{c-2}{24}}}{\eta(\t)^2}\ ,
\end{align}
for a generic representation. For the vacuum module, we simply get
\begin{align}
\chi_0(\tau) = q^{-\frac{c}{24}} \prod_{n=2}^{\infty} \frac{1}{(1-q^n)} \prod_{n=3}^{\infty} \frac{1}{(1-q^n)} = \frac{q^{-\frac{c-2}{24}}(1-q)^2(1-q^2)}{\eta(\tau)^2}
\label{Vaccume ch}
\end{align}
from the following null states:
\begin{align}
\big< 0 \big| L_1 L_{-1} \big| 0 \big> = 0, \quad
\big< 0 \big| W_1 W_{-1} \big| 0 \big> = 0, \quad
\big< 0 \big| W_2 W_{-2} \big| 0 \big> = 0.
\end{align}

\paragraph{$\CW$-algebra associated to a Lie algebra}
The most straight-forward way of constructing a $\CW$-algebra is to start with generators of dimensions $d_1, \ldots d_r$ and then try to fix various structure constants by imposing Jacobi identity. This way is notoriously difficult to perform in practice, which were done only up to 3 generators \cite{Blumenhagen:1990jv}. Instead, more systematic approach is available.

Start with an affine Kac-Moody algebra $\widehat{\mathfrak{g}}_k$. From here, one can obtain an associated $\CW$-algebra via quantum Drinfeld-Sokolov reduction \cite{Drinfeld:1984qv, Bershadsky:1989mf, Feigin:1990pn}. This $\CW$-algebra $\CW(\widehat{\mathfrak{g}}_k, \Lambda)$ is labelled by the choice of an $\mathfrak{su}(2)$ embedding $\Lambda: \mathfrak{su}(2) \hookrightarrow \mathfrak{g}$, where $\mathfrak{g}$ is the finite part of the affine Kac-Moody algebra. When $\fg = \mathfrak{su}(N)$, the $\mathfrak{su}(2)$ embedding is classified in terms of partitions of $N$ or Young tableaux of $N$ boxes. This choice determines how the Virasoro algebra is realized in affine Kac-Moody algebra.

The choice of $\Lambda$ determines the degrees of the $\CW$-algebra generators. It goes as follows. One can decompose the adjoint representation of $\fg$ into the $\mathfrak{su}(2)$ representations as
\begin{align}
 \textrm{adj} (\fg) = \bigoplus_{j} V_j \ ,
\end{align}
where $V_j$ denotes $2j+1$ dimensional spin-$j$ representation under $\mathfrak{su}(2)$.
Once we have the decomposition as above, the $\CW(\fg, \Lambda)$-algebra associated to the affine Lie algebra $\widehat{\fg}$ and the $\mathfrak{su}(2)$ embedding $\Lambda$ will have the generators of spins (dimensions) given by $j+1$.
For example, let us choose $\Lambda$ to be given by the principal embedding $\Lambda_{pr}$. Then the adjoint representation decomposes into $\fg = \oplus_{i=1}^r V_{d_i-1}$ where $r$ is the rank of $\fg$ and $d_i$ are given by the degrees of the Casimir operators. Sometimes we denote the $\CW$-algebra given by the principal embedding as $\CW(\fg) \equiv \CW(\fg, \Lambda_{pr})$. Our main focus in section \ref{sec:Walgebra} will be the modular constraint coming from $\CW(\fg)$-algebra.

For the $\CW(d_1, \ldots, d_r)$-algebra, we have generators with the mode expansions given as
\begin{align}
 W^{(d_i)} (z) = \sum_{n \in \IZ} W^{(d_i)}_n z^{-n-d_i} \ .
\end{align}
Verma module is simply generated by acting negative modes of the generators $W^{(d_i)}_{-n}$ on the highest-weight state. Therefore, the reduced character for a generic module (that does not have any null state) is simply given as
\begin{align} \label{Wpri}
 \chi_h (\t) = q^{h-\frac{c}{24}} \prod_{n \ge 1} \frac{1}{(1-q^n)^r} = \frac{q^{h-\frac{c-r}{24}}}{\eta(\t)^r} \ ,
\end{align}
where we omitted the dependence on the weights for the generators except for the stress-energy tensor.
The vacuum states are defined to be annihilated by all the generators of mode number greater than equal to zero. In addition, there are null states at level $1, 2, \ldots, d-1$ of the form
$ W^{(d)}_{-n_1} \ldots W^{(d)}_{-n_k} \ket{0}$ with $1 \le \sum_k n_k \le d-1$.
Now, we can write the vacuum character for general rank-$r$ $\mathcal{W}(d_1, d_2, \ldots, d_r)$-algebra as
\begin{align} \label{Wvac}
\chi_0(\t) = q^{-\frac{c}{24}} \prod_{i=1}^r \prod_{n=d_i}^{\infty} \frac{1}{(1-q^n)} =
\frac{q^{-\frac{c-r}{24}}}{\eta(\t)^{r}} \prod_{i=1}^r \prod_{j=1}^{d_i-1}(1-q^j) \ .
\end{align}

Now, the partition function of a CFT with $\CW(d_1, \ldots, d_r)$-algebra symmetry should be written in terms of $\CW$-algebra characters, instead of Virasoro characters. For the computational convenience, we mainly focus on the reduced partition function given as
\begin{align}
\widehat{Z}(\tau,\bar{\tau}) = |\tau|^{\frac{r}{2}} |\eta(\tau)|^{2r} Z(\tau,\bar{\tau}).
\label{reduced partition function}
\end{align}
Note that $ |\tau|^{\frac{1}{2}} |\eta(\tau)|^{2} $ is modular invariant, therefore the reduced partition function is also invariant under $\tau \rightarrow - \frac{1}{\tau}$.

\section{Modular Constraint with Virasoro algebra}  
\label{sec:Virasoro}                                   

\subsection{The Modular Bootstrap Equation}

The torus partition function of a two dimensional compact (bosonic) CFT
can be defined as
\begin{align}
  Z(\t,\bar \t)
  = \text{Tr}_{\CH(S^1)}\left[
  q^{L_0-\frac{c_L}{24}} \bar q^{\bar L_0-\frac{c_R}{24}}\right]
  \text{ with } q = e^{2\pi i \t}\ ,
  \label{partition_def}
\end{align}
where $\t$ parametrizes the complex structure of the torus,
and trace is taken over the states of a given CFT on a unit circle.
We focus on CFTs having parity-invariant spectrum and thus free
from the gravitational anomaly, i.e., $c_L=c_R$ in what follows.

One can decompose the partition function $Z(\t,\bar \t)$ of a given
parity-preserving CFT in terms of the Virasoro characters as
\begin{align}   \label{decomposition_Vir}
\begin{split}
  Z(\t,\bar \t) & =  \chi_0(\t) \bar{\chi}_0(\bar \t) + \sum_{h,\bar h}
  d_{h,\bar h} \Big[ \chi_h(\t) \bar{\chi}_{\bar h}(\bar \t) +
  \chi_{\bar h}(\t) \bar{\chi}_{h}(\bar \t) \Big]
  \\ & + \sum_{j=1} d_j \Big[ \chi_j(\t) \bar{\chi}_{0}(\bar \t) +
  \chi_0(\t) \bar{\chi}_{j}(\bar \t) \Big]\ ,
\end{split}
\end{align}
where $\chi_h(\t)$ denotes the Virasoro character for the highest-weight representation with weight $h$.
The vacuum and non-vacuum characters take the forms
\begin{align}
\begin{split}
  \chi_0 (\t) & = q^{-\frac{c}{24}} \prod_{n=2}^\infty \frac{1}{1-q^n} \ ,
   \\
  \chi_{h>0} (\t) & = q^{h-\frac{c}{24}} \prod_{n=1}^\infty \frac{1}{1-q^n}\ .
\end{split}
\end{align}

Unless a given CFT suffers from the large diffeomorphism anomaly,
the torus partition function (\ref{partition_def}) has to be invariant
under the modular transformation $SL(2,\mathbb{Z})$ generated by $T$ and $S$,
\begin{align}
  T : \t \to \t + 1 \ , \qquad
  S : \t \to -\frac{1}{\t}\ .
\end{align}
Invariance of the partition function under the $T$-transformation
requires that all states to carry integer spins, i.e., $j=|h-\bar h| \in \mathbb{Z}_{\ge 0}$.
The invariance under the $S$-transformation implies that
the spectrum $d_{h,\bar h}$ and $d_j$ are further constrained to
satisfy
\begin{align}
  0 = \CZ_\text{vac} (\t,\bar \t) + \sum_{h,\bar h}d_{h,\bar h} \Big[
  \CZ_{h,\bar h}(\t,\bar\t) + {\bar \CZ}_{h,\bar h}(\t,\bar \t) \Big]
  + \sum_{j=1} d_j \Big[ \CZ_j (\t,\bar \t) + {\bar \CZ}_j(\t,\bar \t) \Big] ,
  \label{condition01}
\end{align}
where
\begin{align}
  \CZ_\text{vac}\big( \t,\bar \t \big) & = \chi_0\big( \t \big)\bar \chi_0\big( \bar \t \big) -
  \chi_0\big(-\frac{1}{\t}\big) \bar \chi_0\big(-\frac{1}{\bar \t} \big)\ ,
  \nonumber \\
  \CZ_{h,\bar h}\big( \t,\bar \t \big) & = \chi_h \big( \t \big)\bar \chi_{\bar h} \big(\bar \t \big) -
  \chi_h \big( -\frac{1}{\t} \big)\bar \chi_{\bar h} \big(- \frac{1}{\bar \t} \big)\ ,
  \\
  \CZ_j\big( \t,\bar \t \big) & = \chi_j \big( \t \big)\bar \chi_{0} \big(\bar \t \big) -
  \chi_j \big( -\frac{1}{\t} \big)\bar \chi_{0} \big(- \frac{1}{\bar \t} \big) \ .
  \nonumber
\end{align}

Our goal in this paper is to study the consequences of the equation \eqref{condition01} on the spectrum of operators.
It is in general very difficult to solve the constraint equation (\ref{condition01}) analytically.
However the numerical method of semi-definite programming (SDP)
has been playing a key role in studying the feasibility of 
(\ref{condition01}).
The procedure goes as follows: First, make a hypothesis on the CFT spectrum.
Second, search for a linear functional $\a$ satisfying the conditions below
\begin{align}
  \a\Big[ \CZ_\text{vac}(\t,\bar \t) \Big] =1 \ , \qquad \a\Big[ \CZ_{h,\bar h} (\t,\bar \t)\Big]\geq 0\ ,
  \qquad \a \Big[ \CZ_{j}(\t,\bar \t) \Big] \geq 0\
  \label{SDP01}
\end{align}
for $(h,\bar h)$ subject to the hypothesis. If such an $\a$ exists, the non-negativity of $d_{h,\bar h}$ and
$d_j$ implies that
\begin{align}
  \a\Big[ \CZ_\text{vac}(\t,\bar \t) \Big] + \sum_{h,\bar h} d_{h,\bar h}
  \a \Big[ \CZ_{h,\bar h} (\t,\bar \t) + {\bar \CZ}_{h,\bar h} (\t,\bar \t)  \Big]
  + \sum_{j=1} d_j \a \Big[ \CZ_{j}(\t,\bar \t) + {\bar \CZ}_{j}(\t,\bar \t) \Big] > 0\ .
  \nonumber
\end{align}
We thus find a contradiction that (\ref{condition01}) cannot be satisfied, and
the hypothetical CFT spectrum is ruled out.

In this section, we investigate the numerical upper bounds on the
so-called {\it scalar gap}, {\it overall gap} and {\it twist gap}
defined below for parity-preserving CFTs with and without conserved currents:
\begin{enumerate}
\item {\bf scalar gap problem}: We impose a gap $\D_s$ in the spectrum of scalar primaries. In other words,
the spectrum of a hypothetical CFT are constrained to have scalar primaries with
conformal dimensions $\D\geq \D_s$ while the unitary bound $\D \geq j$
is satisfied for other primaries with spin $j$.

\item {\bf overall gap problem}: There is a gap $\text{max}(\D_o,j)$ in the conformal dimensions
of all non-degenerate primaries. The conformal dimension $\D$ of non-degenerate primaries of spin $j$
are required to satisfy  $\D\geq \text{max}(\D_o,j)$. However, we do not impose the above gap
on the the conserved currents. If $\D_o \leq 1$, this condition is identical
to the scalar gap problem. As the maximal gap $\D_o$ grows, one can expect
to have upper bounds on $\D_o$ different to those of the scalar gap problem.

\item {\bf twist gap problem}: We also study the universal gap $\D_t$
on the twist $t$ defined by $t=\D-j=2 \ \text{min}(h,\bar h)$. The gap is again relaxed for
the conserved currents. A putative CFT are thus constrained to have non-degenerate primaries
with conformal dimensions $\D\geq j+\D_t$ only. Among the three problems, the twist gap problem is expected to show the most stringent upper bound.
\end{enumerate}

It is convenient to use a linear functional $\a$ of the form
\begin{align}
  \a = \sum_{p=0}^N \sum_{m+n=2p+1}
  \left. \left( \t \frac{\partial}{\partial \t} \right)^m
  \left( \bar \t \frac{\partial}{\partial \bar \t} \right)^n
  \right|_{\t=i,\bar \t=-i}
\end{align}
to implement the semi-definite programming problem (\ref{SDP01}).
The spins of primaries are in practice truncated up to
$j \leq j_\text{max}$ where $j_\text{max}$ is carefully chosen such that
the numerical bounds for $\D_s$, $\D_o$ and $\D_t$ are
well-stabilized for a given derivative order $N_{\textrm{max}}=2N+1$. In the present work,
the default value for the derivative order is $41$, but
can be enhanced occasionally up to $81$ if necessary.
In order to make the numerical analysis simple,
we solve a different but equivalent SDP problem (\ref{SDP01}) with
the partition function and characters multiplied by
certain modular-invariant factors to get
\begin{align}
  \widehat Z(\t,\bar t) & = \Big| \t^{\frac 14} \eta(\t) \Big|^2 Z(\t,\bar \t)\ ,
  \nonumber \\
  \widehat \chi(\t) & = \left. \t^{\frac 14} \eta(\t) \chi(\t) \right.\ .
\end{align}
Here $\widehat Z(\t,\bar \t)$ and $\widehat\chi(\t)$
are often referred to as the reduced partition function and reduced characters respectively \cite{Friedan:2013cba,Collier:2016cls}.

\subsection{Numerical Bounds on Spectrum Gap}
%
\begin{figure}[h!]
\begin{center}
\subfigure[Bounds in the region $1 \le c \le 10$. We zoom near the region around $c=8$.]{\includegraphics[width=.84\textwidth]{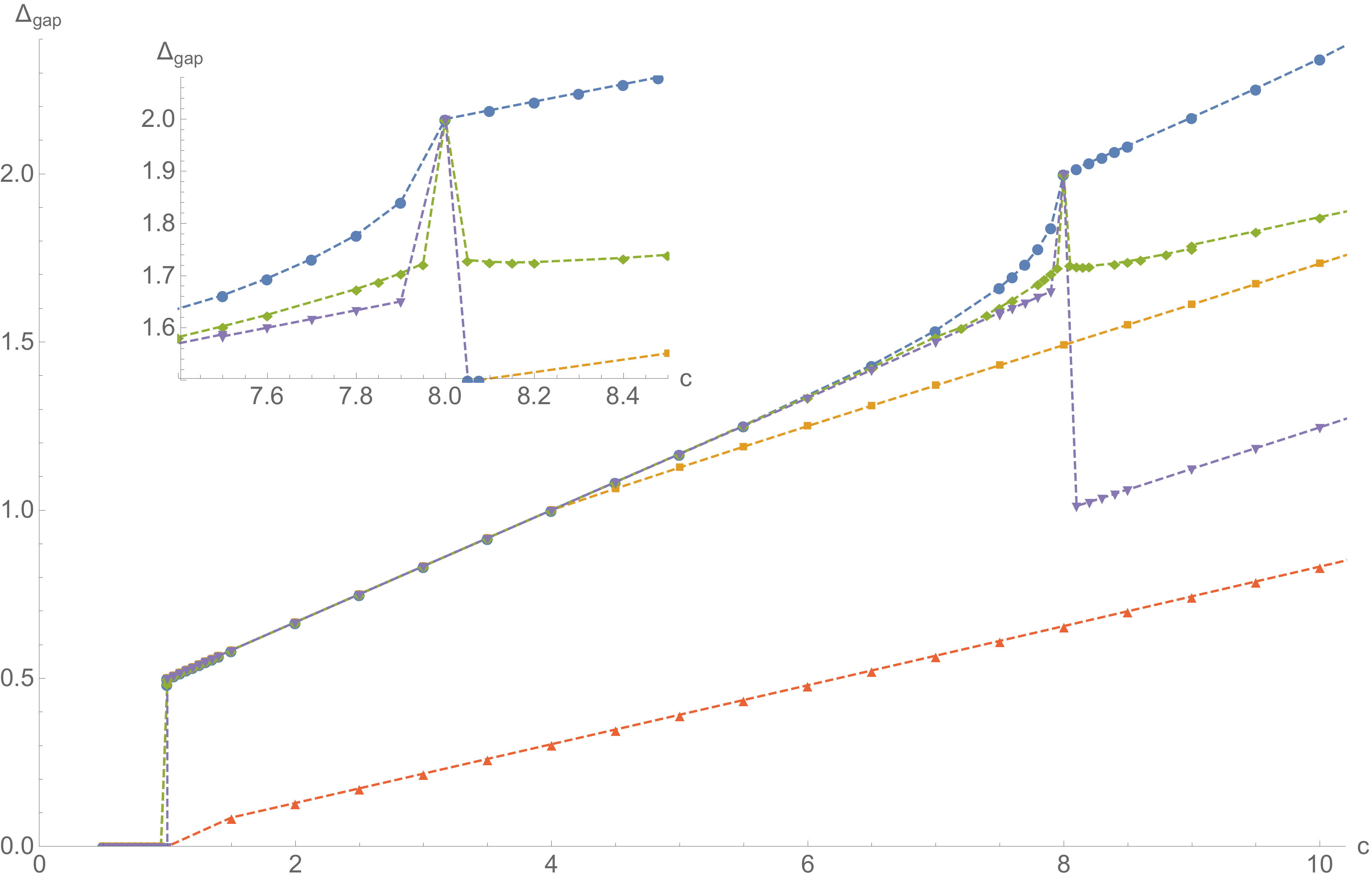}}
\subfigure[Bounds in the region $1 \le c \le 26$]{\includegraphics[width=.84\textwidth]{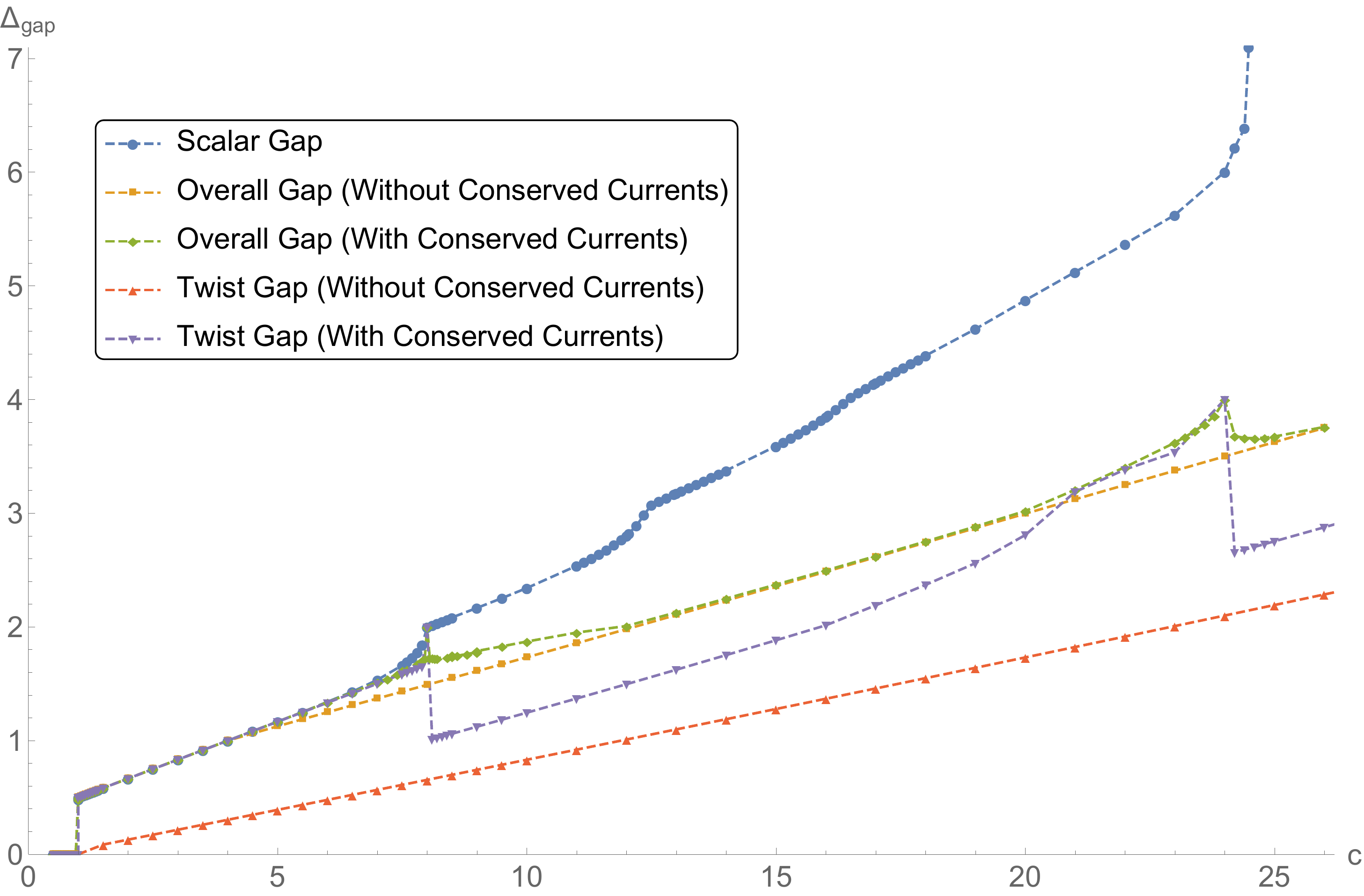}}
\end{center}
\caption{Numerical upper bounds on the scalar gap $\D_s$, the overall gap $\D_o$, and the twist
gap $\D_t$.}
\label{W2_Fig1_normalbound}
\end{figure}

We solve the SDP problems (\ref{SDP01}) with the scalar, overall and twist gaps using the {\tt SDPB}
package\cite{Simmons-Duffin:2015qma}. Figure \ref{W2_Fig1_normalbound} shows the numerical upper bounds on $\D_s$, $\D_o$ and $\D_t$ for parity-preserving CFTs with and without conserved currents.

\begin{figure}[h!]
\begin{center}
\subfigure[$5.9 \le c \le 6.1$]{\includegraphics[width=.85\textwidth]{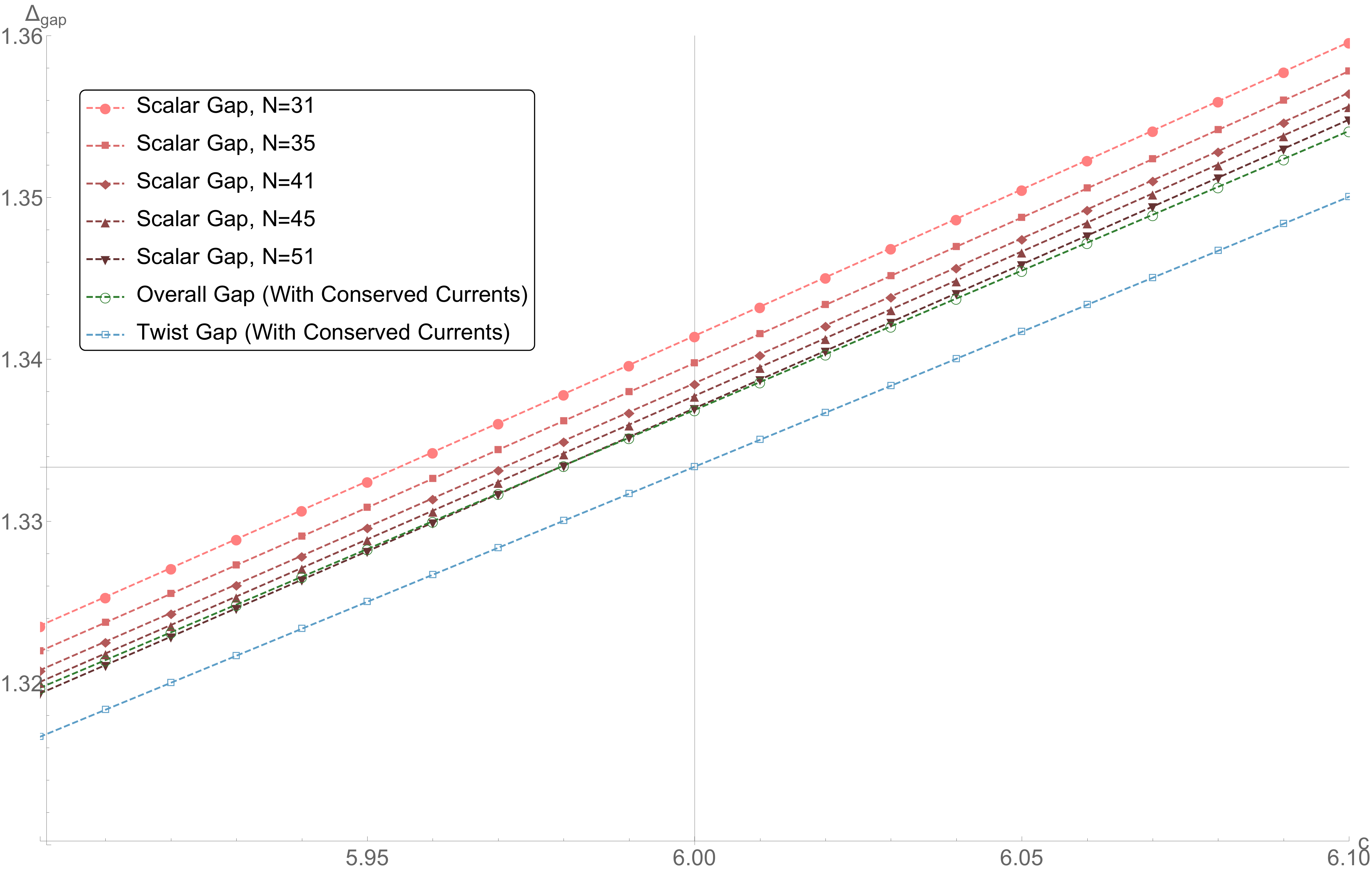}}
\subfigure[$6.8 \le c \le 7.2$]{\includegraphics[width=.85\textwidth]{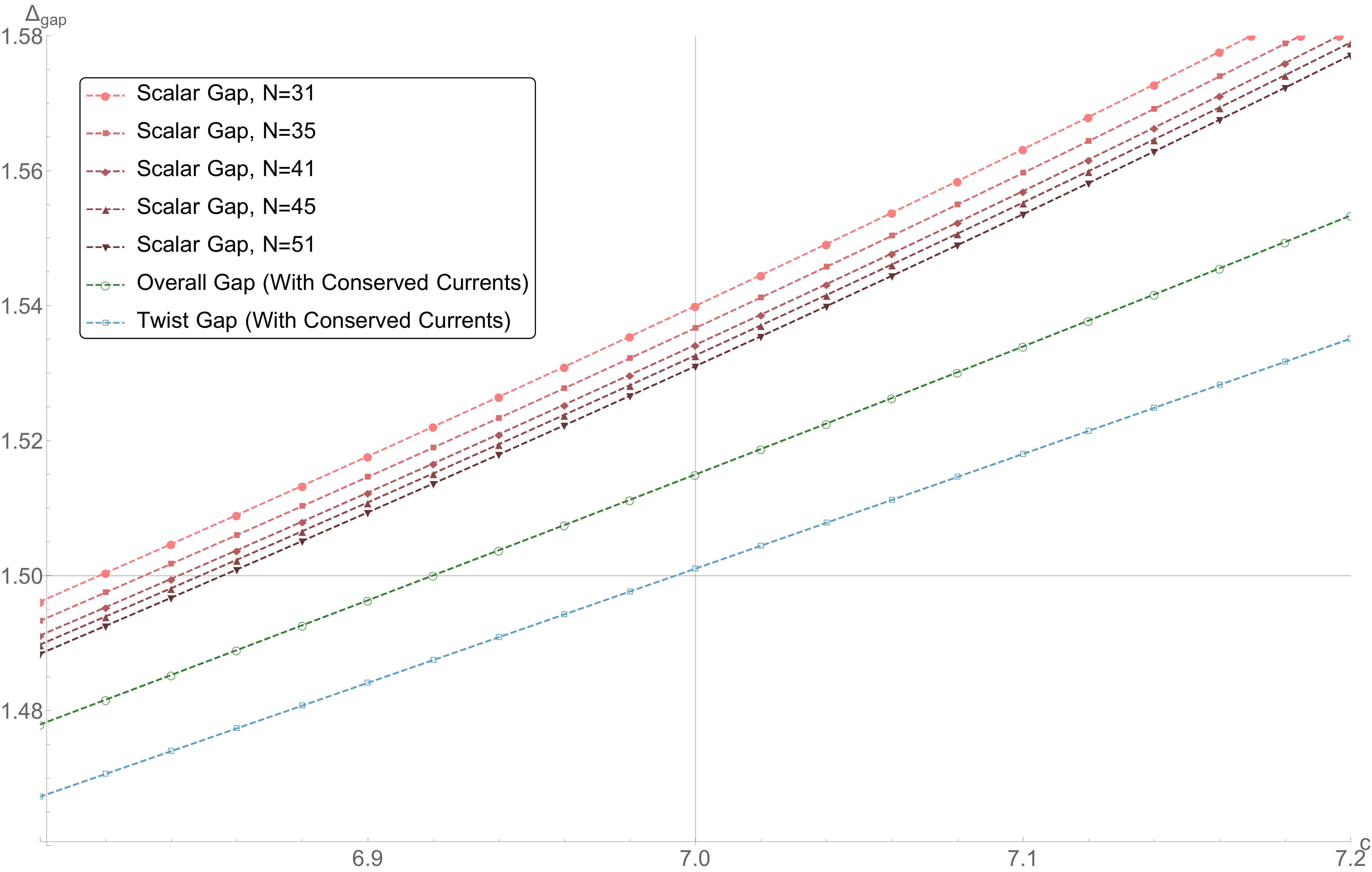}}
\end{center}
\caption{Numerical upper bounds on scalar gap, overall gap and twist gap zoomed in around $c=6$ and $7$.}
\label{W2_Fig1_zoomed_c6}
\end{figure}

As studied in \cite{Collier:2016cls}, there is no essential difference between
the CFTs with and without conserved currents
for the numerical bounds on the scalar gap $\D_s$, especially when $c \le 8$.
The authors of \cite{Collier:2016cls} found that
$(\widehat{A}_1)_1$, $(\widehat{A}_2)_1$, $(\widehat{G}_2)_1$, $(\widehat{D}_4)_1$ and $(\widehat{E}_8)_1$ WZW models
are realized on the numerical bounds at $c=1,2,\frac{14}{5},4,8$.
It is rather tempting to test the possibility of realizing all the WZW models with Deligne's exceptional series $\widehat{A}_1 \subset \widehat{A}_2 \subset \widehat{G}_2 \subset \widehat{D}_4 \subset \widehat{E}_6 \subset \widehat{E}_7 \subset \widehat{E}_8$ at the numerical boundary.
Figure \ref{W2_Fig1_zoomed_c6} however shows that other level one WZW models
for the Deligne's exceptional series would not be realized on the boundary even for the sufficiently large order of derivative.
It was also reported in \cite{Collier:2016cls} that there is no upper bound on the scalar gap $\D_s$
beyond $c\geq25$ where one can easily construct a modular invariant ``partition function''
of a noncompact CFT having no scalar primaries.

Unlike the scalar gap $\D_s$, we find that the numerical
bounds of the overall gap $\D_o$ is sensitive to the existence of conserved currents.
Note that we do not impose the gap condition on the conserved currents.
It turns out that the numerical bound on $\D_o$ for the CFT
with conserved currents start to deviate from that of the CFT without conserved currents at $c=4$.
We find a sharp peak at $c=8$ when the conserved currents are included, which can be identified again with the $(\widehat{E}_8)_1$ WZW model.

\begin{figure}[t]
\begin{center}
\includegraphics[width=.84\textwidth]{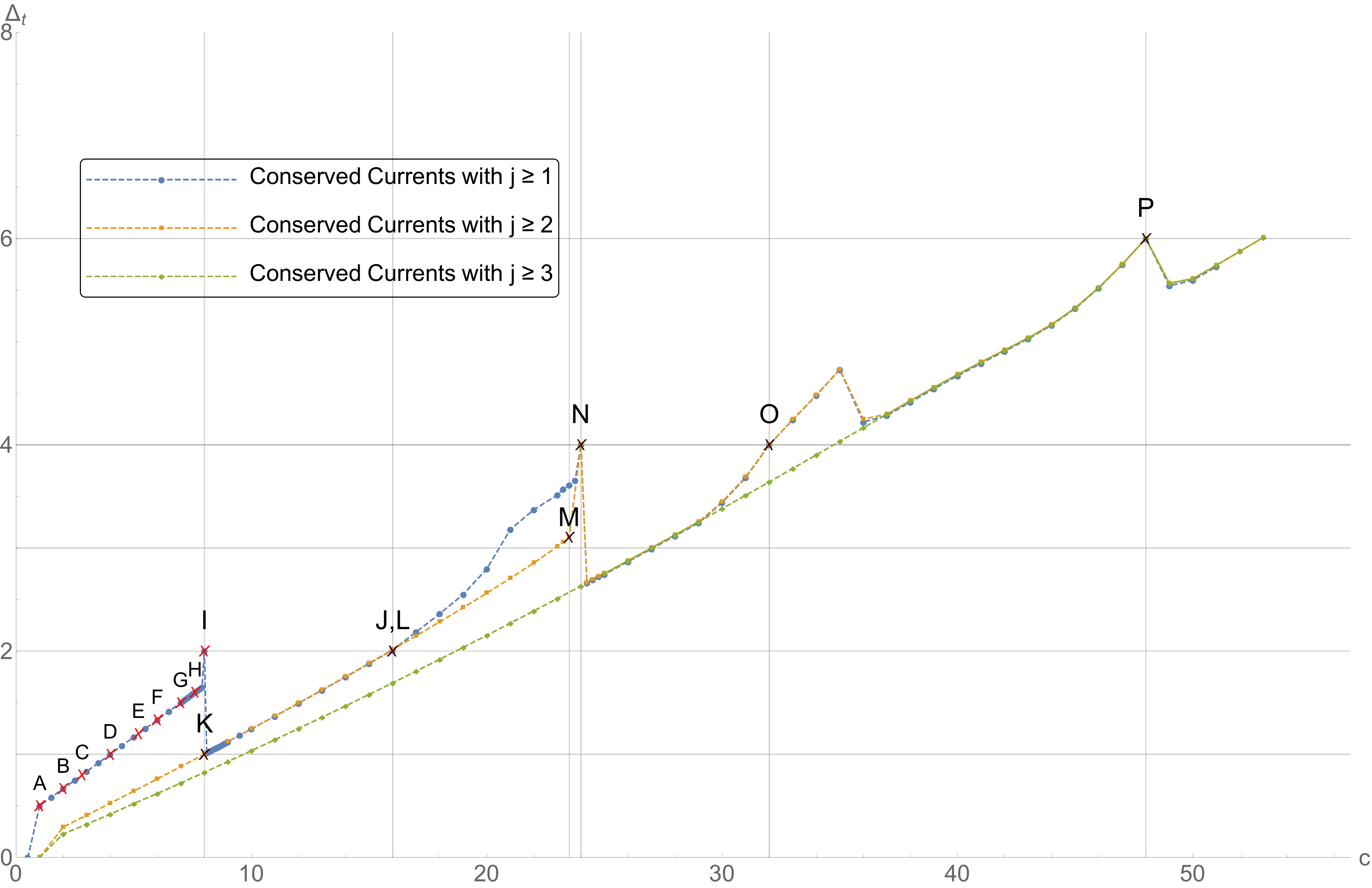}
\end{center}
\caption{Numerical upper bounds on the twist gap in the range of $1 \le c \le 55$ with the conserved currents of $j \ge 1$, $j \ge 2$ and $j \ge 3$ included. Each {\bf{x}}-mark on the boudnary refers to a certain RCFT in Tables \ref{List_j1}, \ref{List_j2}.}
\label{W2_Fig4_largeC2_CC2}
\end{figure}

We observe that the curve for the numerical bounds on the twist gap $\D_t$ with the conserved currents shows the most dramatic pattern.
We find that all the level one WZW models for the Deligne's exceptional
series including $(\widehat{F}_4)_1$, $(\widehat{E}_6)_1$ and $(\widehat{E}_7)_1$
now sit on the numerical bounds.
We also observe that a modular invariant partition function with integer degeneracies is realized at the bound for $c=\frac{38}{5}$.
Interestingly, this value of central charge is identical to the Sugawara central charge for $\widehat{E}_{7\half}$ with level $1$. We discuss this point in more detail in section \ref{subsec:WZW}.

The twist gap problem without the conserved currents also shows a different numerical upper bound.
The asymptotic slope of this bound is compatible with $\frac{c-1}{12}$ \cite{Hartman:2014oaa}.

In Figure \ref{W2_Fig4_largeC2_CC2}, we further explore the numerical upper bounds on the twist gap $\D_t$ for CFTs with conserved currents, by restricting the spins of the conserved currents. We consider the case without the spin-$1$ current (only the spins $j\geq2$ included), and also the case without $j=1, 2$ current (only the spins $j\geq3$ included).

We find $13$ special points on the numerical upper bounds of $\D_t$
that may correspond to RCFTs, some of which are known and the others are conjectured, when
$j\geq1$ conserved currents are included. Those RCFTs are summarized in Table \ref{List_j1}.
We can further show that only these $13$ RCFTs saturate the {\it integer} bounds on
the degeneracies of the scalar primaries with dimension $\D=\D_t$.
We will discuss this point in section \ref{sec:Spec}.
If we exclude $j=1$ conserved current in the spectrum,
the three additional RCFTs conjectured in \cite{Hampapura:2016mmz}, including the Baby Monster CFT, are realized at the numerical bounds.
They are summarized in Table \ref{List_j2}.

\begin{table}[t]
\centering
{
\begin{tabular}{|c|c |c |c| c| }
\hline
  Label &  $c$     & $\Delta_{t}$ & Maximal Degeneracy & Expected CFT \\
\hline\hline
 \rule{0in}{2.5ex}A & $1$  & $1/2$ & deg $= 4.000000$ & $(\widehat{A}_1)_1$ WZW model\\ [0.5ex]
 \hline
 \rule{0in}{2.5ex}B & $2$  & $2/3$ & deg $= 18.000000$ & $(\widehat{A}_2)_1$ WZW model \\ [0.5ex]
 \hline
 \rule{0in}{2.5ex}C & $14/5$  & $4/5$ & deg $= 49.000000$ &$(\widehat{G}_2)_1$ WZW model \\[0.5ex]
 \hline
\rule{0in}{2.5ex}D&  $4$  & $1  $ & deg $= 192.0000000$ & $(\widehat{D}_4)_1$ WZW model \\[0.5ex]
\hline
\rule{0in}{2.5ex}E& $26/5$  & $6/5$ & deg $= 676.000004$ & $(\widehat{F}_4)_1$ WZW model \\[0.5ex]
\hline
 \rule{0in}{2.5ex}F& $6$  & $4/3$ & deg $= 1458.000091$ & $(\widehat{E}_6)_1$ WZW model \\[0.5ex]
\hline
 \rule{0in}{2.5ex}G& $7$  & $3/2$ & deg $= 3136.000011$ & $(\widehat{E}_7)_1$ WZW model \\[0.5ex]
\hline
 \rule{0in}{2.5ex}H&$38/5$  & $8/5$ & deg $= 3249.000405$ & $(\widehat{E}_{7\frac{1}{2}})_1$ WZW model \\[0.5ex]
\hline
 \rule{0in}{2.5ex}I& $8$  & $2  $ & deg $= 61504.00000$ & $(\widehat{E}_8)_1$ WZW model \\[0.5ex]
\hline
 \rule{0in}{2.5ex}J&$16$  & $2  $ & deg $= 246016.0000$ & $(\widehat{E}_8 \times \widehat{E}_8)_1$ WZW model \\[0.5ex]
\hline
\hline
 \rule{0in}{2.5ex}N& $24$  & $4  $ & deg $= 38762915689.0000$ & Monster CFT \\[0.5ex]
\hline
 \rule{0in}{2.5ex}O& $32$  & $4  $ & deg $= 19461087009.0351$ & ``$c=32$ ECFT" \\[0.5ex]
\hline
 \rule{0in}{2.5ex}P& $48$  & $6  $ & deg $= 1847926789775361.00$ & $c=48$ ECFT \\[0.5ex]
\hline
\end{tabular}
\caption{List of theories on the numerical boundary of $\D_t$. We include the conserved currents of $j \ge 1$ in spectrum.}
\label{List_j1}
}
\end{table}

\begin{table}[ht]
\centering
{
\begin{tabular}{|c |c |c |c | c| }
\hline
 Label & $c$  &  $\Delta_{t}$ & Maximal Degeneracy  & Automorphism \\
\hline
\hline
  \rule{0in}{2.5ex}K& $8$  & $1  $ & deg $= 496.000000$ & $O_{10}^+(2).2$ \\ [0.5ex]
\hline
 \rule{0in}{2.5ex}L& $16$  & $2  $ & deg $= 69255.00000$ &$O_{10}^+(2).2^{16}$ \\ [0.5ex]
\hline
  \rule{0in}{2.5ex}M& $47/2$  & $ 3 $ & deg $= 19105641.071$ & Baby Monster \\ [0.5ex]
\hline
\end{tabular}
\caption{List of theories on the numerical boundary of $\D_t$. We include the conserved currents of $j \ge 2$ in spectrum.}
\label{List_j2}
}
\end{table}

\section{Spectroscopy}       
\label{sec:Spec}            

\subsection{Spectrum analysis in modular bootstrap}
\label{section41}

We discuss in this section how to constrain the degeneracy of primary operators above the vacuum
in a hypothetical CFT when the numerical bound on the twist gap $\D_t$ is saturated \cite{Collier:2016cls}.
Let us first start with upper bound on the degeneracy of the lowest scalar primaries
of conformal dimension $\D=\D_t$. As long as $\D_t$ is below the numerical bound,
there can exist certain linear functional $\b$ such that
\begin{align}
  \b\Big[ \CZ_{\frac{\D_t}{2},\frac{\D_t}{2}}(\t,\bar \t) \Big] & = 1\ ,
  \label{opt01}
\end{align}
and
\begin{align}
  \b\Big[ \CZ_{h,\bar h}(\t,\bar \t) \Big] & \geq 0 \text{ for } (h,\bar h)\neq \big(\frac{\D_t}{2},\frac{\D_t}{2}\big)\ , \qquad
  \b\Big[ \CZ_j(\t,\bar \t) \Big]  \geq 0 \ ,
  \label{opt02}
\end{align}
but acts negatively on $\CZ_\text{vac}(\t,\bar \t)$ to satisfy the modular constraint (\ref{condition01}). Then,
(\ref{opt01}) and (\ref{opt02}) imply
\begin{align}
  2 d_{\frac{\D_t}{2},\frac{\D_t}{2}} & = - \b\Big[ \CZ_\text{vac} \Big] -  \left\{
  \sum_{(h,\bar h)\neq(\frac{\D_t}{2},\frac{\D_t}{2})} d_{h,\bar h} \b\Big[ \CZ_{h,\bar h} + \bar{\CZ}_{h,\bar h} \Big]
  + \sum_{j} d_h \b \Big[ \CZ_{j} + \bar{\CZ}_j \Big] \right\}
  \nonumber \\ & \leq - \b\Big[ \CZ_\text{vac} (\t,\bar \t) \Big]\ .
  \label{degbound02}
\end{align}
To obtain the upper bound on $d_{\frac{\D_t}{2},\frac{\D_t}{2}}$, one thus need to solve
an optimization problem of searching a linear functional $\b$ that maximize
\begin{align}
  \b\Big[\CZ_\text{vac}(\t,\bar \t)\Big] \ .
\end{align}
Let us denote such a linear functional by $\b^*$. Then this leads to
\begin{align}
  d_{\frac{\D_t}{2},\frac{\D_t}{2}} \leq d_{\frac{\D_t}{2},\frac{\D_t}{2}}^* = - \frac{1}{2} \b^*\Big[ \CZ_\text{vac}(\t,\bar \t)\Big]\ .
  \label{degbound01}
\end{align}
It will be shown later that the $16$ RCFTs that lie on the numerical bounds
on $\D_t$ saturate the degeneracy bound (\ref{degbound01}).

We can further determine the entire spectrum of
a putative CFT, referred to as the {\it{extremal spectrum}},
when the degeneracy of the lowest primary scalar saturates the bound (\ref{degbound01}).
This is because, when $d_{\frac{\D_t}{2},\frac{\D_t}{2}}=d^\ast_{\frac{\D_t}{2},\frac{\D_t}{2}}$,
the first line of (\ref{degbound02}) becomes
\begin{align}
  0 = & \sum_{(h,\bar h)\neq(\frac{\D_t}{2},\frac{\D_t}{2})} d_{h,\bar h} \left\{
  \b^*\Big[ \CZ_{h,\bar h}(\t,\bar \t) \Big]+
  \b^*\Big[ \bar{\CZ}_{h,\bar h}(\t,\bar \t) \Big]\right\}
  \nonumber \\ &
  + \sum_{j} d_j \left\{
  \b^*\Big[ \CZ_{j}(\t,\bar \t)\Big]  +
  \b^*\Big[ \bar{\CZ}_j(\t,\bar \t)\Big] \right\}.
\end{align}
and thus
\begin{align}
  \b^* \Big[ \CZ_{h,\bar h}(\t,\bar \t) \Big] = 0
  \qquad \left(
  \b^* \Big[ \CZ_{j,0}(\t,\bar \t)] \Big] = 0 \right)
  \label{EFM beta}
\end{align}
for each primary in the extremal spectrum, i.e., $d_{h,\bar h}\neq 0$ ($d_{j}\neq0$). Therefore, by examining the functional $\beta^*$ evaluated at the Virasoro weights $(h, \bar{h})$, one can determine which value of the weights are allowed.
This analysis is initiated in \cite{ElShowk:2012hu}, often called as the extremal functional method.

One can also find the upper bounds on the degeneracies $d_{h',\bar{h}'}$ of
the primaries in the extremal spectrum. This can be done by solving another optimization problem, that is to search for a linear functional $\b^*_{(h',\bar{h}')}$ that
\begin{align}
  \text{maximize } \b^*_{(h',\bar{h}')}\Big[\CZ_\text{vac}(\t,\bar \t)\Big]
  \label{beta_star_1}
\end{align}
such that
\begin{align}
  \b^*_{(h',\bar{h}')}\Big[ \CZ_{h',\bar{h}'} (\t,\bar \t) \Big] = 1 \ ,
\label{beta_star_2}
\end{align}
and
\begin{align}
  \b^*_{(h',\bar{h}')}\Big[ \CZ_{h,\bar h}(\t,\bar \t) \Big] \geq 0 \text{ for } (h,\bar h)\neq (h',\bar{h}')\ ,
  \label{beta_star_3}
\end{align}
Although it is not guaranteed
that a CFT having the extremal spectrum always maximizes
the degeneracies at all weights, it is still interesting
to ask if there exist such CFTs. We will see soon in the next subsection
that $16$ RCFTs that are realized on the numerical bounds of $\D_t$
indeed saturate the upper bounds on the degeneracies of all the primaries in the extremal spectrum.

\subsection{WZW models with Deligne's exceptional series} \label{subsec:WZW}
In this subsection, we utilize the extremal functional method (EFM) to investigate hypothetical  CFTs on $\D_t = \D_t^*$ having the maximal degeneracies at all weights in the extremal spectrum.
It turns out that the WZW models with level one for Deligne's exceptional series are the CFTs of such type. Among them, the spectrum of WZW models for $\fg=\widehat{A}_1, \widehat{A}_2, \widehat{G}_2,\widehat{D}_4$ and $\widehat{E}_8$ are also shown to agree with the extremal spectrum of CFTs on $\D_s = \D_s^*$ at $c=1, 2, \frac{14}{5}, 4$ and $8$ \cite{Collier:2016cls}. We present below our numerical spectral data at $c= \frac{26}{5}, 6, 7, \frac{38}{5}$ from which we identify the CFTs of our interest with the WZW models for $\fg = \widehat{F}_4, \widehat{E}_6, \widehat{E}_7$ and also with the mysterious $\widehat{E}_{7 \frac{1}{2}}$.

\begin{itemize}
\item {\bf Spectrum Analysis for the $(\widehat{F}_4)_1$ WZW model}

\begin{figure}[h]
\begin{center}
\includegraphics[width=.40\textwidth]{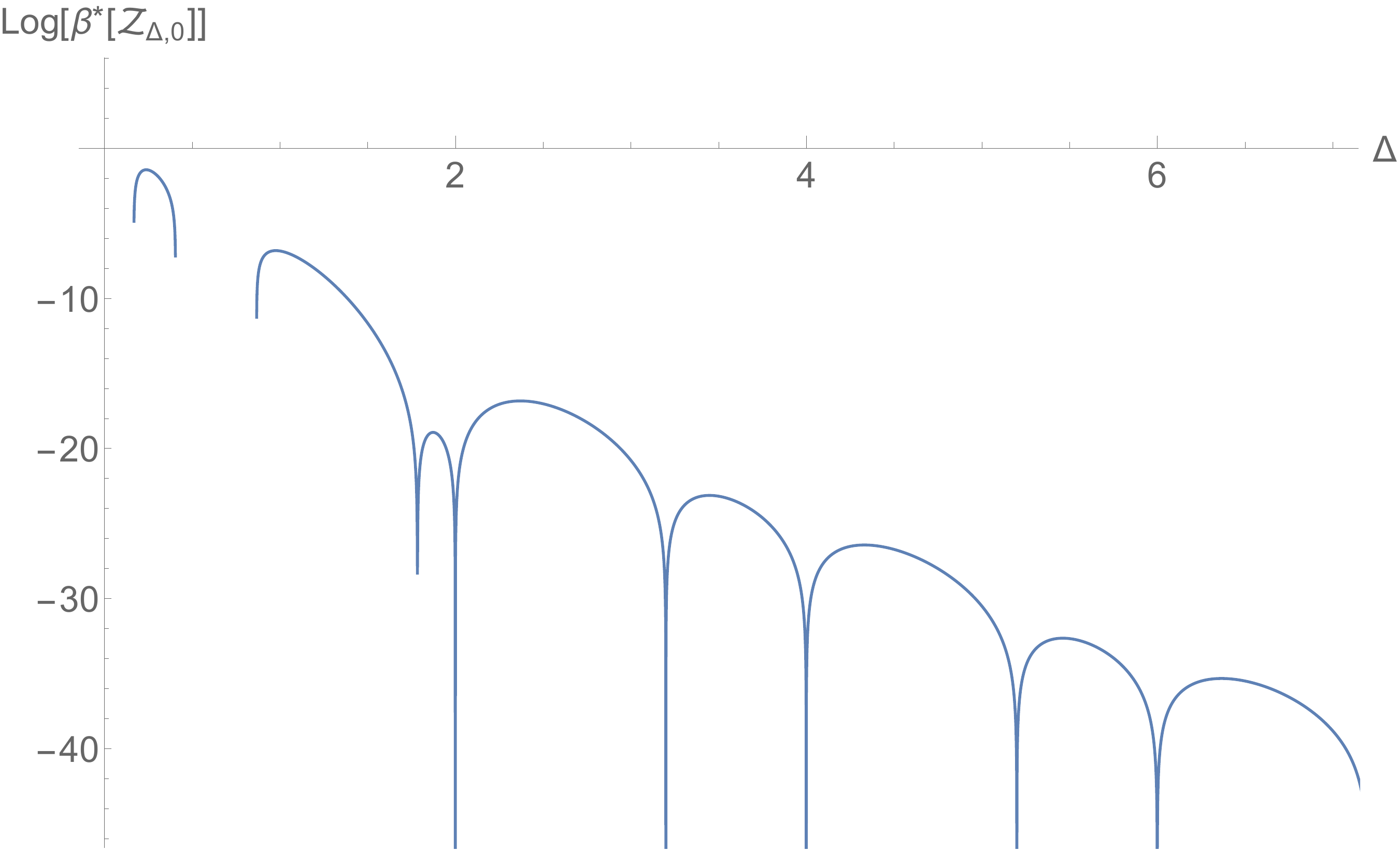} \qquad
\includegraphics[width=.40\textwidth]{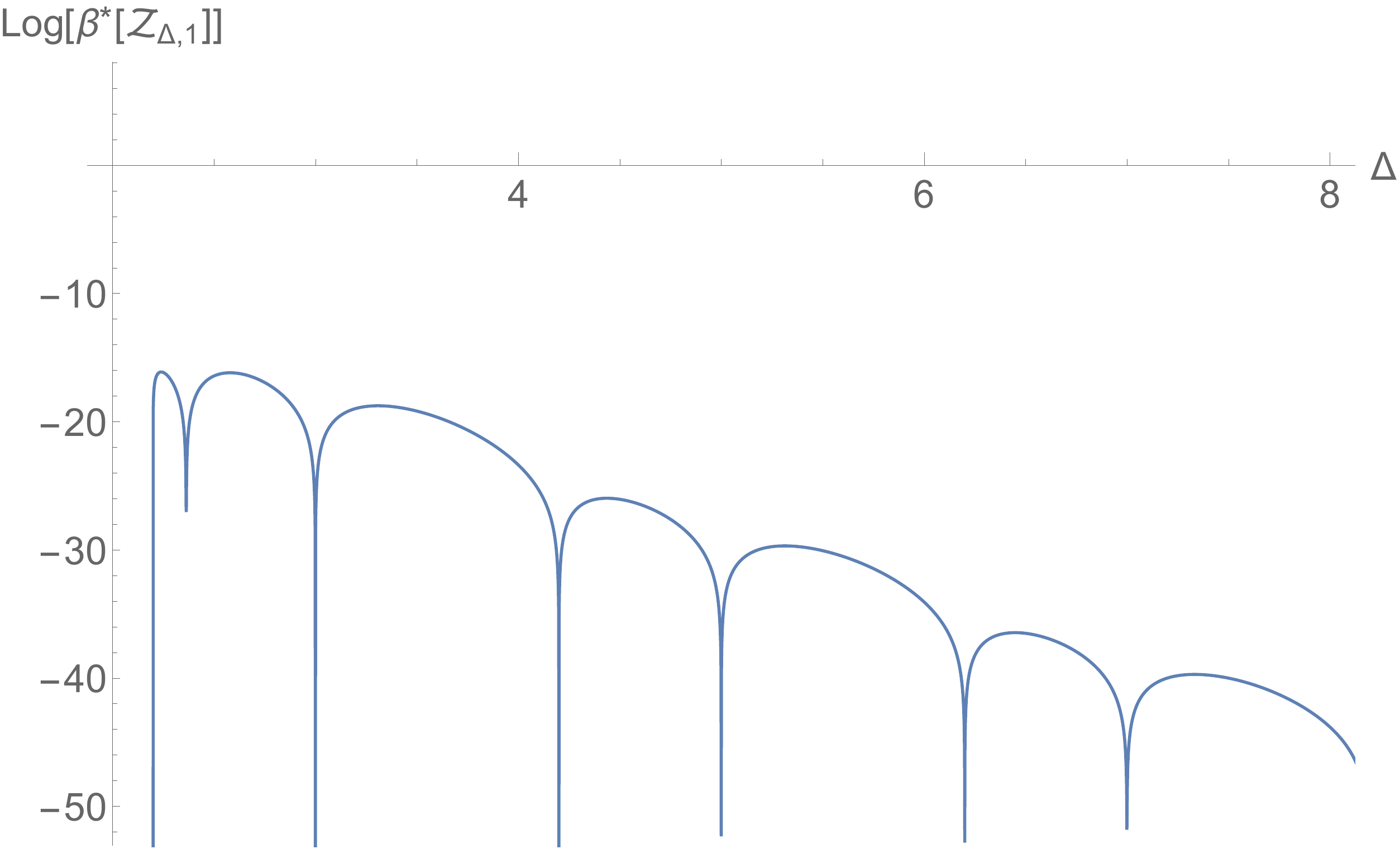}
\end{center}
\caption{The extremal functional $\beta^*$ acting on spin-0 (left) and spin-1 (right) primaries at $c=\frac{26}{5}$ and $\D_t = \frac{6}{5}$.}
\label{F4 deg}
\end{figure}

\begin{table}[ht]
\centering
{
\begin{tabular}{|c |c || c| c || c|c|}
\hline
 \rule{0in}{3ex}     $(h,\bar{h})$         & Max. Deg & $(h,\bar{h})$ &  Max. Deg & $(h,\bar{h})$ & Max. Deg\\
\hline
\rule{0in}{3ex} $(\frac{3}{5}, \frac{3}{5})$ &     676.0000 & $(1, 1)$ & 2704.0000  & $(1, 0)$ &      52.00028 \\
\hline
\rule{0in}{3ex} $(\frac{3}{5}, \frac{8}{5})$ &  7098.0001 & $(2, 1)$ & 16848.001  & $(2, 0)$ &  324.0007 \\
\hline
\rule{0in}{3ex} $(\frac{3}{5}, \frac{13}{5})$ &  35802.002 & $(3, 1)$ & 80444.061 & $(3, 0)$ &   1547.0091 \\
 \hline
\rule{0in}{3ex} $(\frac{8}{5}, \frac{8}{5})$ &  74529.0001   & $(2, 2)$ & 104976.005  & $(4, 0)$ &  5499.0126\\
 \hline
\end{tabular}
\caption{ The maximum value of degeneracies for low-lying states in a putative CFT with $c=\frac{26}{5}$.}
\label{F4degWZW}
}
\end{table}

We apply the EFM to a hypothetical CFT with $c=\frac{26}{5}$. The results illustrated in Figure \ref{F4 deg} suggest that the extremal spectrum of spin-0 and spin-1 have the conformal dimensions $\Delta_{j=0} = \{\frac{6}{5}+2n, 2+2n \}$ and $\Delta_{j=1} = \{\frac{11}{5}+2n, 3+2n \}$ for $n \in \mathbb{Z}_{\ge 0} $.

We utilize the linear fuctionals $\beta^*_{h,\bar{h}}$ in \cref{beta_star_1,beta_star_2,beta_star_3} to obtain the maximal degeneracies of low-lying primaries in the extremal spectrum, listed in the Table \ref{F4degWZW}. It implies that in terms of the Virasoro characters, the partition function of a putative CFT of our interest can be decomposed into the following form.
\begin{align}
{Z}_{c=\frac{26}{5}}(\t, \bar{\t}) &= {\chi}_{0}(\tau) \bar{{\chi}}_{0}(\bar{\tau}) + 676 {\chi}_{\frac{3}{5}}(\tau) \bar{{\chi}}_{\frac{3}{5}}(\bar{\tau}) +  7098 \left({\chi}_{\frac{3}{5}}(\tau) \bar{{\chi}}_{\frac{8}{5}}(\bar{\tau}) + \text{c.c.} \right) \nonumber \\
& + 2704 {\chi}_{1}(\tau) \bar{{\chi}}_{1}(\bar{\tau})+ 16848 \Big({\chi}_{2}(\tau) \bar{{\chi}}_{1}(\bar{\tau}) + \text{c.c.} \Big) + 104976 {\chi}_{2}(\tau) \bar{{\chi}}_{2}(\bar{\tau}) \nonumber \\
&  +52 \Big( {\chi}_{1}(\tau)  \bar{{\chi}}_{0}(\bar{\tau}) + \text{c.c.}\Big) +324 \Big({\chi}_{2}(\tau)  \bar{{\chi}}_{0}(\bar{\tau}) + \text{c.c.} \Big)+ \cdots.
\label{F4 result}
\end{align}

The affine character
of $(\widehat{F}_4)_1$ is known to agree with the solution \eqref{primary_ch} to the second order modular differential equation \eqref{MDE02} with $c=\frac{26}{5}$,
\begin{align}
\begin{split}
f_{\text{vac}}^{c=\frac{26}{5}} (\t)&\equiv \chi^{c=\frac{26}{5}}_{[1;0,0,0,0]} (\t)= q^{-\frac{13}{60}}\left(1 + 52 q + 377 q^2 +1976 q^3 + \mathcal{O}(q^4)\right),  \\
f_{\frac{3}{5}}^{c=\frac{26}{5}} (\t)&\equiv \chi^{c=\frac{26}{5}}_{[0;0,0,0,1]} (\t)= q^{\frac{3}{5}-\frac{13}{60}} \left(26 + 299 q + 1702 q^2 + 7475 q^3  + \mathcal{O}(q^4)\right),
\end{split}
\end{align}
where the overall constant $a_0$ of \eqref{primary_ch} is now fixed by the dimension of fundamental representation of $F_4$.

Using this affine character, one can simplify the modular invariant partition function \eqref{F4 result} as
\begin{equation}
Z_{c=\frac{26}{5}}(\t, \bar{\t}) = |f_{\text{vac}}^{F_4} (\t)|^2 + |f_{\frac35}^{F_4} (\t)|^2,
\label{F4wzwkp}
\end{equation}
which perfectly agree with the modular invariant partition function of $(\widehat{F}_4)_1$ WZW model\cite{Gannon:1992nq}. Therefore, we identify the putative CFT at $c=\frac{26}{5}$ with the $(\widehat{F}_4)_1$ WZW model.

\item {\bf Spectrum Analysis for the $(\widehat{E}_6)_1$ WZW model}

\begin{figure}[h]
\begin{center}
\includegraphics[width=.40\textwidth]{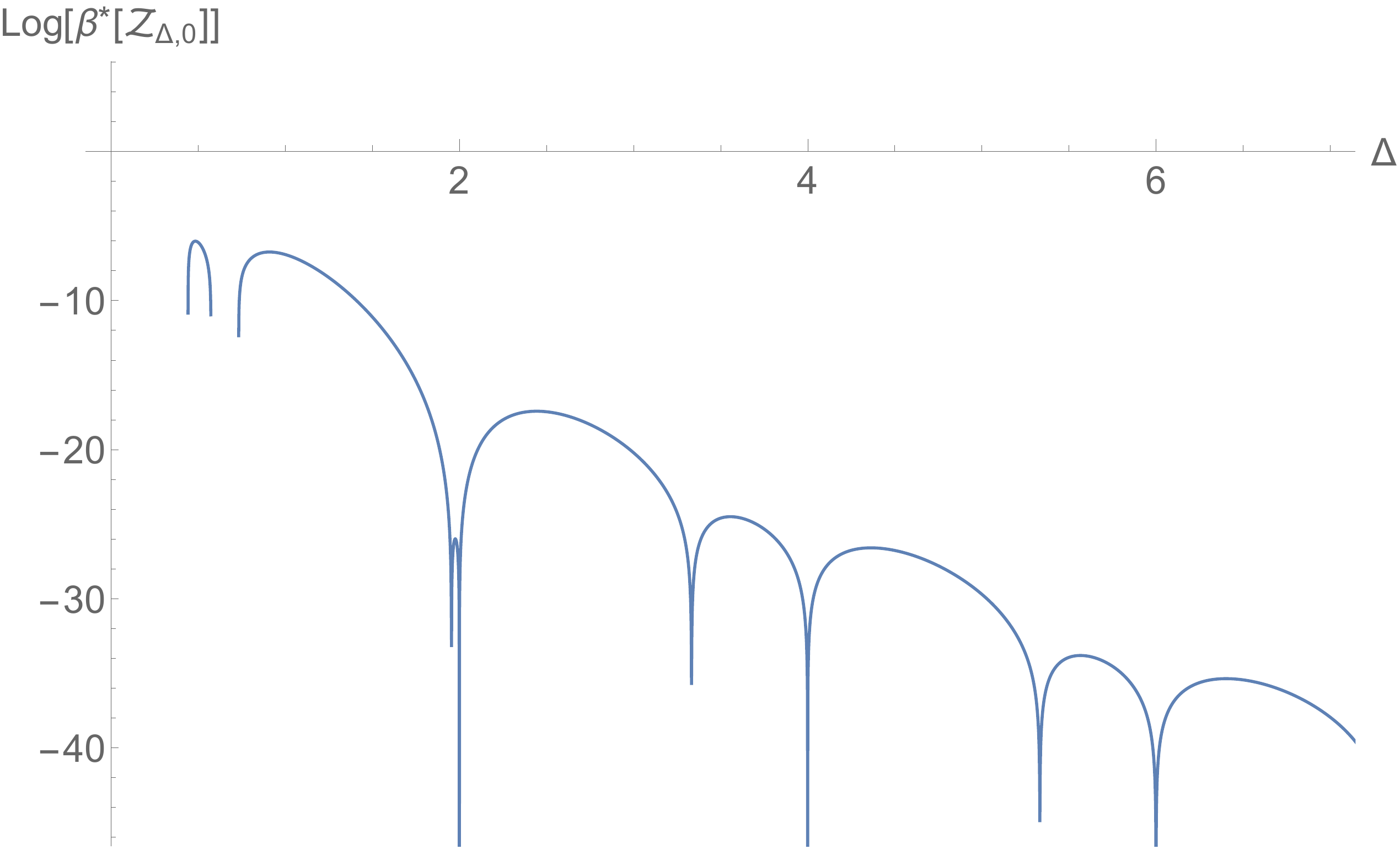} \qquad
\includegraphics[width=.40\textwidth]{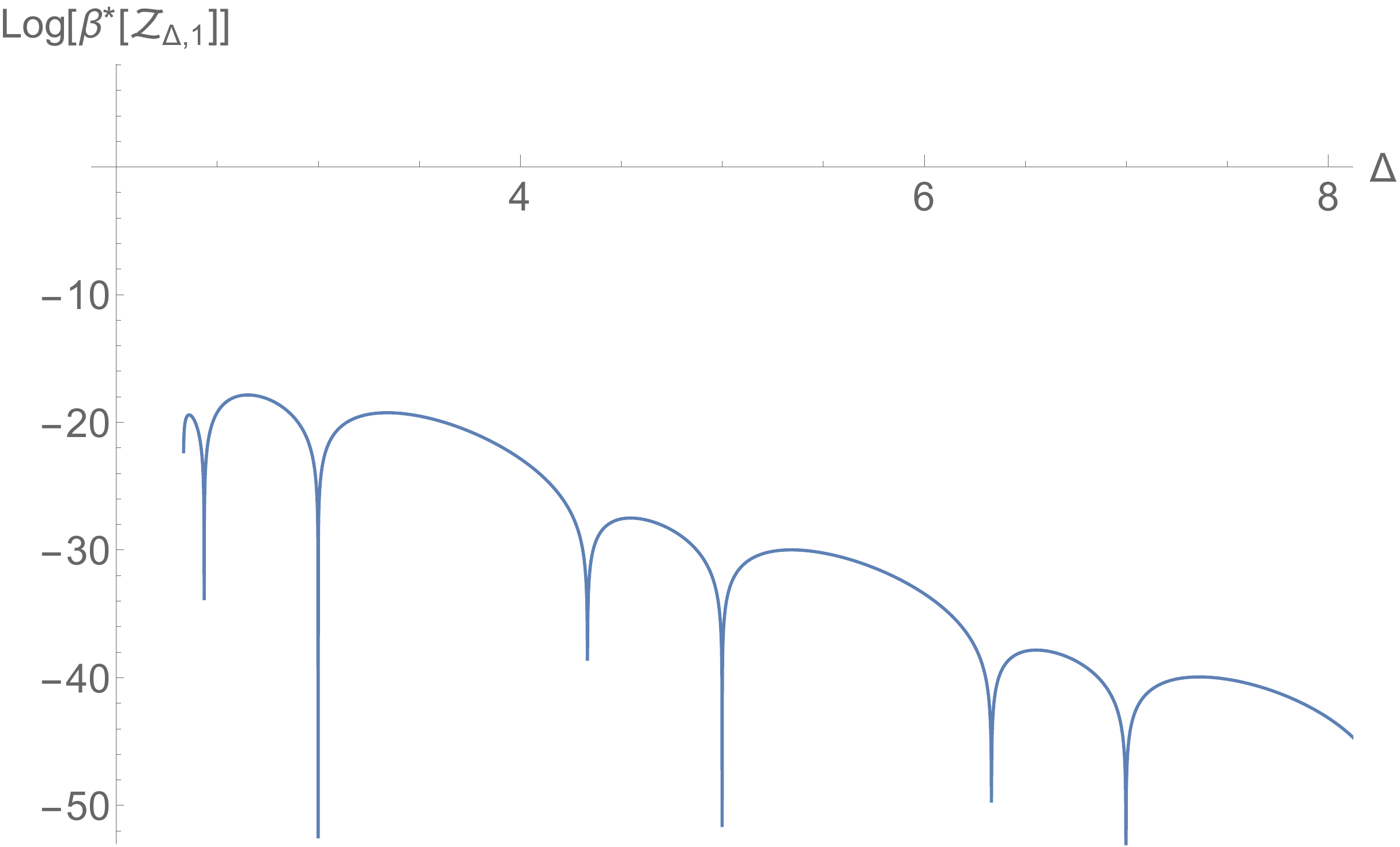}
\end{center}
\caption{The extremal functional $\beta^*$ acting on spin-0 (left) and spin-1 (right) primaries at $c=6$ and $\D_t = \frac{4}{3}$.}
\label{EFM_E6}
\end{figure}

\begin{table}[ht]
\centering
{
\begin{tabular}{|c |c || c| c || c|c|}
\hline
 \rule{0in}{3ex}     $(h,\bar{h})$         & Max. Deg & $(h,\bar{h})$ &  Max. Deg & $(h,\bar{h})$ & Max. Deg\\
\hline
\rule{0in}{3ex} $(\frac{2}{3}, \frac{2}{3})$ &  1458.0001 & $(1, 1)$ & 6084.0001  & $(1, 0)$ &  78.00023 \\
\hline
\rule{0in}{3ex} $(\frac{2}{3}, \frac{5}{3})$ &  18954.003 & $(2, 1)$ & 50700.004  & $(2, 0)$ &  650.0012 \\
\hline
\rule{0in}{3ex} $(\frac{2}{3}, \frac{8}{3})$ &  112266.08 & $(3, 1)$ & 278850.00 & $(3, 0)$ &  3575.010  \\
 \hline
\rule{0in}{3ex} $(\frac{5}{3}, \frac{5}{3})$ &  246402.001  & $(2, 2)$ & 422500.05 & $(4, 0)$ &  14806.03 \\
 \hline
\end{tabular}
\caption{ The maximum value of degeneracies for low-lying states in a putative CFT with $c=6$ .}
\label{E6deg}
}
\end{table}

Using the linear functional \eqref{EFM beta}, we can learn that the extremal spectrum with $c=6$ and $\Delta_t= \frac{4}{3}$ contains the scalar primaries of $\Delta_{j=0}  = \{\frac{4}{3}+2n, 2+2n \} $  and spin-one primaries of  $\Delta_{j=1} = \{\frac{7}{3}+2n, 3+2n \}(n \ge 0)$, as depicted in Figure \ref{EFM_E6}. We also summarize the maximal degeneracies of various primaries in the extremal spectrum in Table \ref{E6deg}.

We can express the partition function of a CFT that contains primaries in Table \ref{E6deg}
in terms of two solutions to \eqref{MDE02} with $c=6$ as follows,
\begin{align}
    Z_{c=6}(\t, \bar{\t}) = f_{\text{vac}}^{c=6} (\t) \bar{f}_{\text{vac}}^{c=6} (\bar{\t}) + 2 f_{\frac{2}{3}}^{c=6} (\t) \bar{f}_{\frac{2}{3}}^{c=6} (\bar{\t}),
    \label{E6 partition}
\end{align}
where
\begin{align}
\begin{split}
    f_{\text{vac}}^{c=6} (\t)&= q^{-\frac{1}{4}} \left(1 + 78 q + 729 q^2 +  4382 q^3 + \mathcal{O}(q^4) \right),  \\
    f_{\frac{2}{3}}^{c=6} (\t)&= q^{\frac{2}{3}-\frac{1}{4}} \left(27 + 378 q + 2484 q^2  +  12312 q^3 +  \mathcal{O}(q^4) \right).
\end{split}
\end{align}
Furthermore, two solutions $f_\text{vac}^{c=6}(\t)$ and $f_{\frac23}^{c=6}(\t)$
can be identified as the affine characters of $\widehat{E}_6$
\begin{align}
\begin{split}
f_{\text{vac}}^{c=6} (\t)&= \chi^{c=6}_{[1;0,0,0,0,0,0]} (\t),  \\
f_{\frac{2}{3}}^{c=6} (\t)&= \chi^{c=6}_{[0;1,0,0,0,0,0]} (\t) = \chi^{c=6}_{[0;0,0,0,0,1,0]} (\t).
\end{split}
\end{align}
Here, two representations $[0;1,0,0,0,0,0]$ and $[0;0,0,0,0,1,0]$ are complex conjugate to each other, and their characters are indistinguishable unless we turn on additional chemical potentials for the Cartan parts of the current algebra.

The partition function (\ref{E6 partition}) then becomes the partition function of
$(\widehat{E}_6)_1$ WZW model \eqref{modular inv ptn}.

\item {\bf Spectrum Analysis for the $(\widehat{E}_7)_1$ WZW model}

\begin{figure}[h]
\begin{center}
\includegraphics[width=.40\textwidth]{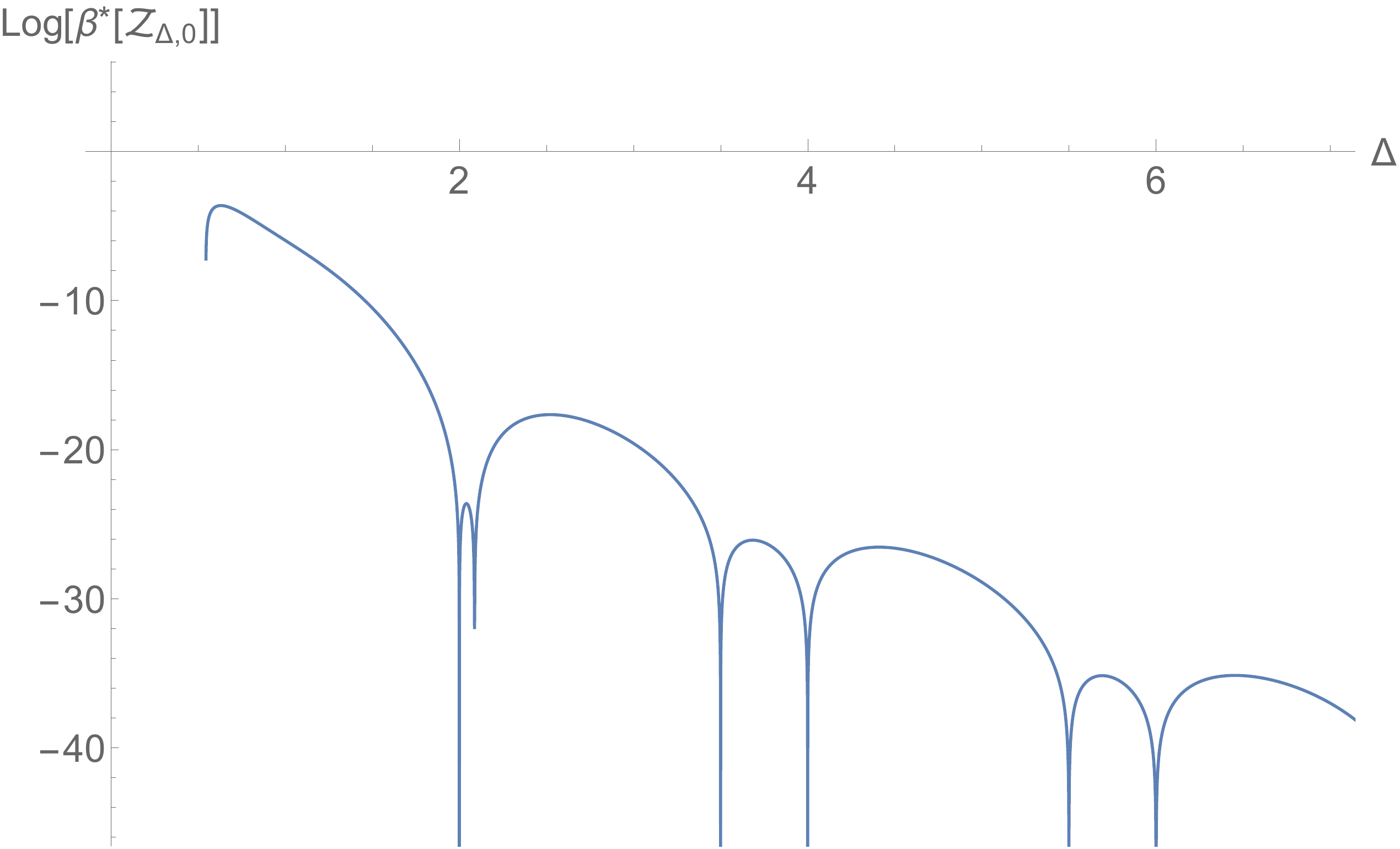} \qquad
\includegraphics[width=.40\textwidth]{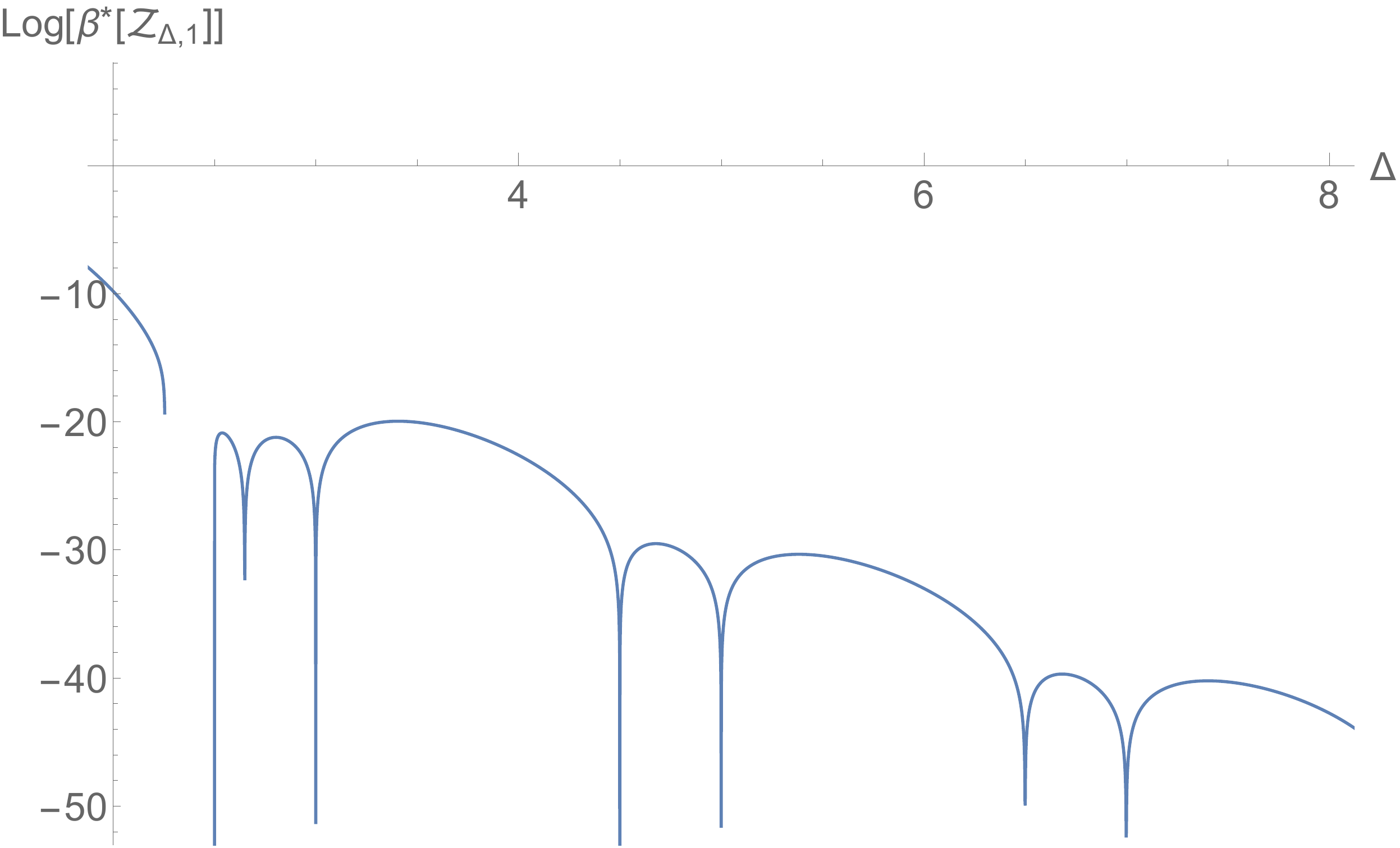}
\end{center}
\caption{The extremal functional $\beta^*$ acting on spin-0 (left) and spin-1 (right) primaries at $c=7$ and $\D_t = \frac{3}{2}$.}
\label{EFM_E7}
\end{figure}

\begin{table}[ht]
\centering
{
\begin{tabular}{|c |c || c| c || c|c|}
\hline
 \rule{0in}{3ex}     $(h,\bar{h})$         & Max. Deg & $(h,\bar{h})$ &  Max. Deg & $(h,\bar{h})$ & Max. Deg\\
\hline
\rule{0in}{3ex} $(\frac{3}{4}, \frac{3}{4})$ &  3136.0000 & $(1, 1)$ & 17689.000  & $(1, 0)$ &  133.00116  \\
\hline
\rule{0in}{3ex} $(\frac{3}{4}, \frac{7}{4})$ &  51072.000 & $(2, 1)$ & 204687.00  & $(2, 0)$ &  1539.0104  \\
\hline
\rule{0in}{3ex} $(\frac{3}{4}, \frac{11}{4})$ &  362880.00 & $(3, 1)$ & 1344364.01  & $(3, 0)$ &     10108.085 \\
 \hline
\rule{0in}{3ex} $(\frac{7}{4}, \frac{7}{4})$ &  831744.01  & $(2, 2)$ & 2368521.01 & $(4, 0)$ & 49665.351   \\
 \hline
\end{tabular}
\caption{ The maximum value of degeneracies for low-lying states in a putative CFT with $c=7$ .}
\label{E7deg}
}
\end{table}

As depicted in Figure \ref{EFM_E7}, the extremal spectrum with $c=7$ and $\Delta_t = \frac{3}{2}$ contains spin-0 primaries of $\Delta_{j=0} = \{\frac{3}{2}+2n, 2+2n\}$ and spin-1 primaries of $\Delta_{j=1} = \{\frac{5}{2}+2n, 3+2n \}$ ($n \in \mathbb{Z}_{\ge 0}$). The maximal degeneracies at various weights in the extremal spectrum are listed in Table $\ref{E7deg}$.

The solutions to \eqref{MDE02} with $c=7$ are known to agree with the $\widehat{E}_7$ affine character with $a_0 = 56$
\begin{align}
\begin{split}
f_{\text{vac}}^{c=7} (\t) = \chi^{c=7}_{[1;0,0,0,0,0,0,0]} (\t)&= q^{-\frac{7}{24}}\left(1 +133 q+1673 q^2+11914 q^3 + \ldots \right), \\
f_{\frac34}^{c=7} (\t) = \chi^{c=7}_{[0;0,0,0,0,0,1,0]} (\t)&= q^{\frac{3}{4}-\frac{7}{24}} \left(56 +968 q +7504 q^2 +42616 q^3 + \ldots \right).
\end{split}
\end{align}
It is straightforward to see that the partition function of $(\widehat{E}_7)_1$ WZW model  \eqref{modular inv ptn},
\begin{align}
Z_{c=7}(\t, \bar{\t}) = |f_{\text{vac}}^{c=7} (\t)|^2 + |f_{\frac34}^{c=7} (\t)|^2,
\label{E7 partition function}
\end{align}
is consistent to the maximal degeneracies in Table \ref{E7deg}.

\end{itemize}

\begin{itemize}
\item {\bf Spectrum Analysis for the $(\widehat{E}_{7\frac{1}{2}})_1$ WZW model}
\begin{figure}[h]
\begin{center}
\includegraphics[width=.40\textwidth]{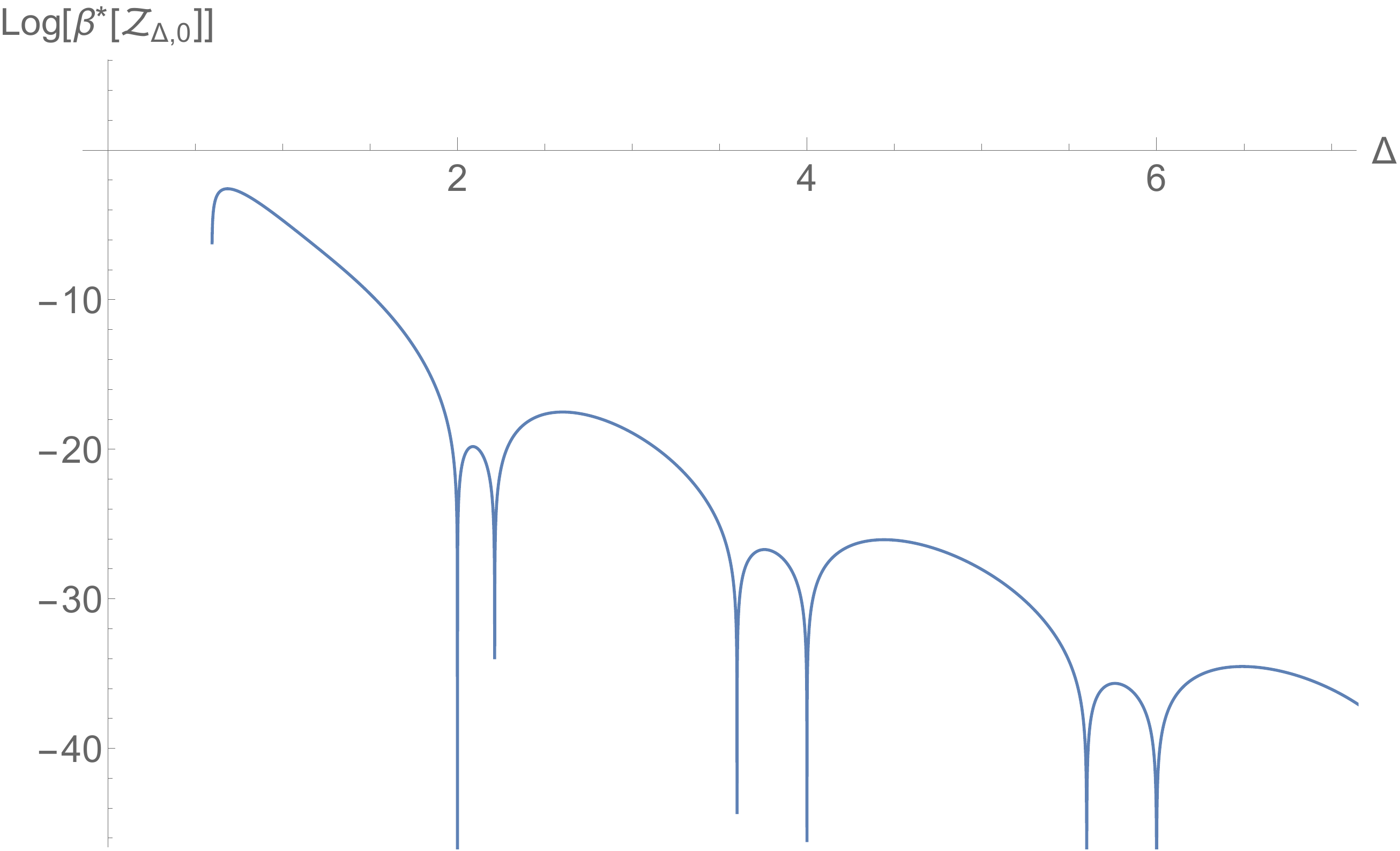} \qquad
\includegraphics[width=.40\textwidth]{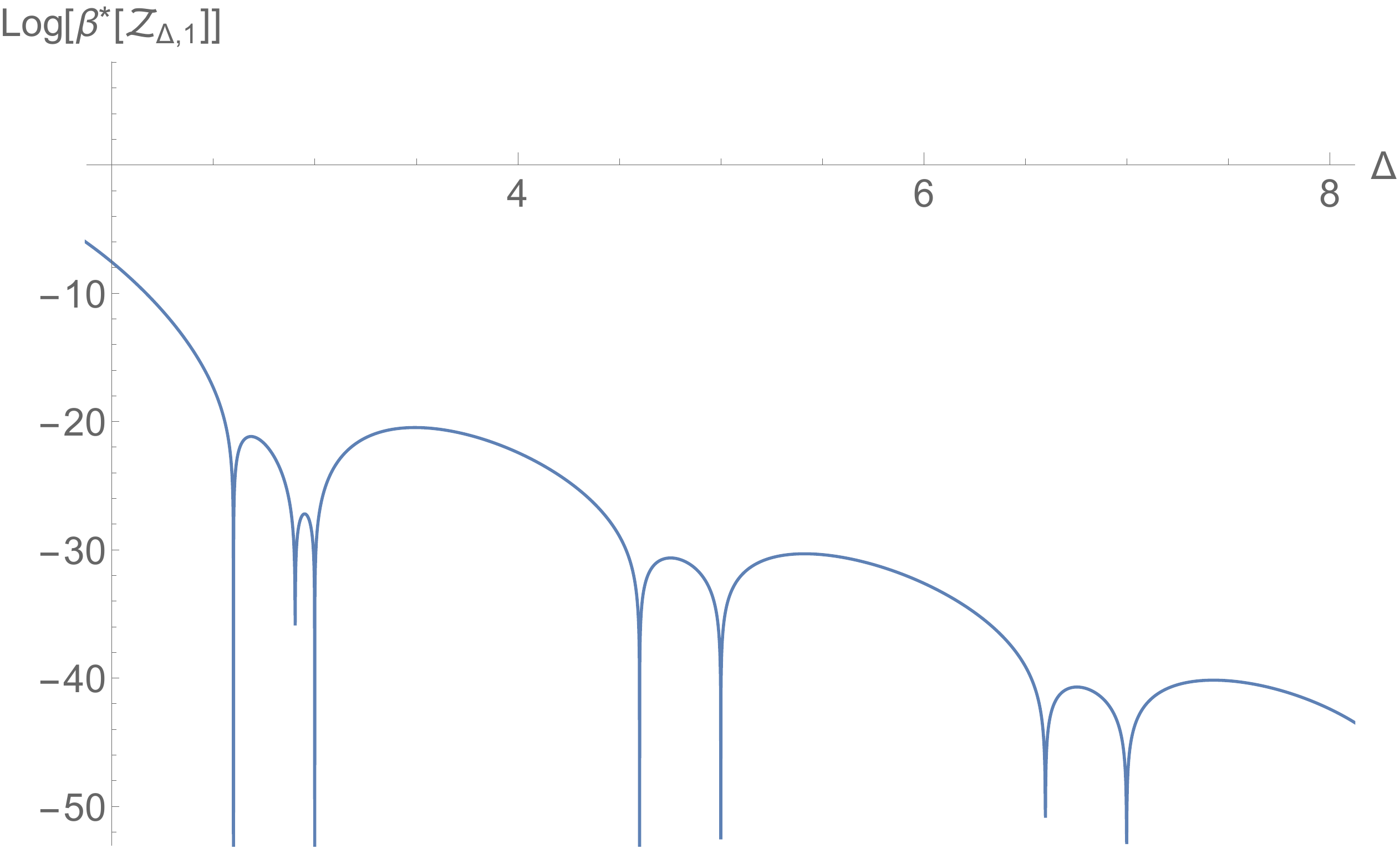}
\end{center}
\caption{The extremal functional $\beta^*$ acting on spin-0 (left) and spin-1 (right) primaries at $c=\frac{38}{5}$ and $\D_t = \frac{8}{5}$.}
\label{EFM_E7h}
\end{figure}

\begin{table}[ht]
\centering
{
\begin{tabular}{|c |c || c| c || c|c|}
\hline
 \rule{0in}{3ex}     $(h,\bar{h})$         & Max. Deg & $(h,\bar{h})$ &  Max. Deg & $(h,\bar{h})$ & Max. Deg\\
\hline
\rule{0in}{3ex} $(\frac{4}{5}, \frac{4}{5})$ &  3249.0004 & $(1, 1)$ & 36100.000 & $(1, 0)$ &  190.00412  \\
\hline
\rule{0in}{3ex} $(\frac{4}{5}, \frac{9}{5})$ &  59565.012 & $(2, 1)$ & 501600.00  & $(2, 0)$ & 2640.0481  \\
 \hline
\rule{0in}{3ex} $(\frac{9}{5}, \frac{9}{5})$ &  1092025.06  & $(2, 2)$ & 6969600.01 & $(3, 0)$ & 19285.021   \\
 \hline
\end{tabular}
\caption{The maximum value of degeneracies for low-lying states in a putative CFT at $c=\frac{38}{5}$. For these results, the maximum number of derivative is set to $N_{\textrm{max}}=55$, while spin is truncated at $j_{\textrm{max}}=40$.}
\label{E7hdeg}
}
\end{table}

It is shown in \cite{Mathur:1988na} that there is one more value of central charge $c=\frac{38}{5}$ where the all the coefficients of the two solutions to \eqref{MDE02} become positive integers,
\begin{align}
\begin{split}
f_{\text{vac}}^{c=\frac{38}{5}} (\t)&= q^{-\frac{19}{60}} \left(1 +190q + 2831q^2 + 22306 q^3+129276 q^4 + \mathcal{O}(q^5) \right),  \\
f^{c=\frac{38}{5}}_{\frac{4}{5}} (\t)&= q^{\frac{4}{5}-\frac{19}{60}} \left(57 + 1102q + 9367 q^2 + 57362 q^3 + 280459 q^4 + \mathcal{O}(q^5) \right).
\label{E7h char}
\end{split}
\end{align}
The non-vacuum character $f_{\text{vac}}^{c=\frac{38}{5}} (\t)$ in \eqref{E7h char} can arise from  an affine Lie algebra if there is one with dimension 190. Interestingly, mathematicians have discovered that there is indeed such a Lie algebra called $E_{7\frac12}$, in an attempt to fill in a certain gap in the Deligne's exceptional series \cite{landsberg2006sextonions}.

As illustrated in Figure \ref{EFM_E7h}, the extremal spectrum contains scalar primaries of conformal dimension $\Delta_{j=0} = \{\frac{8}{5}+2n, 2+2n \}$ and spin-one primaries of conformal dimension $\Delta_{j=1} = \{\frac{13}{5}+2n, 3+2n \}$($n \in \mathbb{Z}_{\ge 0}$). As summarized in Table \ref{E7hdeg}, they have positive integer maximal degeneracies.

It turns out that the modular invariant partition function of the a CFT that contains primaries in Table \ref{E7hdeg} can be simply expressed as,
\begin{align}
Z_{c=\frac{38}{5}}(\t, \bar{\t}) = f_{\text{vac}}^{c=\frac{38}{5}} (\t) \bar{f}_{\text{vac}}^{c=\frac{38}{5}} (\bar{\t}) + f^{c=\frac{38}{5}}_{\frac{4}{5}} (\t) \bar{f}^{c=\frac{38}{5}}_{\frac{4}{5}} (\bar{\t}).
\label{E7h_partition}
\end{align}
Based on this observation, we suspect that there may exist a RCFT at $c=\frac{38}{5}$ where $f_{\text{vac}}^{c=\frac{38}{5}} (\t)$ and $f^{c=\frac{38}{5}}_{\frac{4}{5}} (\t)$ in \eqref{E7h char} can be understood as the characters of $(\widehat{E}_{7\frac{1}{2}})_1$.\footnote{One of the fusion rule of $c=38/5$ RCFT appears to be negative\cite{Mathur:1988gt}. One can circumvent this inconsistency by interchanging two characters $f_{\text{vac}}^{c=\frac{38}{5}} (\t)$ and $f^{c=\frac{38}{5}}_{\frac{4}{5}} (\t)$ in \eqref{E7h char} which means allowing the non-unitary primary in the spectrum. The new theory with $c=-58/5$ has 57-fold degenerate identity character, thus not interpreted as a consistent CFT\cite{Mathur:1988gt}.}

\item {\bf Spectrum Analysis for $(\widehat{E}_8 \times \widehat{E}_8)_1$ WZW model}
\begin{figure}[h]
\begin{center}
\includegraphics[width=.40\textwidth]{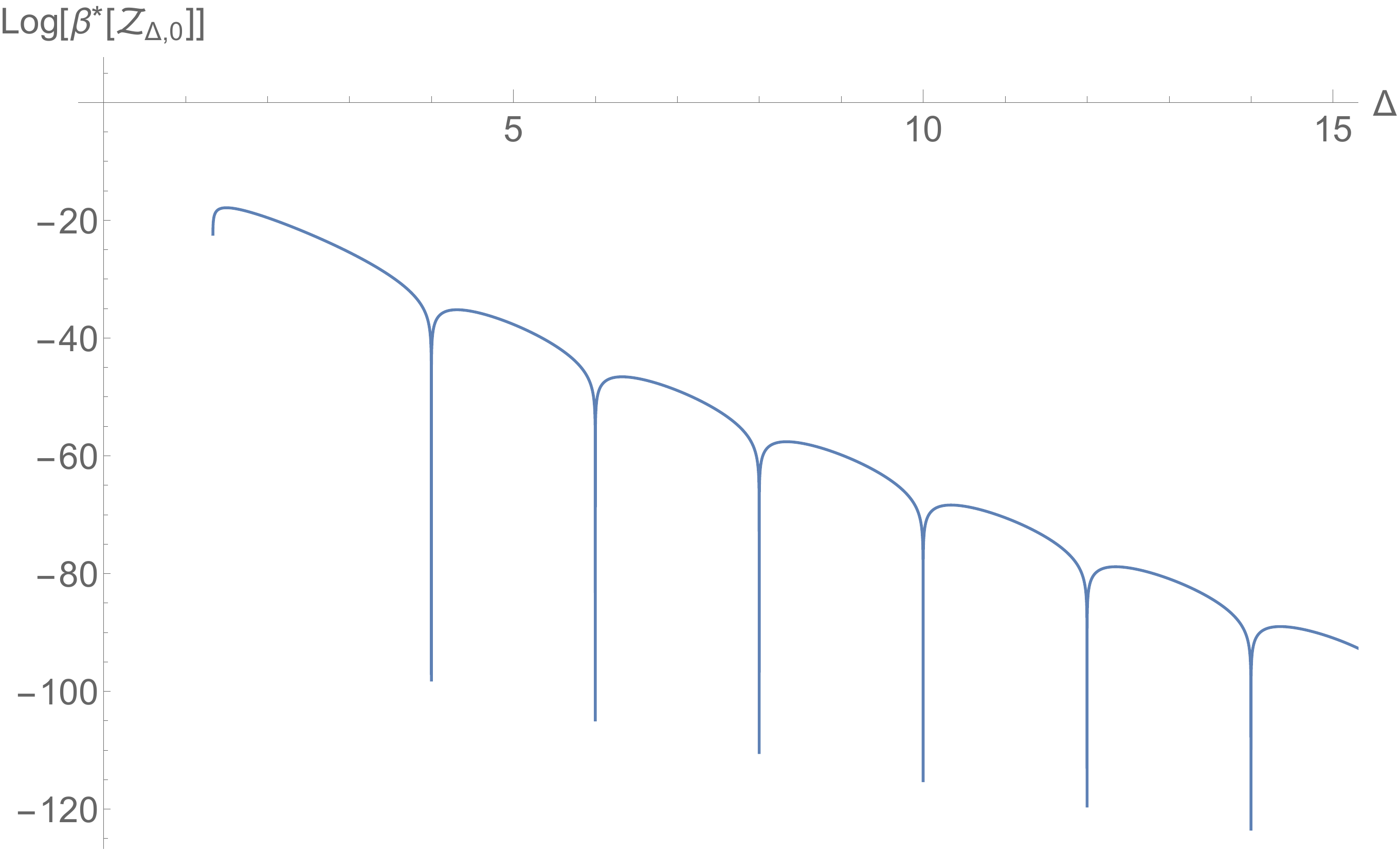} \qquad
\includegraphics[width=.40\textwidth]{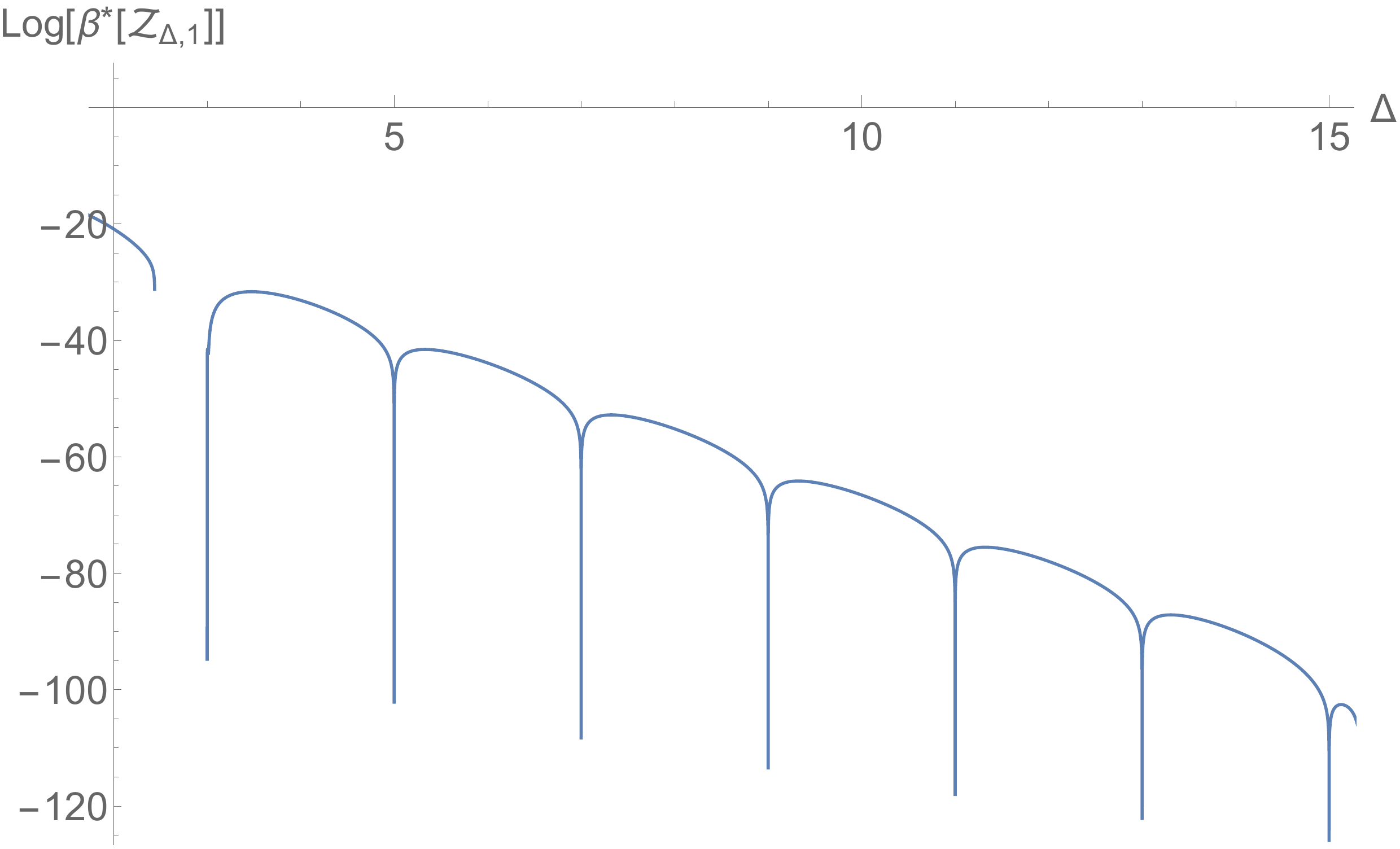}
\end{center}
\caption{The extremal functional $\beta^*$ acting on spin-0 (left) and spin-1 (right) primaries at $c=16$ and $\D_t =2$.}
\label{EFM_E8E8}
\end{figure}

\begin{table}[ht]
\centering
{
\begin{tabular}{|c |c || c| c |}
\hline
 \rule{0in}{3ex}     $(h,\bar{h})$         & Max. Deg & $(h,\bar{h})$ &  Max. Deg \\
\hline
\rule{0in}{3ex} $(1, 1)$ &  246016.0000000000 & $(1, 0)$ &  496.000000000000 \\
\hline
\rule{0in}{3ex} $(2, 1)$ &  34350480.00000000 & $(2, 0)$ &  69255.0000000000   \\
 \hline
\rule{0in}{3ex} $(3, 1)$ &  1014200960.000000  & $(3, 0)$ & 2044760.00000000  \\
 \hline
\rule{0in}{3ex} $(2, 2)$ &  4796255025.000000  & $(4, 0)$ & 32485860.0000000   \\
 \hline
\rule{0in}{3ex} $(3, 2)$ &  283219707600.0000  & $(5, 0)$ & 357674373.000000   \\
 \hline
\end{tabular}
\caption{ The maximum value of degeneracies for low-lying states in a putative CFT with $c=16$. Here the maximum number of derivative is set to $N_{\textrm{max}}=55$, while the spin was truncated at $j_{\textrm{max}}=40$.}
\label{E8E8deg}
}
\end{table}

The extremal spectrum of a putative CFT with $(c=16, \Delta_t=2)$ and their maximal degeneracies are presented in Figure \ref{EFM_E8E8} and Table \ref{E8E8deg}, respectively.
The partition function read off from Table \ref{E8E8deg}. It can be written in a simple form
\begin{align}
Z_{c=16}(\tau, \bar{\tau}) = j(\t)^{\frac{2}{3}} \bar{j}(\bar{\t})^{\frac{2}{3}} \ ,
\label{z16}
\end{align}
where $j(\t)$ denotes the modular invariant $j$-function \eqref{j inv}. The result \eqref{z16} perfectly agrees with the partition function of the $(\widehat{E}_8 \times \widehat{E}_8)_1$ WZW model.


\end{itemize}

\subsection{$c \ge 8$ RCFTs without Kac-Moody symmetry}

In the previous subsection, we uncovered that the WZW models with level one for Deligne's exceptional series can maximize
the degeneracies of the
extremal spectrum at ten among sixteen special points
on the numerical bound of $\D_t$, as depicted in Figure \ref{W2_Fig4_largeC2_CC2}.
It is known that the characters of such WZW models are the solutions to (\ref{MDE02}).
In this subsection, we will show that the degeneracies of every extremal spectrum of certain RCFTs saturate their upper bound at the remaining $3+3$ points.
It turns out that such RCFTs have characters that agree with the solutions to \eqref{3rd MDE}, and have no Kac-Moody symmtery but finite group symmetry of very large order. Such finite groups include the Monster and the Baby Monster groups.

\begin{itemize}
\item {\bf $c=24$ Monster CFT}
\begin{figure}[h]
\begin{center}
\includegraphics[width=.40\textwidth]{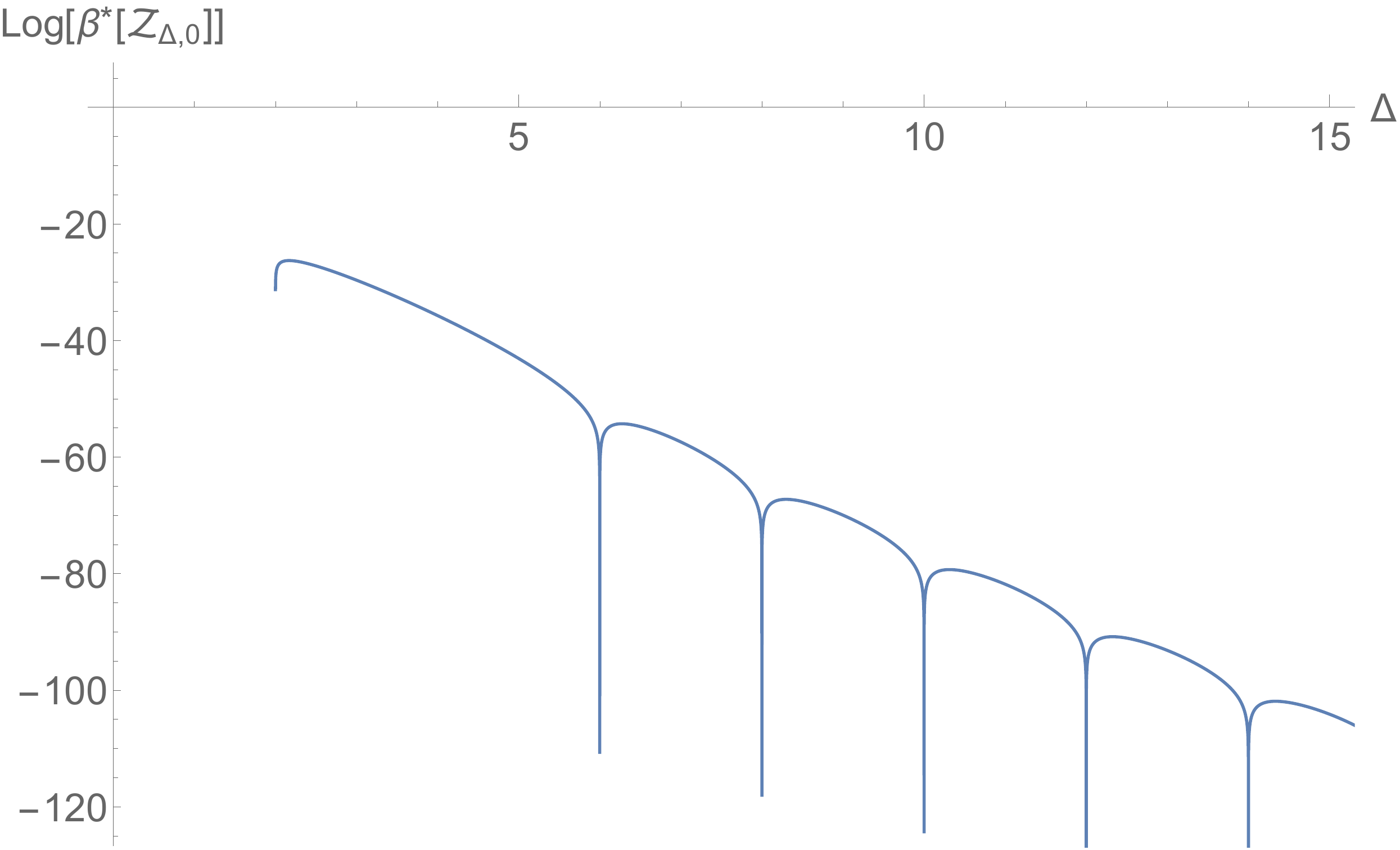} \qquad
\includegraphics[width=.40\textwidth]{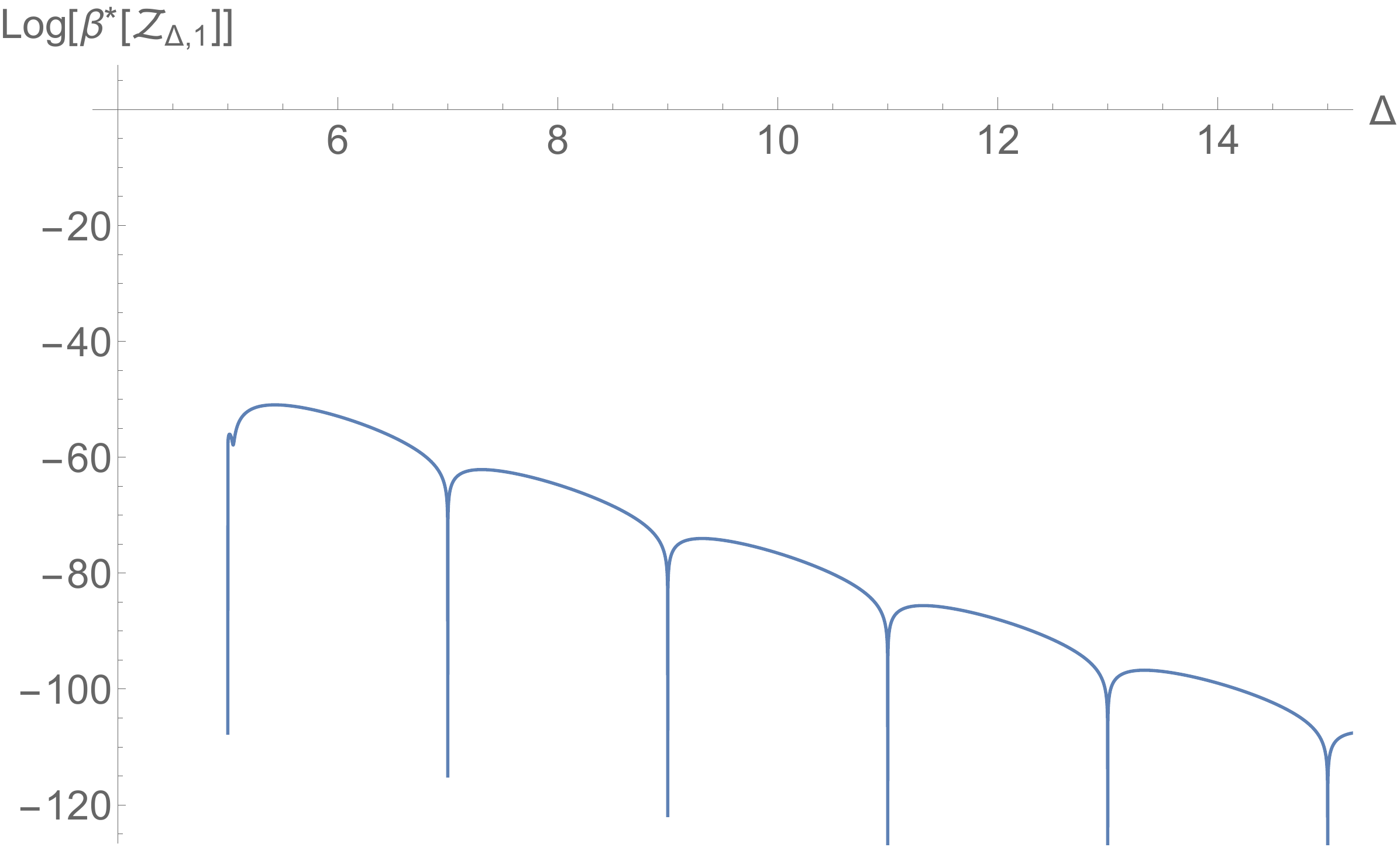}
\end{center}
\caption{The extremal functional $\beta^*$ acting on spin-0 (left) and spin-1 (right) primaries at $c=24$ and $\D_t =4$.}
\label{EFM_Monster}
\end{figure}

\begin{table}[ht]
\centering
{
\begin{tabular}{|c |c |}
\hline
 \rule{0in}{3ex}     $(h,\bar{h})$         & Max. Deg  \\
\hline
\rule{0in}{3ex} $(2, 2)$ &  38762915689.000000   \\
\hline
\rule{0in}{3ex} $(2, 3)$ &  4192992837508.0000      \\
 \hline
\rule{0in}{3ex} $(2, 4)$ &  165895451930858.000    \\
 \hline
\rule{0in}{3ex} $(3, 3)$ &  453556927359376.000      \\
 \hline
\rule{0in}{3ex} $(3, 4)$ &  17944946332265576.00     \\
 \hline
\rule{0in}{3ex} $(4, 4)$ &  709990476262174276.00     \\
 \hline
\end{tabular}
\caption{The maximum value of degeneracies for low-lying states in a putative CFT with $c=24$. The maximum number of derivatives is set to $N_{\textrm{max}}=55$, while the spin was truncated at $j_{\textrm{max}}=40$.}
\label{Monsterdeg}
}
\end{table}

Let us search for a hypothetical CFT with $c=24$ and $\Delta_t=4$ that contains the extremal spectrum, illustrated in Figure \ref{EFM_Monster}, with the maximal degeneracies. From Table \ref{Monsterdeg}, we find its partition function can be written as
\begin{align}
\begin{split}
{Z}_{c=24}(\t, \bar{\t}) =  & \ {\chi}_{0}(\tau) \bar{{\chi}}_{0}(\bar{\tau})  + \Big( 4192992837508 {\chi}_{2}(\tau) \bar{{\chi}}_{3}(\bar{\tau}) + \text{c.c.} \Big)   \\
&   + 38762915689 {\chi}_{2}(\tau) \bar{{\chi}}_{2}(\bar{\tau}) + 453556927359376 {\chi}_{3}(\tau) \bar{{\chi}}_{3}(\bar{\tau})  \\
& + \Big( 165895451930858 {\chi}_{2}(\tau) \bar{{\chi}}_{4}(\bar{\tau}) + \text{c.c.} \Big)  + \cdots \\
&+   \Big( 196883 {\chi}_{2}(\tau) \bar{{\chi}}_{0}(\bar{\tau}) + \text{c.c.} \Big) + \cdots \\
= &\ (j(\t)-744) (\bar{j}(\bar{\t})-744).
\label{k1ECFTmpartition}
\end{split}
\end{align}
We therefore identify the putative CFT of our interest as the Monster CFT of \cite{frenkel1984natural, Frenkel:1988xz}.

\item {\bf ``$c=32$ Extremal CFT"}
\begin{figure}[h]
\begin{center}
\includegraphics[width=.40\textwidth]{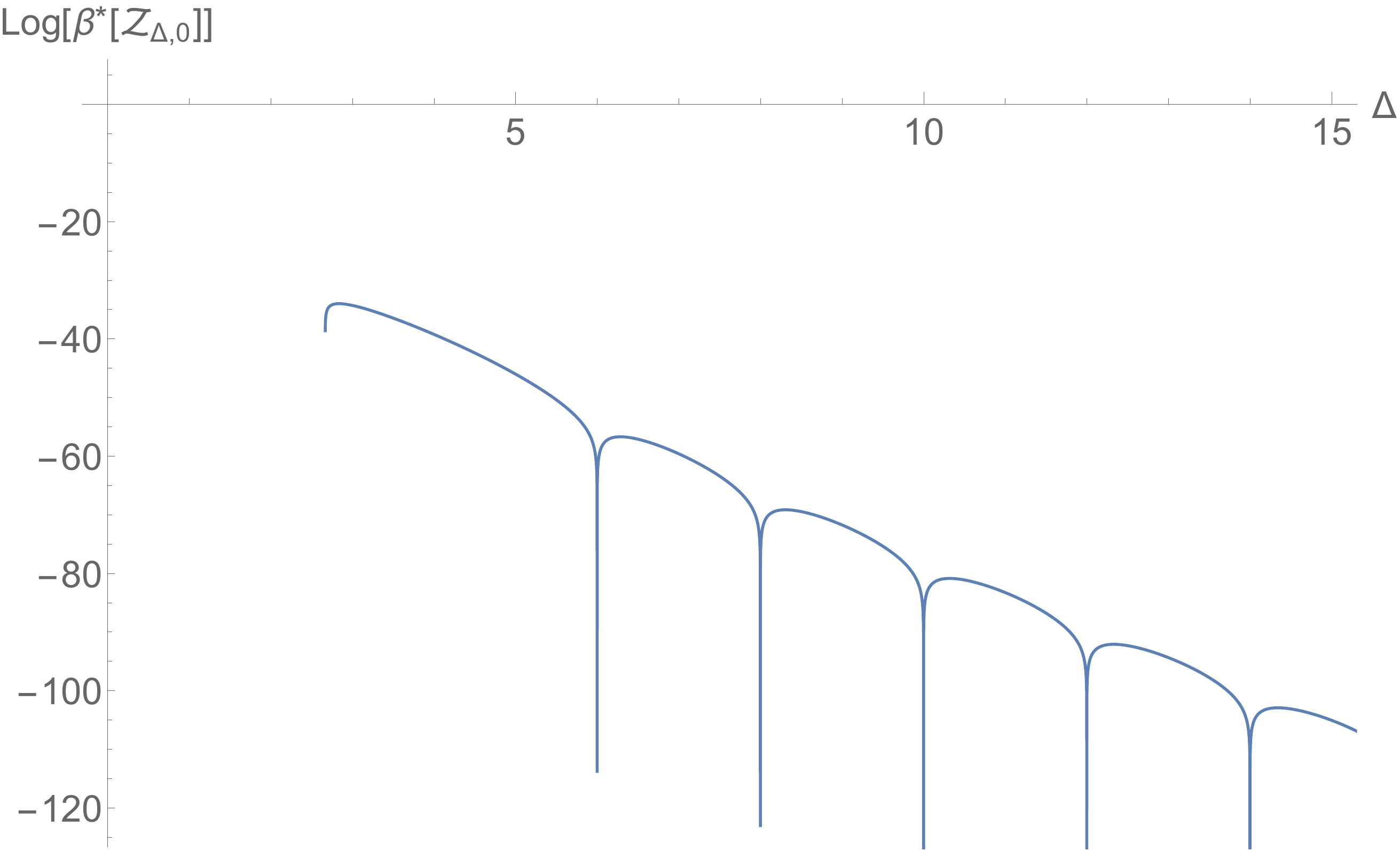} \qquad
\includegraphics[width=.40\textwidth]{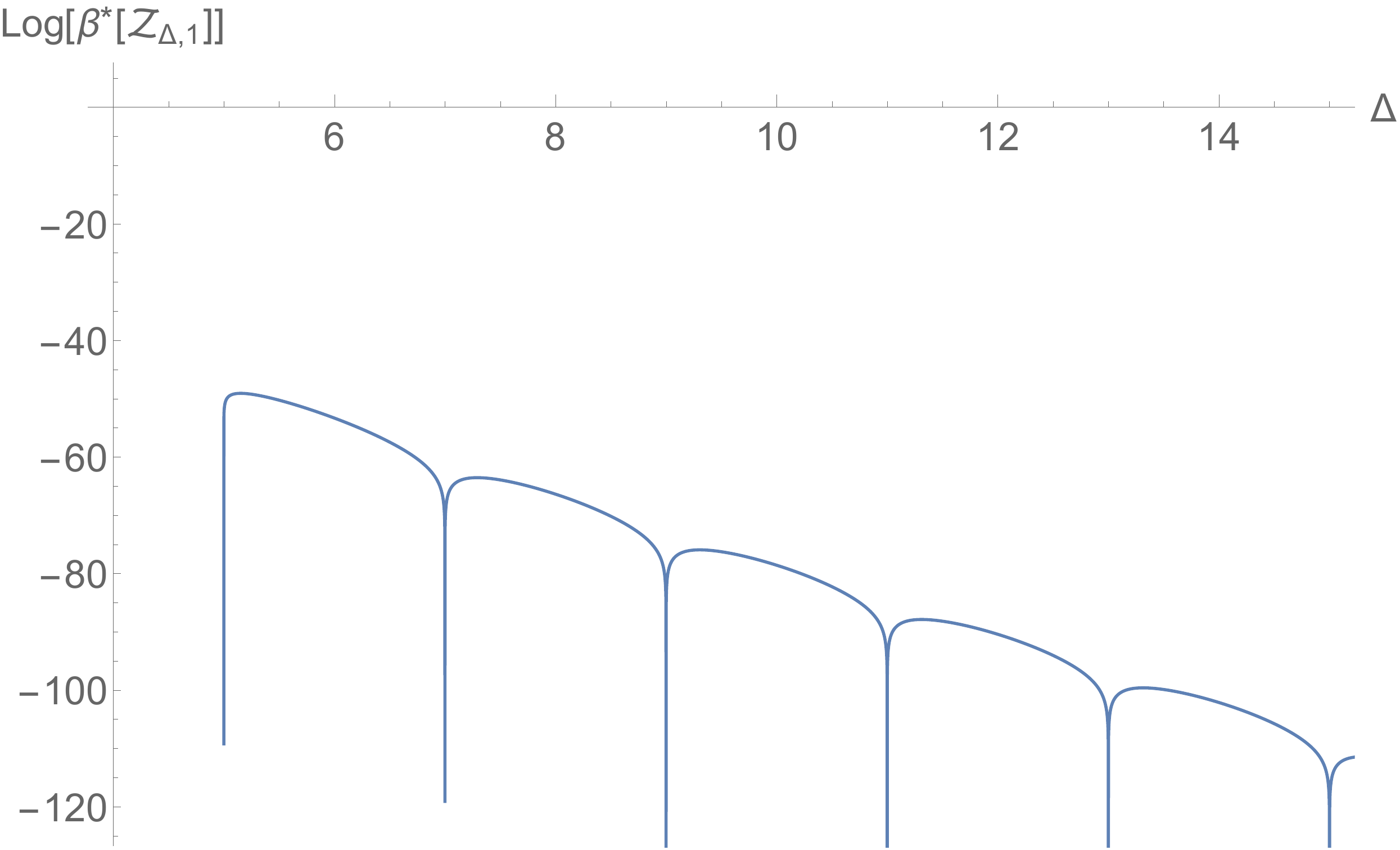}
\end{center}
\caption{The extremal functional $\beta^*$ acting on spin-0 (left) and spin-1 (right) primaries at $c=32$ and $\D_t =4$.}
\label{EFM_c32}
\end{figure}

\begin{table}[ht]
\centering
{
\begin{tabular}{|c |c |}
\hline
 \rule{0in}{3ex}     $(h,\bar{h})$         & Max. Deg  \\
\hline
\rule{0in}{3ex} $(2, 2)$ &  19461087009.000000000   \\
\hline
\rule{0in}{3ex} $(2, 3)$ &  9652699156464.0000004    \\
 \hline
\rule{0in}{3ex} $(2, 4)$ &  966591762372600.00002    \\
 \hline
\rule{0in}{3ex} $(3, 3)$ &  4787738781606144.0000   \\
 \hline
\rule{0in}{3ex} $(3, 4)$ &  479429514136809600.01    \\
 \hline
\rule{0in}{3ex} $(4, 4)$ &  48008604794505640003.18     \\
 \hline
\end{tabular}
\caption{ The maximum value of degeneracies for low-lying states in putative CFT with $c=32$. The maximum number of derivatives is set to $N_{\textrm{max}}=55$, while the spin is truncated at $j_{\textrm{max}}=40$.}
\label{c32deg}
}
\end{table}

The extremal spectrum with $c=32$ and $\Delta_t=4$ have scalar primaries of $\Delta_{j=0} = \{4+2n\}$
and spin-one primaries of $\Delta_{j=1} = \{5+2n\}$ for $n \in \mathbb{Z}_{\ge 0}$, as depicted in Figure \ref{EFM_c32}.
The upper bounds on degeneracies of the extremal spectrum
are summarized in Table \ref{c32deg}.
One can easily see that the partition function of the $c=32$ extremal CFT given below,
\begin{align}
    Z_{c=32}(\t,\bar{\t}) = \left(j(\t)^{\frac{4}{3}}-992 j(\t)^{\frac{1}{3}}\right)
    \left( \bar{j}(\bar{\t})^{\frac{4}{3}} - 992\bar{j}(\bar{\t})^{\frac{1}{3}}\right),
\end{align}
is consistent with the maximal degeneracies as in Table \ref{c32deg}.

As discussed in \cite{Tuite:2008pt}, the vertex operator algebra giving the above partition function can be constructed
as a $\mathbf{Z}_2$ orbifold of free bosons on an extremal self-dual lattice
of rank $32$. However, the classification and the automorphism group
of these lattices have been poorly understood.

\end{itemize}

\begin{itemize}
\item {\bf $c=8$ RCFT without Kac-Moody symmetry} :
\begin{figure}[h]
\begin{center}
\includegraphics[width=.40\textwidth]{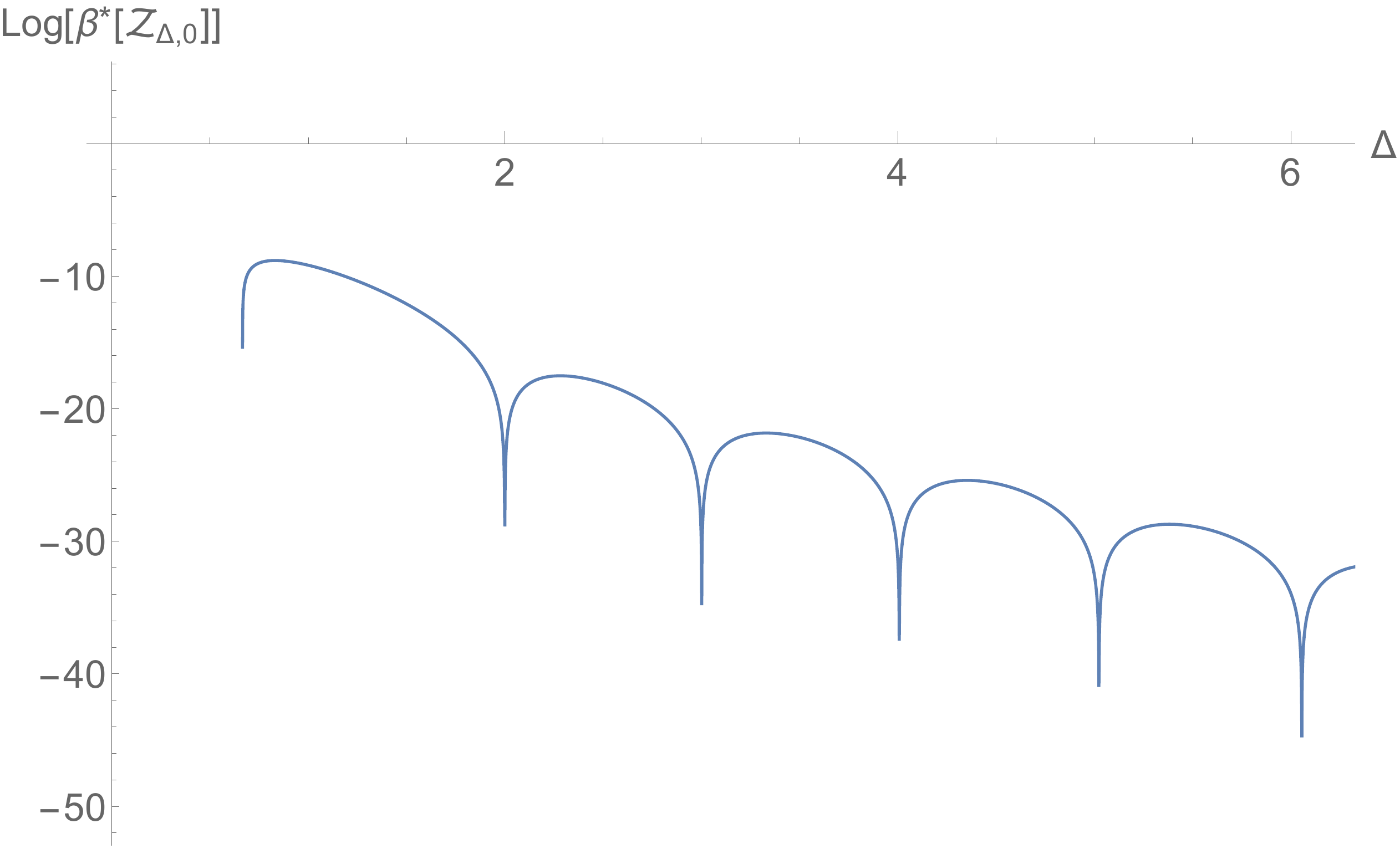} \qquad
\includegraphics[width=.40\textwidth]{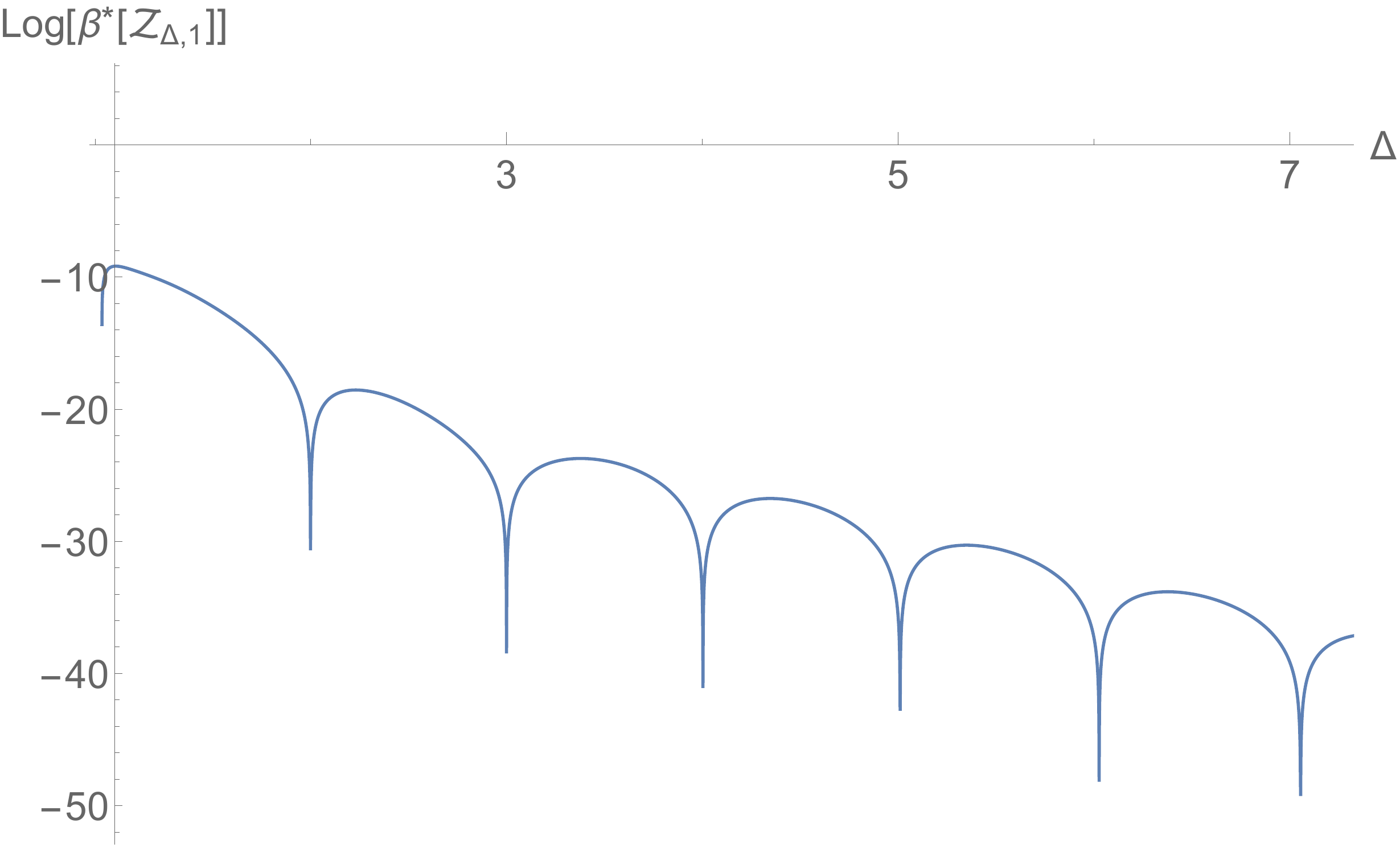}
\label{NKMc8_spectrum}
\end{center}
\caption{The extremal functional $\beta^*$ acting on spin-0 (left) and spin-1 (right) primaries at $c=8$ and $\D_t =1$, without the spin-1 conserved current.}
\label{EFM_c8gapped}
\end{figure}

\begin{table}[ht]
\centering
{
\begin{tabular}{|c |c || c| c || c|c|}
\hline
 \rule{0in}{3ex}     $(h,\bar{h})$         & Max. Deg & $(h,\bar{h})$ &  Max. Deg & $(h,\bar{h})$ & Max. Deg\\
\hline
\rule{0in}{3ex} $(\frac{1}{2}, \frac{1}{2})$ &  496.0000000 & $(1, 1)$ & 33728.00000  & $(2, 0)$ &  155.000000  \\
\hline
\rule{0in}{3ex} $(\frac{1}{2}, \frac{3}{2})$ &  17360.00000 & $(2, 1)$ & 505920.0000  & $(3, 0)$ &  868.000000  \\
\hline
\rule{0in}{3ex} $(\frac{3}{2}, \frac{3}{2})$ &  607600.0009 & $(2, 2)$ & 7612825.000  & $(4, 0)$ &  5610.00000   \\
 \hline
\end{tabular}
\caption{ The maximum value of degeneracies for low-lying states in putative CFT with $c=8$, without the spin-1 conserved current. The maximum number of derivatives is set to $N_{\textrm{max}}=55$, while the spin is truncated at $j_{\textrm{max}}=40$.}
\label{c8gapped}
}
\end{table}

Note first that the CFT of our interest does not contain
spin-one currents at all. Applying the EFM, we investigate the spin-0 and
spin-1 extremal spectrum of a CFT with ($c=8$, $\D_t$=1).
Their conformal dimensions can be read from Figure \ref{EFM_c8gapped},
$\Delta_{j=0}  = \{1+n \} $ and $\Delta_{j=1}  = \{2+n \} $ for $n \in \mathbb{Z}_{\ge 0}$.
We also analyze the maximal degeneracies of the extremal spectrum, summarized
in Table \ref{c8gapped}.
These results suggest that the partition function of a putative CFT with ($c=8,\D_t=1)$
but no Kac-Moody symmetry admits the character decomposition as
\begin{align}
{Z}_{c=8}(\t, \bar{\t}) &= {\chi}_{0}(\tau) \bar{{\chi}}_{0}(\bar{\tau})+ 496 {\chi}_{\frac{1}{2}}(\tau) \bar{{\chi}}_{\frac{1}{2}}(\bar{\tau}) + 17360 \Big( {\chi}_{\frac{1}{2}}(\tau) \bar{{\chi}}_{\frac{3}{2}}(\bar{\tau}) + \text{c.c.}\Big)  \nonumber \\
&+ 33728 {\chi}_{1}(\tau) \bar{{\chi}}_{1}(\bar{\tau}) + 505920 \Big( {\chi}_{2}(\tau) \bar{{\chi}}_{1}(\bar{\tau}) + \text{c.c.}\Big) +  7612825 {\chi}_{2}(\tau) \bar{{\chi}}_{2}(\bar{\tau}) \nonumber \\
& +155 \Big( {\chi}_{2}(\tau) \bar{{\chi}}_{0}(\bar{\tau})+ \text{c.c.} \Big) + 868 \Big( {\chi}_{3}(\tau) \bar{{\chi}}_{0}(\bar{\tau})+ \text{c.c.}\Big) + \cdots .
\label{c8NKM_reduced}
\end{align}

Using the solutions to (\ref{3rd MDE}) with $c=8$,
\begin{align}
\label{O102}
\begin{split}
    f_{\text{vac}}^{c=8}(\tau) &= q^{-\frac{1}{3}}
    \left(1 + 156 q^2 + 1024 q^3 + 6780 q^4 + \mathcal{O}(q^5) \right),
    \\
    f_{\frac{1}{2}}^{c=8}(\tau) &= q^{\frac{1}{2}-\frac{8}{24}}
    \left(1 +36 q +394 q^2 +2776 q^3 +15155 q^4 + \mathcal{O}(q^5) \right),
     \\
    f_{1}^{c=8}(\tau) &=  q^{1-\frac{8}{24}}
    \left(1 +16 q +136 q^2 +832 q^3+4132 q^4  + \mathcal{O}(q^5) \right)
    \end{split}
\end{align}
it is straightforward to check that \eqref{c8NKM_reduced} can be recast into the following form,
\begin{equation}
    Z_{c=8}(\t,\bar\t) = f_{\text{vac}}^{c=8}(\tau) f_{\text{vac}}^{c=8}(\bar{\tau}) +
    496 f_{\frac{1}{2}}^{c=8}(\tau) \bar{f}_{\frac{1}{2}}^{c=8}  +
    33728 f_{1}^{c=8}(\tau) \bar{f}_{1}^{c=8}(\bar{\tau})\ .
\label{c8g partition}
\end{equation}
It is discussed in \cite{Tuite:2008pt} that the automorphism group of
the chiral CFT (more precisely, vertex operator algebra) having
the vacuum character $f_\text{vac}^{c=8}(\t)$ of (\ref{O102})
is the finite group of Lie type, $ O^+_{10}(2).2$. We use
the GAP package \cite{GAP4} to obtain the dimensions of
the irreducible representations of $ O^+_{10}(2)$, some of which
are $155, 340, 868, 2108, 7905, 14756, 31620, 55335, 73780, 505920, 1048576, 1422900$.
Indeed, one can see that various coefficients in
the character decomposition (\ref{c8NKM_reduced})
can be expressed as sums of those dimensions
\begin{gather}
155, \quad 496 = 1 + 155 + 340, \quad 868, \nonumber \\
 17360 = 1+ 155+340+2108+14756, \\
33728 = 2108 + 31620, \quad 505920. \nonumber
\label{c8 decomposition}
\end{gather}

\item {\bf $c=16$ RCFT without Kac-Moody symmetry} :
\begin{figure}[h]
\begin{center}
\includegraphics[width=.40\textwidth]{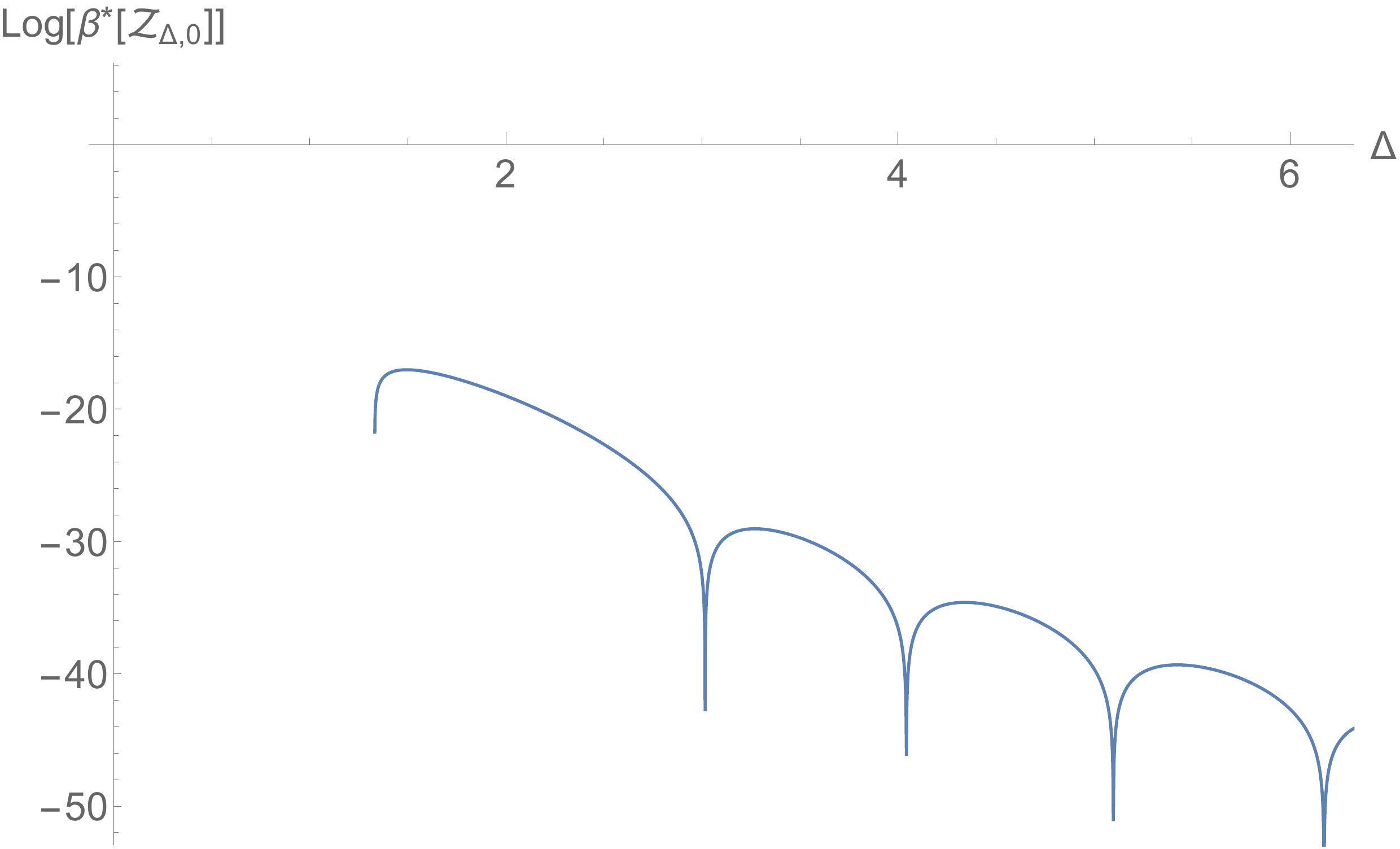} \qquad
\includegraphics[width=.40\textwidth]{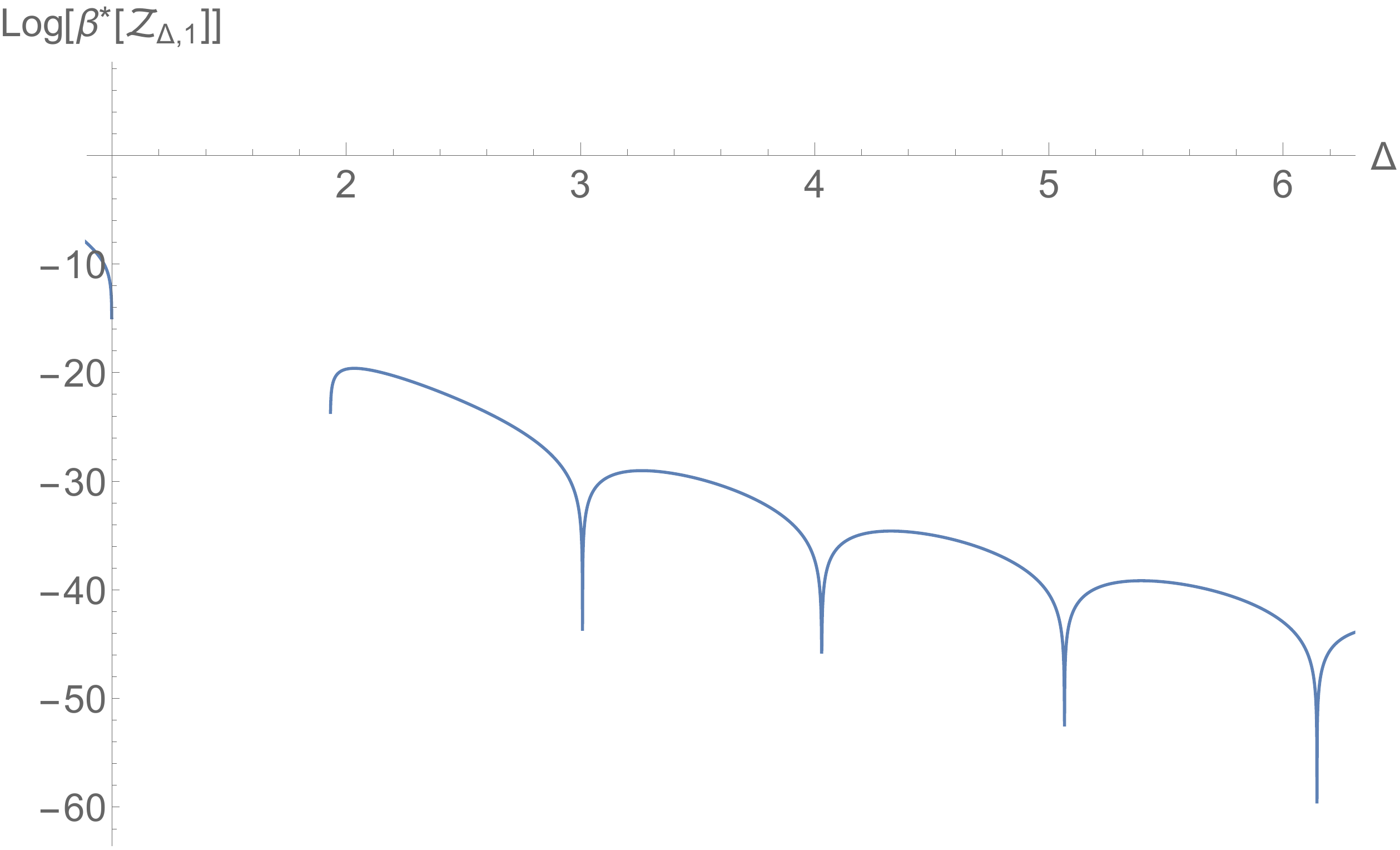}
\end{center}
\caption{The extremal functional $\beta^*$ acting on spin-0(left) and spin-1(right) primaries at $c=16$ and $\D_t =2$, without the spin-1 conserved current.}
\label{EFM_c16gapped}
\end{figure}

\begin{table}[ht]
\centering
{
\begin{tabular}{|c |c || c| c || c|c||}
\hline
 \rule{0in}{3ex}     $(h,\bar{h})$         & Max. Deg & $(h,\bar{h})$ &  Max. Deg & $(h,\bar{h})$ & Max. Deg\\
\hline
\rule{0in}{3ex} $(\frac{3}{2}, \frac{3}{2})$ &  32505856.0032 & $(1, 1)$ & 134912.0000  & $(2, 0)$ &  2295.00000  \\
\hline
\rule{0in}{3ex} $(\frac{3}{2}, \frac{5}{2})$ &  1657798656.0001 & $(2, 1)$ & 18213120.00  & $(3, 0)$ &  63240.0000  \\
\hline
\rule{0in}{3ex} $(\frac{3}{2}, \frac{7}{2})$ &  34228666368.005 & $(2, 2)$ & 2464038225.003 & $(4, 0)$ &  1017636.00    \\
 \hline
\end{tabular}
\caption{ This table summarize the maximal degeneracy of first few states in gapped $c=16$ CFT. The number of derivative is set by 55, while the spin is truncated at 40.}
\label{c16gapped}
}
\end{table}
Let us search for a hypothetical CFT with $(c=16, \D_t=2)$ that does not
have the Kac-Moody symmetry. The extremal spectrum of such a CFT and their upper bound of degeneracies are illustrated in Figure \ref{EFM_c16gapped} and Table \ref{c16gapped}.
These results imply that the partition function of the CFT of our interest
can be expanded as
\begin{align}
    {Z}_{c=16}(q, \bar{q}) = & \ {\chi}_{0}(\tau) \bar{{\chi}}_{0}(\bar{\tau})
    + 32505856 {\chi}_{\frac{3}{2}}(\tau) \bar{{\chi}}_{\frac{3}{2}}(\bar{\tau}) + 1657798656 \Big( {\chi}_{\frac{3}{2}}(\tau) \bar{{\chi}}_{\frac{5}{2}}(\bar{\tau}) + \text{c.c.} \Big) \nonumber \\
    &+ 134912 {\chi}_{1}(\tau) \bar{{\chi}}_{1}(\bar{\tau}) + 18213120 \Big({\chi}_{2}(\tau) \bar{{\chi}}_{1}(\bar{\tau}) + \text{c.c.} \Big)
    \label{c16NKM_reduced} \\
    & +2295 \Big( {\chi}_{2}(\tau)\bar{{\chi}}_{0}(\bar{\tau}) + \text{c.c.}\Big) + 63240 \Big({\chi}_{3}(\tau)
    \bar{{\chi}}_{0}(\bar{\tau})+ \text{c.c.} \Big) + \cdots \ ,
    \nonumber \\ = &
    f_{\text{vac}}^{c=16}(\tau) \bar{f}_{\text{vac}}^{c=16}(\bar{\tau}) + 134912 f_{1}^{c=16}(\tau) \bar{f}_1^{c=16}(\bar{\tau})
    +  32505856 f_{\frac23}^{c=16}(\tau) \bar{f}_{\frac23}^{c=16}(\bar{\tau}),
    \nonumber
\end{align}
where $f_{\text{vac}}^{c=16}(\tau)$, $f_1^{c=16}(\tau)$ and
$f_{\frac23}^{c=16}(\tau)$ are solutions to \eqref{3rd MDE} with $c=16$.
Note that various coefficients in the above decomposition of \eqref{c16NKM_reduced}
can be written as sum of the dimensions of irreducible representations of $O^{+}_{10}(2)$,
\begin{gather}
2295 = 1 + 186 + 2108, \quad 63240 = 55335 + 7905, \nonumber \\
134912 = 186 + 340 + 868 + 22848 + 110670, 
\\
18213120 = 12 \times 1422900 + 1048576 + 73780+ 14756 + 868 +  340 \ . \nonumber
\end{gather}
This observation suggests that the above putative CFT with $(c=16, \D_t=2)$ but no Kac-Moody symmetry may have the $O^{+}_{10}(2)$ symmetry.

\item {\bf Baby Monster CFT with $c=\frac{47}{2}$}

\begin{table}[ht]
\centering
{
\begin{tabular}{|c |c|c |}
\hline
 \rule{0in}{3ex}$N_{\textrm{max}}$ & $\Delta_t$    & Max. degeneracy of $(h,\bar{h}) = (\frac{3}{2}, \frac{3}{2})$     \\
\hline
\rule{0in}{3ex}$41$ & 3.102 & 20633319.646029717379   \\
\hline
\rule{0in}{3ex}$51$ & 3.058 & 20060048.798539029429      \\
\hline
\rule{0in}{3ex}$61$ & 3.034 & 19728535.695677188476      \\
\hline
\rule{0in}{3ex}$71$ & 3.023 & 19597158.910830818660          \\
 \hline
 \rule{0in}{3ex}$81$ & 3.016 & 19499859.838240040877   \\
 \hline
\end{tabular}
\caption{ The degeneracy upperbound of weight $(h,\bar{h}) = (\frac{3}{2}, \frac{3}{2})$ primary, for various $N_{\textrm{max}}$.}
\label{BM_lowest_MD}
}
\end{table}
\begin{table}[ht]
\centering
{
\begin{tabular}{|c |c || c| c ||}
\hline
 \rule{0in}{3ex}     $(h,\bar{h})$         & Max. Deg   $(h,\bar{h})$   &  $(h,\bar{h})$   & Max. Deg  \\
\hline
\rule{0in}{3ex} $(\frac{3}{2}, \frac{3}{2})$ &  19105641.026984403127  & $(\frac{5}{2}, \frac{3}{2})$ &  4980203754.2560961756     \\
\hline
\rule{0in}{3ex} $(\frac{5}{2}, \frac{5}{2})$ &  1298173112605.3499336 & $(2, 2)$ &  9265025041.3227338031     \\
\hline
\rule{0in}{3ex} $(3, 2)$ & 919296372501.31519351   &  $(3, 3)$ &  91214629887092.699664      \\
 \hline
 \rule{0in}{3ex} $(\frac{31}{16}, \frac{31}{16})$ &  9265217540.6086142750 & $(\frac{47}{16}, \frac{31}{16})$ & 1011288637613.8107313 \\
 \hline
\end{tabular}
 \caption{ This table summarize the maximal degeneracy of first few states in gapped $c=47/2$ CFT. The maximum number of derivative is set to $N_{\textrm{max}}=65$, $\delta=10$, while the spin was truncated at $j_{\textrm{max}}=40$.}
\label{babymonster_result}
}
\end{table}

In \cite{Hampapura:2016mmz}, the three solutions to \eqref{3rd MDE} with $c=\frac{47}{2}$ are shown to be same as the characters of the Baby Monster vertex operator algebra\cite{Hoehn:2007aa},
\begin{align}
\begin{split}
f^{c=\frac{47}{2}}_{\text{vac}}(\tau) = \chi_{\text{VB}_{(0)}^{\natural}}&= q^{-\frac{47}{48}} ( 1 + 96256 q^2 + 9646891 q^3 + 366845011 q^4  + \cdots )  \\
f^{c=\frac{47}{2}}_{\frac{3}{2}}(\tau) = \chi_{\text{VB}_{(1)}^{\natural}}&= q^{\frac{3}{2}-\frac{47}{48}} ( 4371 + 1143745 q + 64680601 q^2  + \cdots) \\
f^{c=\frac{47}{2}}_{\frac{31}{16}}(\tau) = \chi_{\text{VB}_{(2)}^{\natural}}&= q^{\frac{31}{16}-\frac{47}{48}} ( 96256  + 10602496 q + 420831232 q^2  + \cdots).
\label{character_VB}
\end{split}
\end{align}
We observed in Figure \ref{W2_Fig4_largeC2_CC2} that the numerical bound at $c=\frac{47}{2}$ is given by $\Delta_t^*=3$\footnote{As shown in Table \ref{BM_lowest_MD}, the upper bound on $\Delta_t$ at $c=\frac{47}{2}$ is approaching to 3 as we increase the total number of derivatives $N_{\text{max}}$.}, which is consistent with the character $f^{c=47/2}_{3/2}(\tau)$ of \eqref{character_VB}. 

On can thus naturally expect that the corresponding CFT with $c=\frac{47}{2}$ and $\D_t=3$ has the Baby Monster symmetry. To prove this hypothesis, we further analyze the maximal degeneracies of the scalar primary of $\Delta=3$, as summarized in Table \ref{BM_lowest_MD}. It appears that the upper bound is indeed converging to $4371^2 = 19105641$ predicted from the character. However the convergence of bound is not fast enough as the order of derivative $N_{\textrm{max}}$ is increased. 

To circumvent the above numerical difficulty, we employ an alternative strategy of adding low-lying discrete spectrum in the semi-definite programming \cref{beta_star_1,beta_star_2,beta_star_3} by hand.
More precisely, let us assume that the partition function of the Baby Monster CFT\footnote{The characters of the Baby Monster modules $\chi_{\text{VB}_{(0, 1, 2)}^{\natural}}$ are given by
$$j(\t) -744= \chi_{\text{VB}_{(0)}^{\natural}}\hspace*{-0.2cm}(\t) \chi_\text{vac}^\text{Ising}(\t)+
\chi_{\text{VB}_{(1)}^{\natural}}\hspace*{-0.2cm}(\t) \chi_{h=\frac12}^\text{Ising}(\t)
+ \chi_{\text{VB}_{(3)}^{\natural}}\hspace*{-0.2cm}(\t)\chi_{h=\frac{1}{16}}^\text{Ising}(\t),$$
where $\chi^\text{Ising}_h(\t)$ denote the Virasoro characters of the Ising model. From this,
it is obvious that the ansatz (\ref{BM_ansatz}) is invariant under the modular transformation.
} can be expressed as
\begin{equation}
Z_{c=47/2} = \left|f^{c=\frac{47}{2}}_{\text{vac}}(\tau)\right|^2 + \left|f^{c=\frac{47}{2}}_{\frac{3}{2}}(\tau)\right|^2
+ \left|f^{c=\frac{47}{2}}_{\frac{31}{16}}(\tau)\right|^2\ .
\label{BM_ansatz}
\end{equation}
Now, we add discrete set of primaries having the conformal weights as below to the SDP problem:
\begin{align}
\begin{split}
 &\ \{ (\Delta, j) \big| \Delta = 3+j, 5+j, 7+j, \cdots ,2\d+3+j \}  \\
              \cup &\ \{ (\Delta, j) \big| \Delta = 4+j, 6+j, 8+j, \cdots ,2\d+4+j \}  \\
              \cup &\ \{ (\Delta, j) \big| \Delta = \frac{31}{8}+j, \frac{47}{8}+j,  \frac{63}{8}+j,\cdots ,2\d+\frac{31}{8}+j \} ,
\end{split}
\end{align}
for $ 0\le j \le j_{\textrm{max}}$ and an arbitrary positive integer $\d$.
For instance, the SDP problem with $\d=10$ can result in
the maximal degeneracies of the extremal spectrum summarized in
Table \ref{babymonster_result}. It is easy to show that
the partition function (\ref{BM_ansatz}) saturates
these upper bounds. We can thus expect that the putative CFT of our
interest has the Baby Monster symmetry.
\end{itemize}

\section{Bootstrapping with $\mathcal{W}$-algebra}
\label{sec:Walgebra}

The two-dimensional CFTs with $\CW(A_2)=\mathcal{W}(2,3)$ symmetry have been studied recently in \cite{Apolo:2017xip,Afkhami-Jeddi:2017idc}. To investigate the universal constraints on the spectrum of higher-spin ``irrational'' CFTs, i.e., $c>2$, the authors of \cite{Apolo:2017xip,Afkhami-Jeddi:2017idc} apply the modular bootstrap method to the torus two-point function
\begin{align}
\text{Tr}\left[ W_0^2 q^{L_0-\frac{c}{24}} \bar{q}^{\bar{L}_0-\frac{c}{24}} \right] \ .
\end{align}
On the other hand, we will focus on the torus partition function
\begin{align}
 Z(\t, \bar{\t}) = \tr \left[ q^{L_0 - \frac{c}{24}} \bar{q}^{\bar{L}_0 - \frac{c}{24}} \right] \ ,
\end{align}
under the assumption that there is a $\CW(d_1, d_2, \ldots, d_r)$-algebra symmetry. This allows us to expand the partition function in the form
\begin{align}
 \begin{split}
  Z(\t,\bar \t) & =  \chi_0(\t) \bar{\chi}_0(\bar \t) + \sum_{h,\bar h}
  d_{h,\bar h} \Big[ \chi_h(\t) \bar{\chi}_{\bar h}(\bar \t) +
  \chi_{\bar h}(\t) \bar{\chi}_{h}(\bar \t) \Big]
  \\ & + \sum_{j=1} d_j \Big[ \chi_j(\t) \bar{\chi}_{0}(\bar \t) +
  \chi_0(\t) \bar{\chi}_{j}(\bar \t) \Big]\ ,
\end{split}
\end{align}
where the characters are given by (assuming the non-vacuum module is non-degenerate)
\begin{align}
 \chi_0(\t) = \frac{q^{-\frac{c-r}{24}}}{\eta(\t)^{r}} \prod_{i=1}^r \prod_{j=1}^{d_i-1}(1-q^j) \ , \qquad
  \chi_h (\t) = \frac{q^{h-\frac{c-r}{24}}}{\eta(\t)^r} \ .
\end{align}
In this section, we examine how the modular invariance for the theories with various $\mathcal{W}$-symmetries constrains the spectrum.

\subsection{Numerical Bounds}


\begin{figure}[t]
\begin{center}
\includegraphics[width=.84\textwidth]{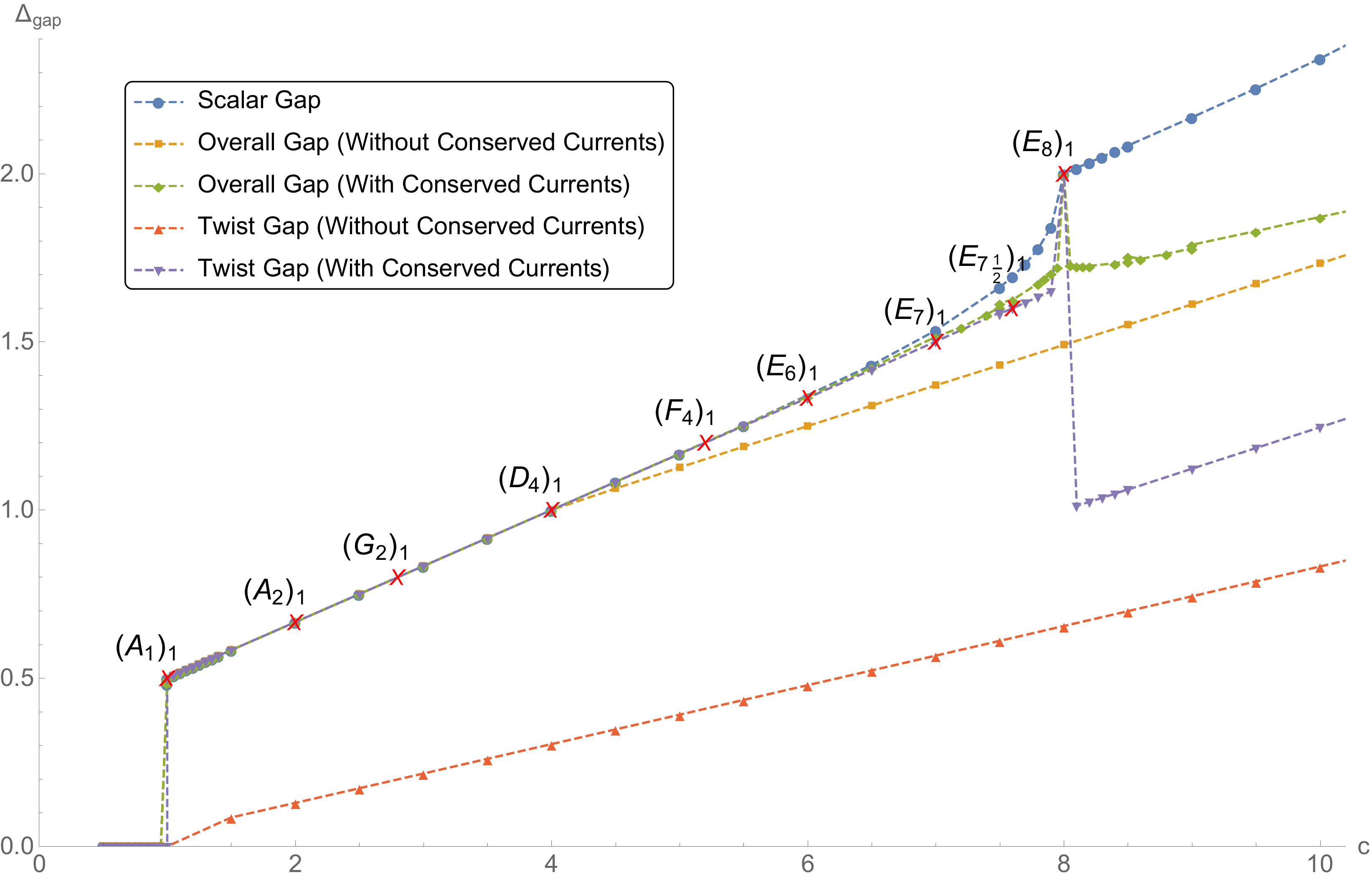}
\end{center}
\caption{Numerical upper bounds on scalar gap, overall gap and twist gap with $\CW(F_4) = \mathcal{W}(2,6,8,12)$ algebra in the range of $1 \le c \le 10$. }
\label{W26812_Fig2_various_normalbound}
\end{figure}

\begin{figure}[t]
\begin{center}
\includegraphics[width=.80\textwidth]{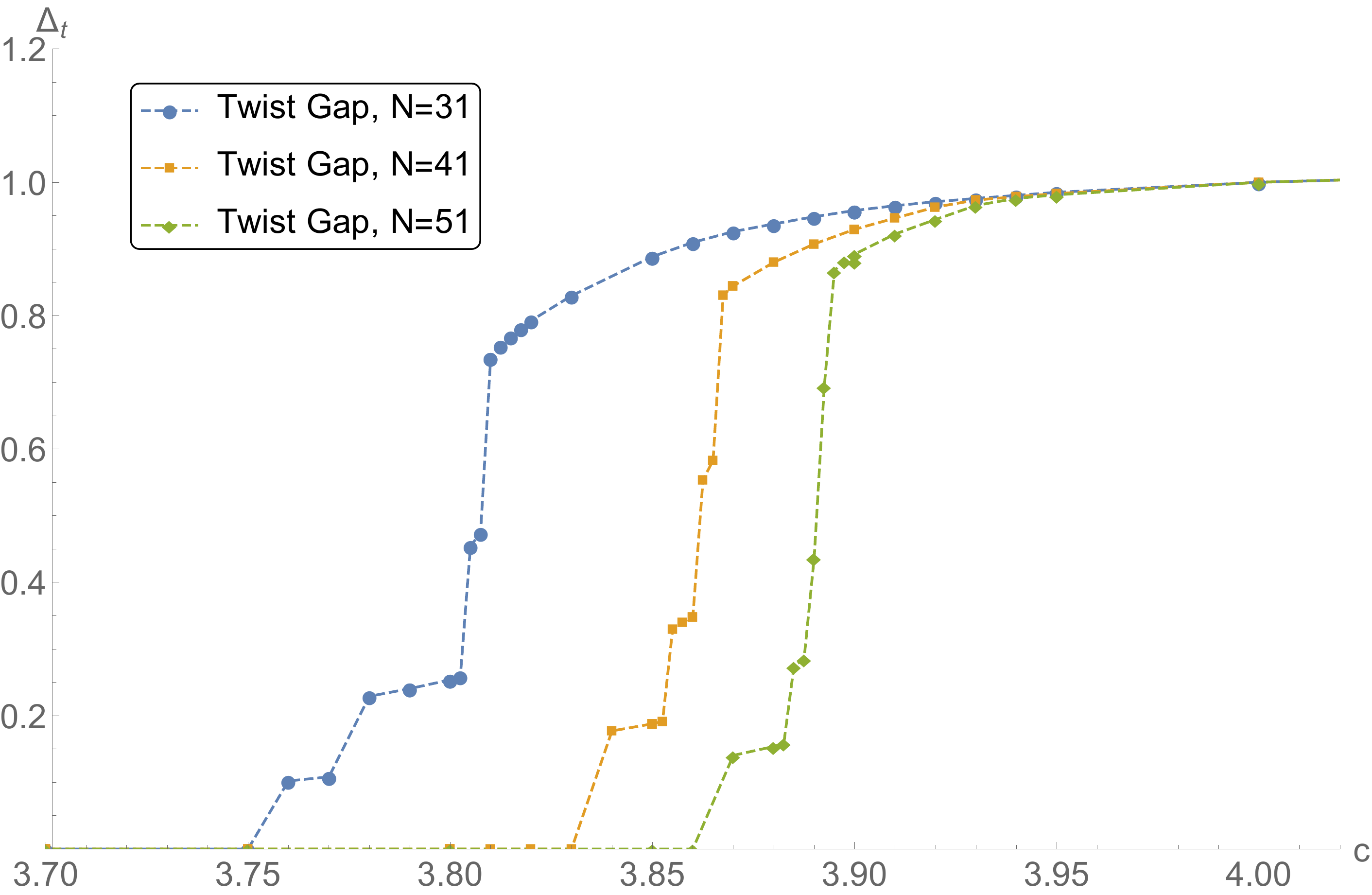}
\end{center}
\caption{Numerical upper bounds on twist gap with $\CW(F_4)=\mathcal{W}(2,6,8,12)$ algebra in the range of $3.7 \le c \le 4.0$. The sharp cliff is pushed towards $c=4$ as the number of derivatives gets increased.}
\label{Tale}
\end{figure}

\begin{figure}[h!]
\begin{center}
\includegraphics[width=.80\textwidth]{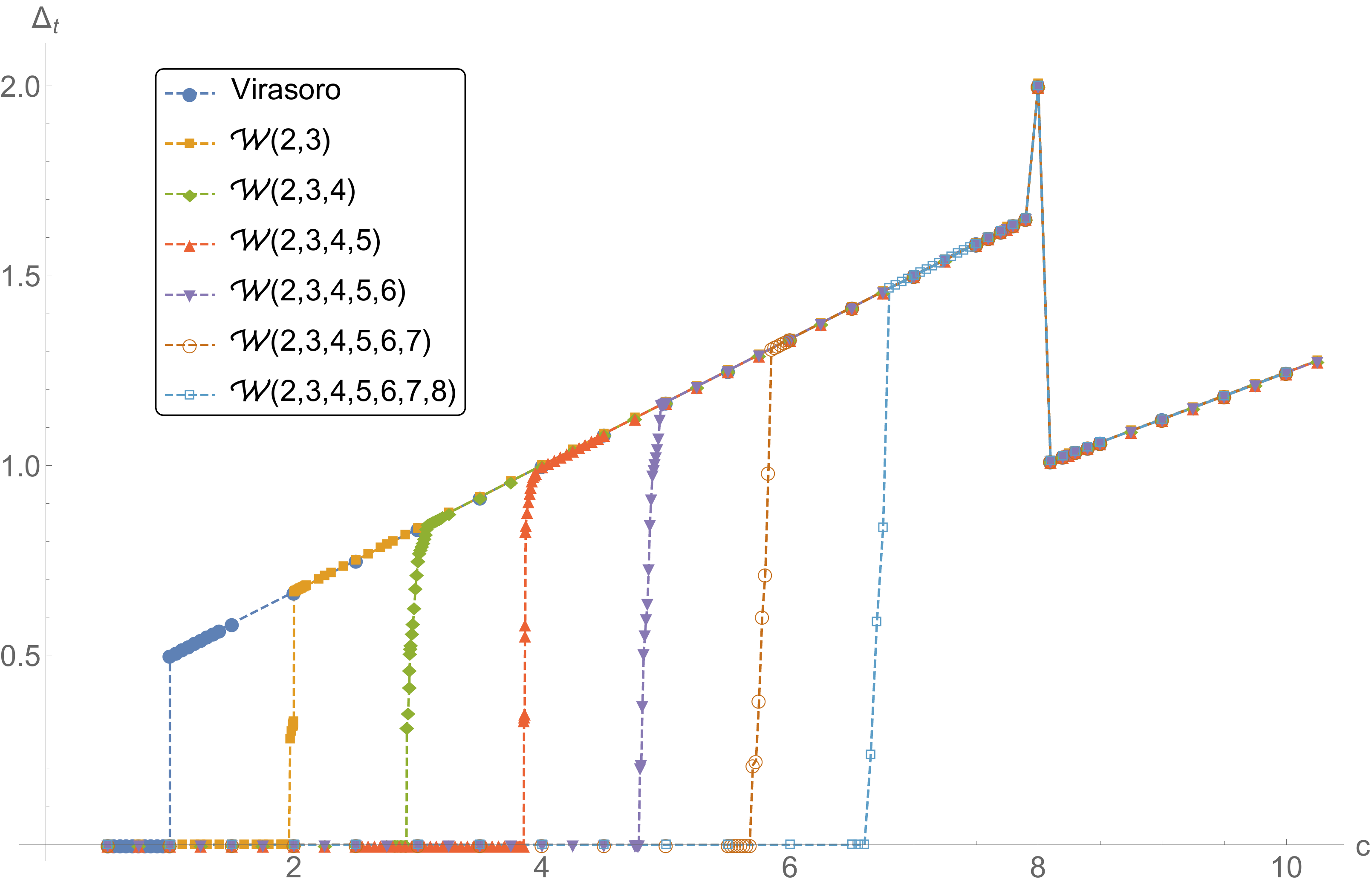}
\end{center}
\caption{Numerical bounds on twist gap with various $\mathcal{W}$-algebra. We assume the presence of the conserved currents of $j \ge 1$ in the spectrum.}
\label{Overallgap}
\end{figure}

In this subsection, we examine modular constraints on the scalar gap $\Delta_s$,
overall gap $\Delta_o$ and twist gap $\Delta_t$ for CFTs with $\mathcal{W}$-symmetry.
To this end, we use the $\mathcal{W}$-algebra characters given in \eqref{Wpri} and \eqref{Wvac}
to decompose the modular invariant partition function \eqref{decomposition_Vir}.

It turns out that the numerical upper bounds with $\mathcal{W}(d_1,d_2, \cdots ,d_r)$-algebra
exhibit the same bounds as in the case for the Virasoro algebra
except for the region  $c \lesssim r$. For instance, we summarize the numerical results with the $\CW(F_4)=\mathcal{W}(2,6,8,12)$-algebra
in Figure \ref{W26812_Fig2_various_normalbound}. Unlike the case of the Virasoro symmetry,
the numerical bounds sharply rise near $c \sim 4$. As is depicted in Figure \ref{Tale}, the sharp cliff
is pushed towards $c=4$ as we increase the total number of derivatives $N_\text{max}$.
We further investigate how the position of the sharp cliff
depends on the rank of the $\mathcal{W}$-algebra. As shown in Figure \ref{Overallgap}, these values
are placed around the rank of the corresponding $\CW$-algebras.
Based on these observations, we can legitimize the assumption in \eqref{Wpri} and \eqref{Wvac}
that the unitary irreducible representations of $\mathcal{W}(d_1,d_2, \cdots ,d_r)$-algebra do not
contain any nontrivial null states when $c \ge r$.

\begin{figure}[t!]
\begin{center}
\includegraphics[width=.84\textwidth]{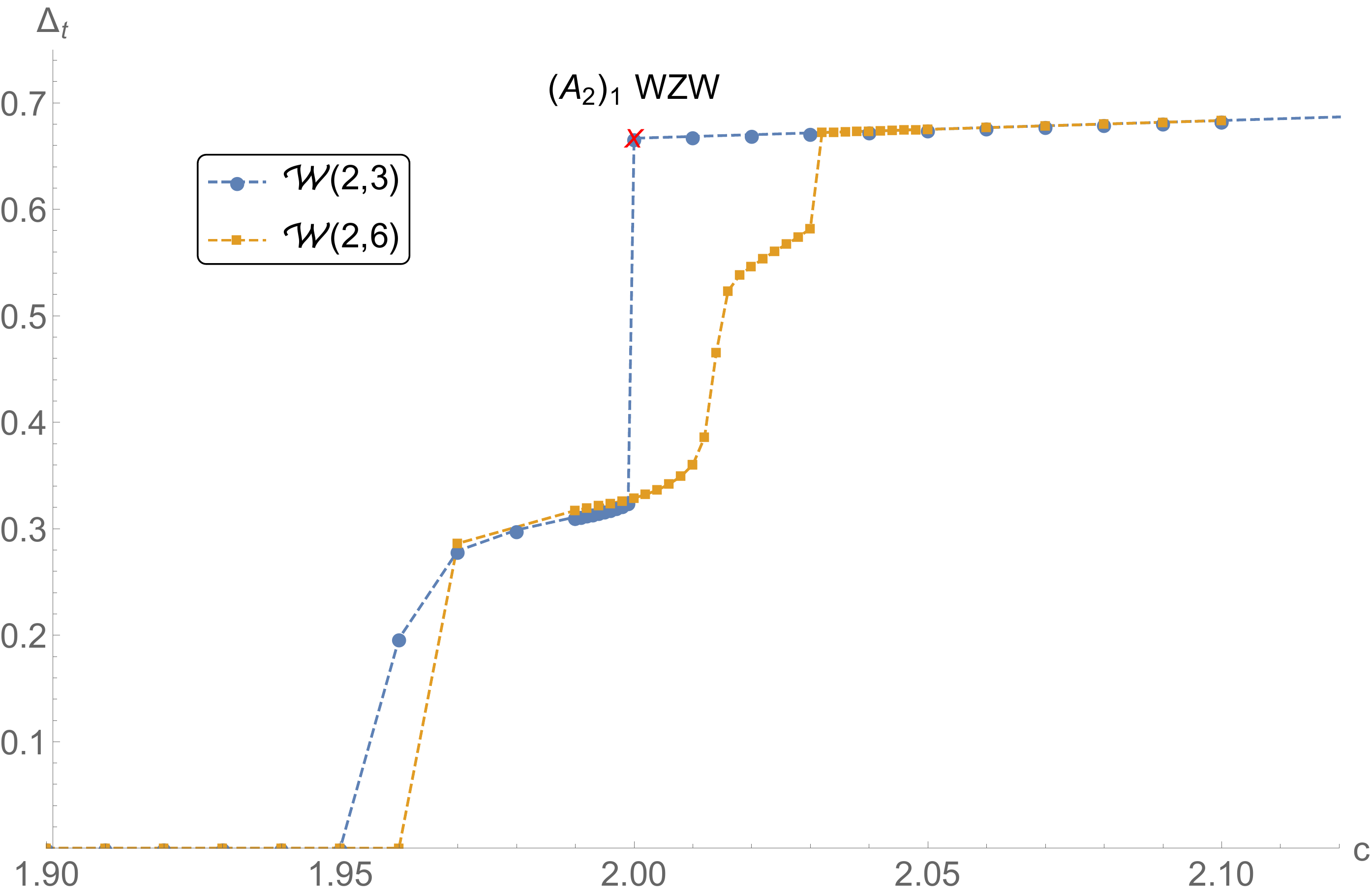}
\end{center}
\caption{Numerical bounds on the twist gap for $\mathcal{W}(2,3)$ and $\mathcal{W}(2,6)$-algebra. The upper bound obtained from the $\mathcal{W}(2,3)$-algebra realizes the $(A_2)_1$ WZW model at the numerical boundary, while $\mathcal{W}(2,6)$ does not.}
\label{Rk2_232426}
\end{figure}

We also find that the $\widehat{\fg}_{k=1}$ WZW model can be placed
at the numerical upper bound on $\D_t$ for the $\CW(\fg)$-algebra.
To illustrate this, let us consider the numerical bounds for the
$\CW(A_2)=\CW(2,3)$ and $\CW(G_2)=\CW(2,6)$.
Figure \ref{Rk2_232426} shows that the $(\widehat{A}_2)_1$ WZW model indeed sits at the numerical bound for the
$\CW(2,3)$-algebra as expected. However, we notice that the model violates the modular constraint
for the $\CW(2,6)$-algebra. This is because when we decompose the partition function
of the $(\widehat{A}_2)_1$ WZW model (\ref{modular inv ptn})(with $h=\frac{1}{3}$ and $N(\widehat{\fg}_1)=1$) using the $\CW(2,6)$ characters,
some of the multiplicities $d_{h,\bar h}$ become negative.

\begin{figure}[t!]
\begin{center}
\includegraphics[width=.47\textwidth]{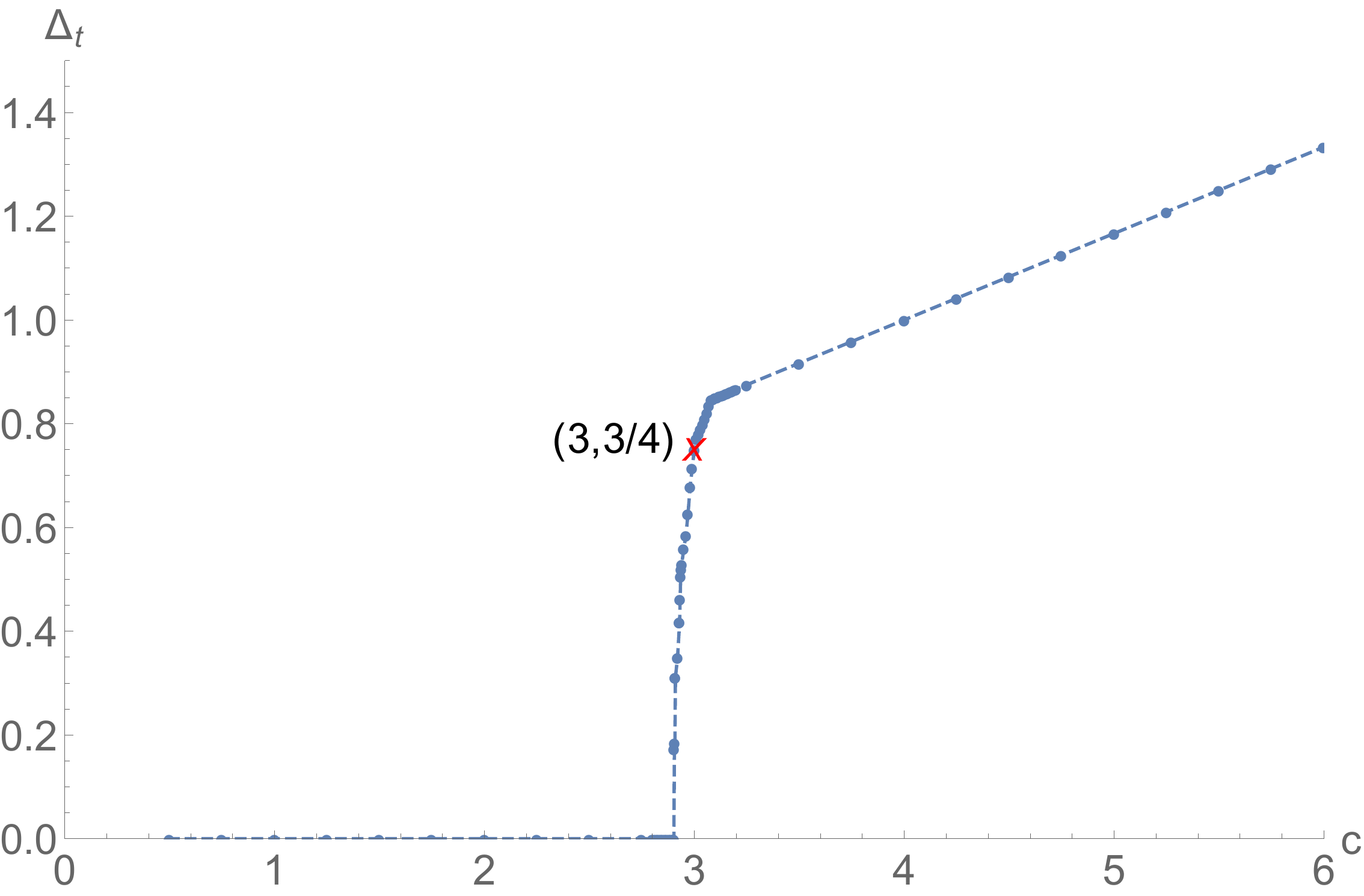} \quad
\includegraphics[width=.47\textwidth]{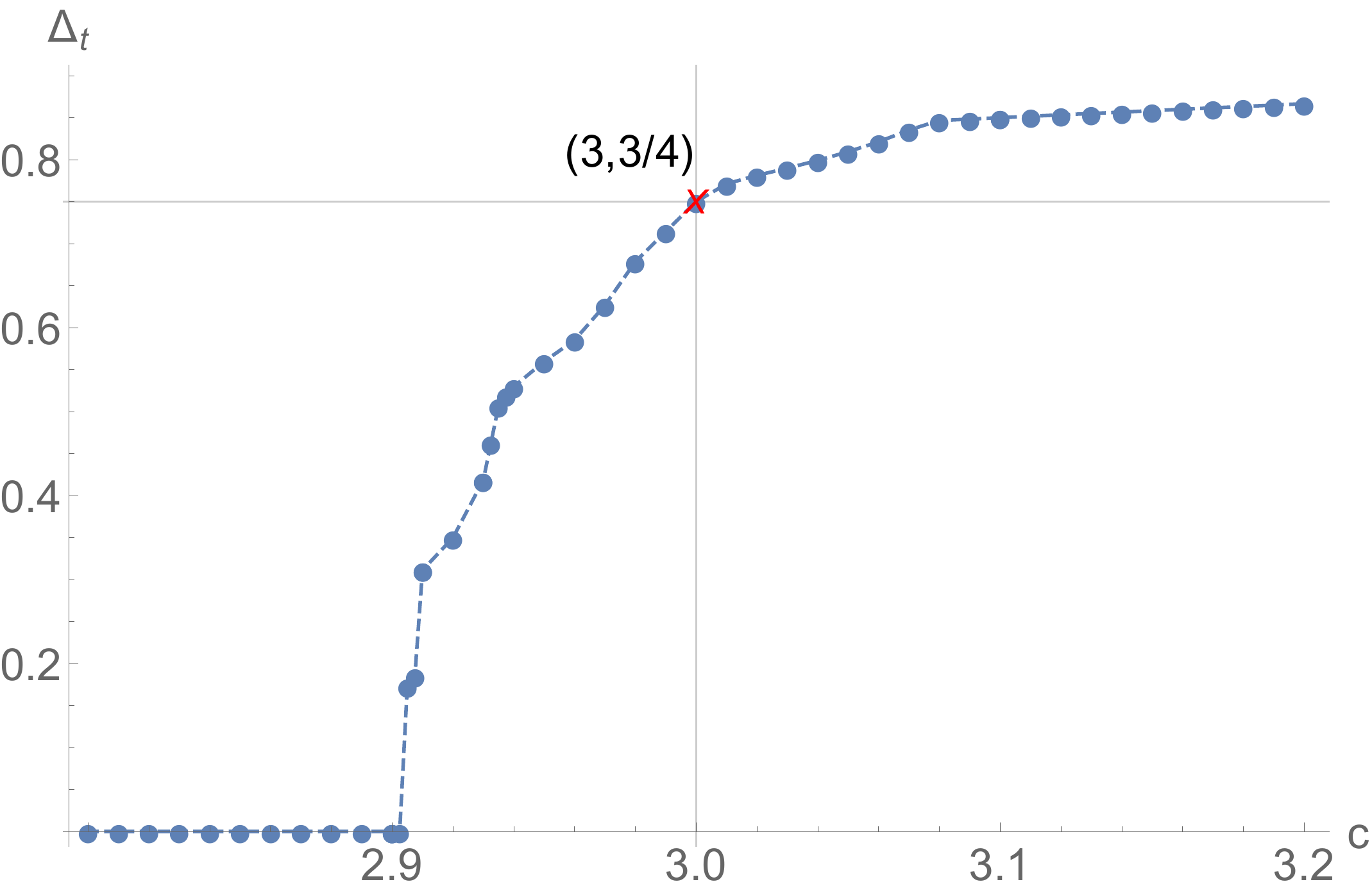}
\end{center}
\caption{Numerical bound on twist gap with $\mathcal{W}(2,3,4)$ algebra. It turns out that the $(\widehat{A}_3)_1$ WZW model is realized on the numerical boundary.}
\label{W234_c3_zoomed}
\end{figure}

Figure \ref{W234_c3_zoomed} shows the numerical bound on $\Delta_t$ for the $\CW(A_3) = \mathcal{W}(2,3,4)$-algebra. Again, $(\widehat{A}_3)_1$ WZW model is located at $(c=3, \Delta_t = \frac{3}{4})$ on the numerical upper bound. Later we verify that the spectrum of $(\widehat{A}_3)_1$ WZW model maximizes the degeneracies of the extremal spectrum for $c=3$ and $\D_t=\frac{3}{4}$. It is interesting to see that the $(\widehat{A}_3)_1$ WZW model is realized at the numerical boundary only using the $\mathcal{W}(2,3,4)$-algebra, but not using the other $\CW$-algebras (including Virasoro). Note that $A_3$ does not belong to the Deligne's exceptional series.


\subsection{Spectral Analysis}
\begin{itemize}
\item {\bf Spectrum Analysis for the $(\widehat{F}_4)_1$ WZW}
\begin{figure}[h]
\begin{center}
\includegraphics[width=.45\textwidth]{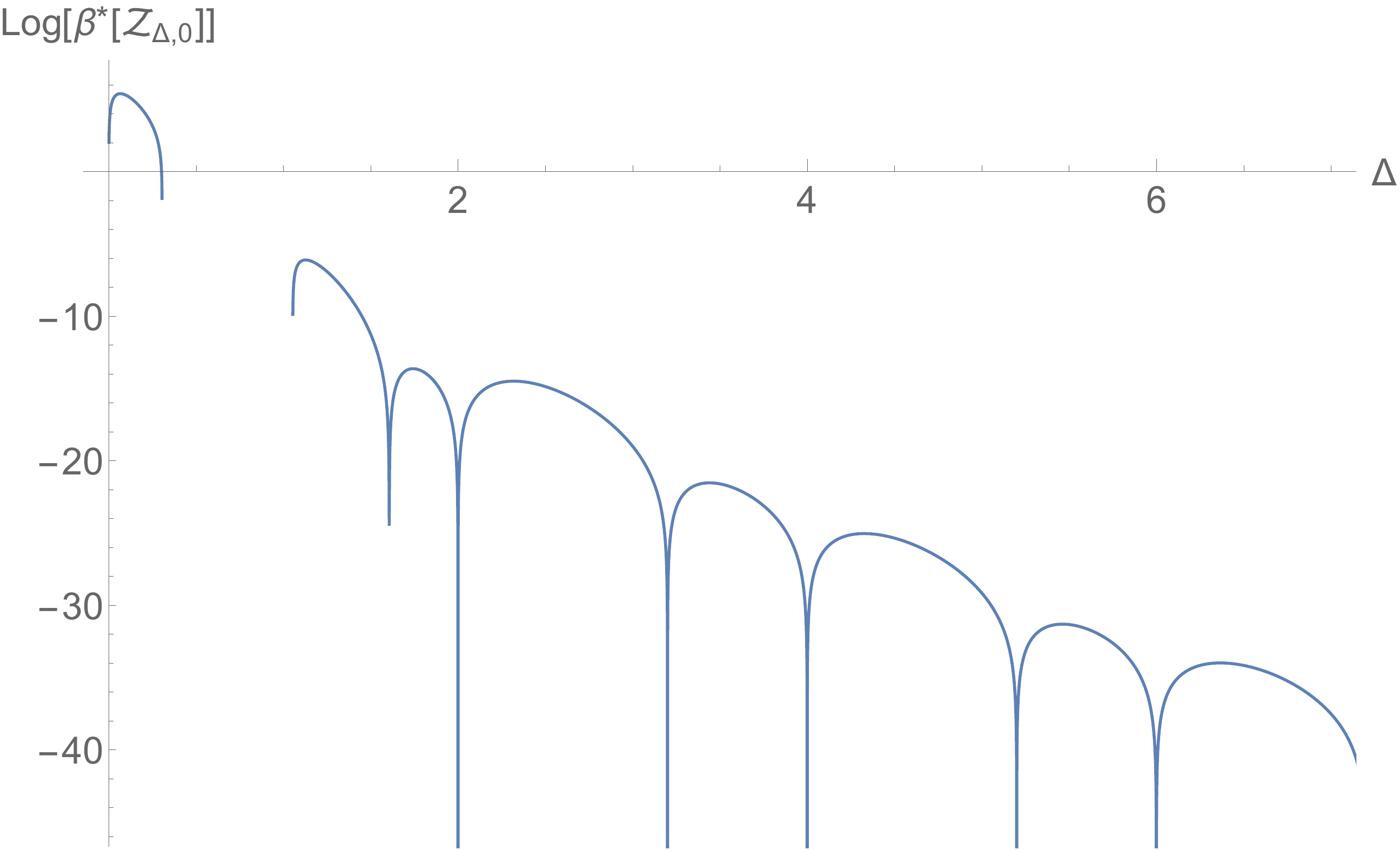} \quad
\includegraphics[width=.45\textwidth]{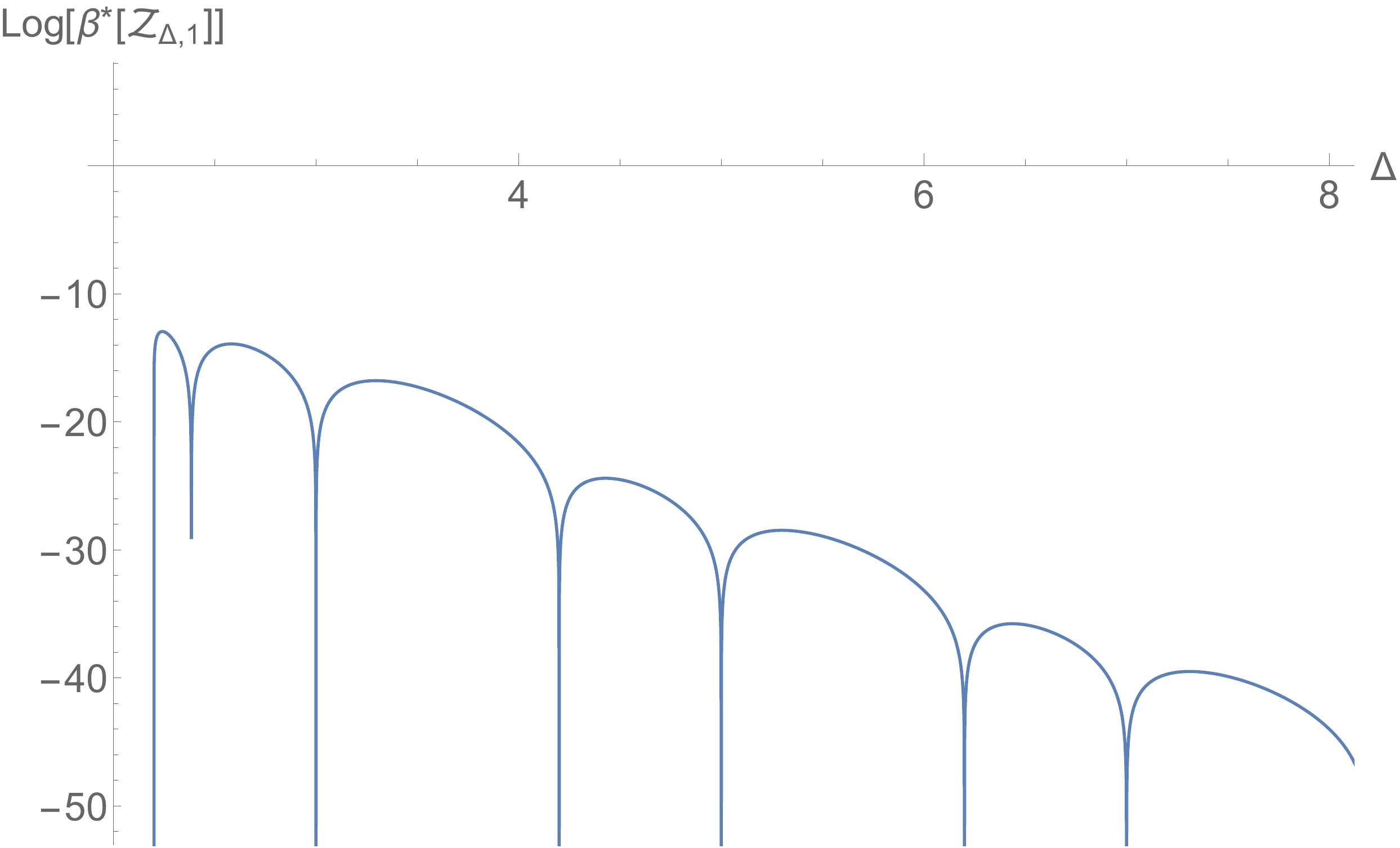}
\end{center}
\caption{The Extremal Functional Method applied for $\mathcal{W}(2,6,8,12)$, $c=\frac{26}{5}$. In this plot, we set $N_\text{max}=51$ and $j_{\textrm{max}}=30$. The left figure is the spin-0 sector, while the right one is the spin-1 sector.}
\label{EFM_F4}
\end{figure}

\begin{table}[ht]
\centering
{
\begin{tabular}{|c |c || c| c || c|c||}
\hline
 \rule{0in}{3ex}     $(h,\bar{h})$         & Max. Deg & $(h,\bar{h})$ &  Max. Deg & $(h,\bar{h})$ & Max. Deg\\
\hline
\rule{0in}{3ex} $(\frac{3}{5}, \frac{3}{5})$ &  676.00000 & $(1, 1)$ & 2704.0000  & $(1, 0)$ &  52.00000  \\
\hline
\rule{0in}{3ex} $(\frac{3}{5}, \frac{8}{5})$ &  5070.0000 & $(2, 1)$ & 8736.000  & $(2, 0)$ &   168.0000  \\
\hline
\rule{0in}{3ex} $(\frac{3}{5}, \frac{13}{5})$ &  14508.000   & $(3, 1)$ & 29900.000 & $(3, 0)$ &  575.0000    \\
 \hline
\rule{0in}{3ex} $(\frac{8}{5}, \frac{8}{5})$ &  38025.000   & $(2, 2)$ & 28224.000 & $(4, 0)$ &  1118.0000   \\
 \hline
\rule{0in}{3ex} $(\frac{8}{5}, \frac{13}{5})$ &  108810.00   & $(2, 3)$ & 96600.008 & $(5, 0)$ &  2700.0000    \\
 \hline
\rule{0in}{3ex} $(\frac{13}{5}, \frac{13}{5})$ &   311364.001    & $(3, 3)$ & 330625.00 & $(6, 0)$ &   4780.0000   \\
 \hline
\end{tabular}
\caption{  Degeneracies of the first few states in $c=\frac{26}{5}$ CFT. We set $N_\text{max}=55$ and $j_\text{max}=40$. We used the $\mathcal{W}(2,6,8,12)$ character with $c=\frac{26}{5}$.}
\label{F4deg W2345}
}
\end{table}
Let us show that the $\widehat{\fg}_{k=1}$ WZW model saturates the upper bound not only on $\D_t$ but also on the degeneracies of the extremal spectrum for the $\mathcal{W}(\fg)$-algebra. As an example, let us consider a hypothetical CFT with $c=\frac{26}{5}$ and $\D_t = \frac{6}{5}$ which lies at the numerical boundary on $\D_t$ for the $\CW(F_4) = \mathcal{W}(2,6,8,12)$-algebra.

To study the extremal spectrum of a hypothetical CFT, we apply the EFM analysis. The conformal dimensions of spin-0 and spin-1 extremal spectrum can be read off from Figure \ref{EFM_F4}, given by $\Delta_{j} = \{ \frac{6}{5} + j + 2n, 2 + j+2n \big \}$ for $j=0,1$ and $n \in \mathbb{Z}_{n \ge 0}$. The maximal degeneracies of various primaries in the extremal spectrum are summarized in Table \ref{F4deg W2345}. It implies that the partition function of the putative CFT of our interest can be expressed as follow,
\begin{align}
\begin{split}
{Z}^{\mathcal{W}(2,6,8,12)}_{c=\frac{26}{5}}(q, \bar{q}) &= {\chi}_{0}(\tau) \bar{{\chi}}_{0}(\bar{\tau}) +  676 {\chi}_{\frac{3}{5}}(\tau) \bar{{\chi}}_{\frac{3}{5}}(\bar{\tau}) +  5070 \left( {\chi}_{\frac{3}{5}}(\tau) \bar{{\chi}}_{\frac{8}{5}}(\bar{\tau}) + \text{c.c.} \right)   \\
& + 2704 {\chi}_{1}(\tau) \bar{{\chi}}_{1}(\bar{\tau}) + 8736 \left( {\chi}_{2}(\tau) \bar{{\chi}}_{1}(\bar{\tau}) + \text{c.c.} \right)  \\
&+ 52 \left( {\chi}_{1}(\tau) \bar{{\chi}}_{0}(\bar{\tau}) + \text{c.c.}) \right) + 168 \left( {\chi}_{2}(\tau) \bar{{\chi}}_{0}(\bar{\tau})+ \text{c.c.} \right)+ \cdots,
\label{W6812:partition}
\end{split}
\end{align}
where $\chi_{h}$ denotes a $\mathcal{W}(2,6,8,12)$ character. One can easily show that \eqref{W6812:partition} agrees with the partition function of the $(\widehat{F}_4)_1$ WZW model \eqref{F4wzwkp}.

\item {\bf Spectrum Analysis for the $(\widehat{A}_3)_1$ WZW}

\begin{figure}[h]
\begin{center}
\includegraphics[width=.40\textwidth]{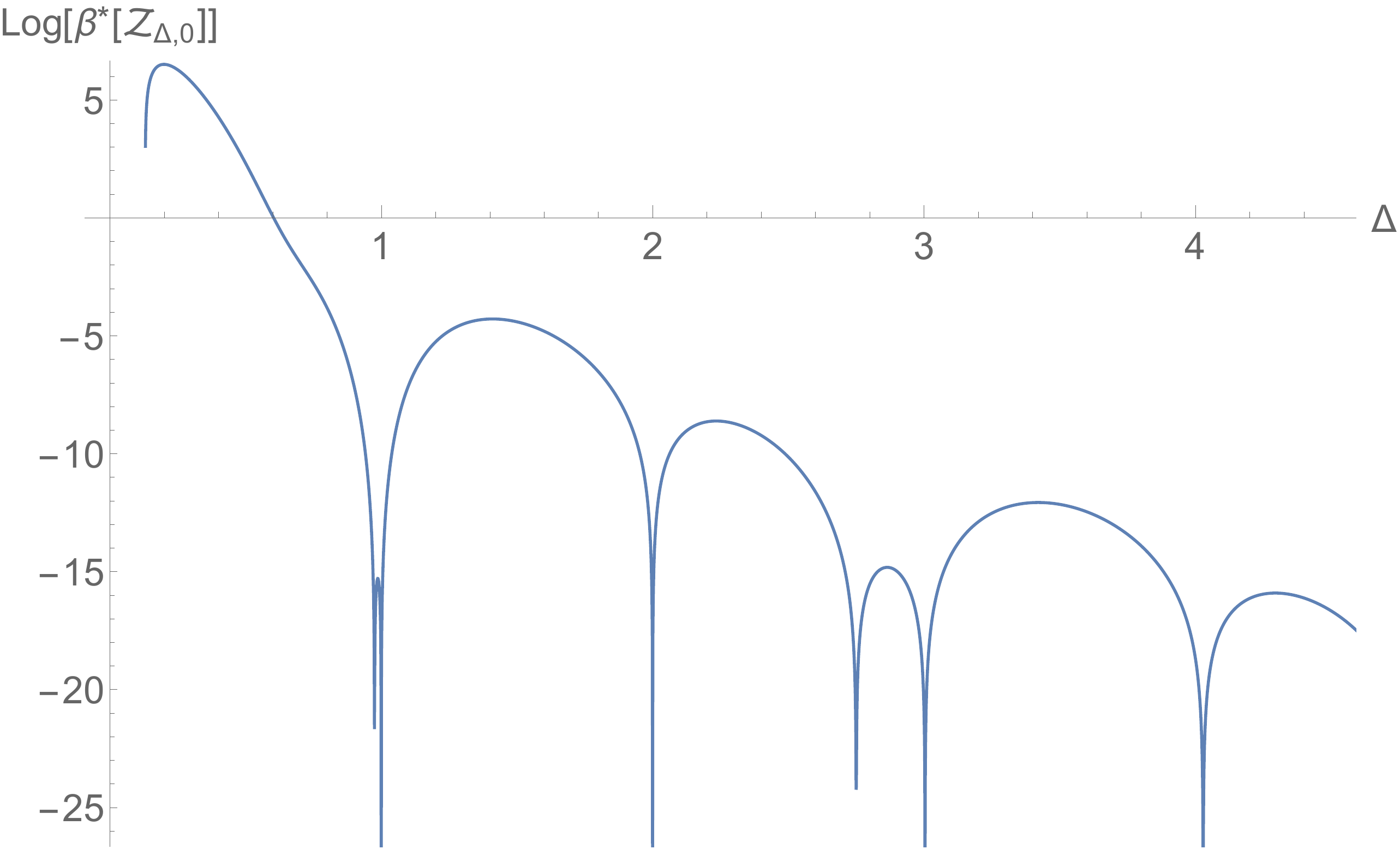} \qquad
\includegraphics[width=.40\textwidth]{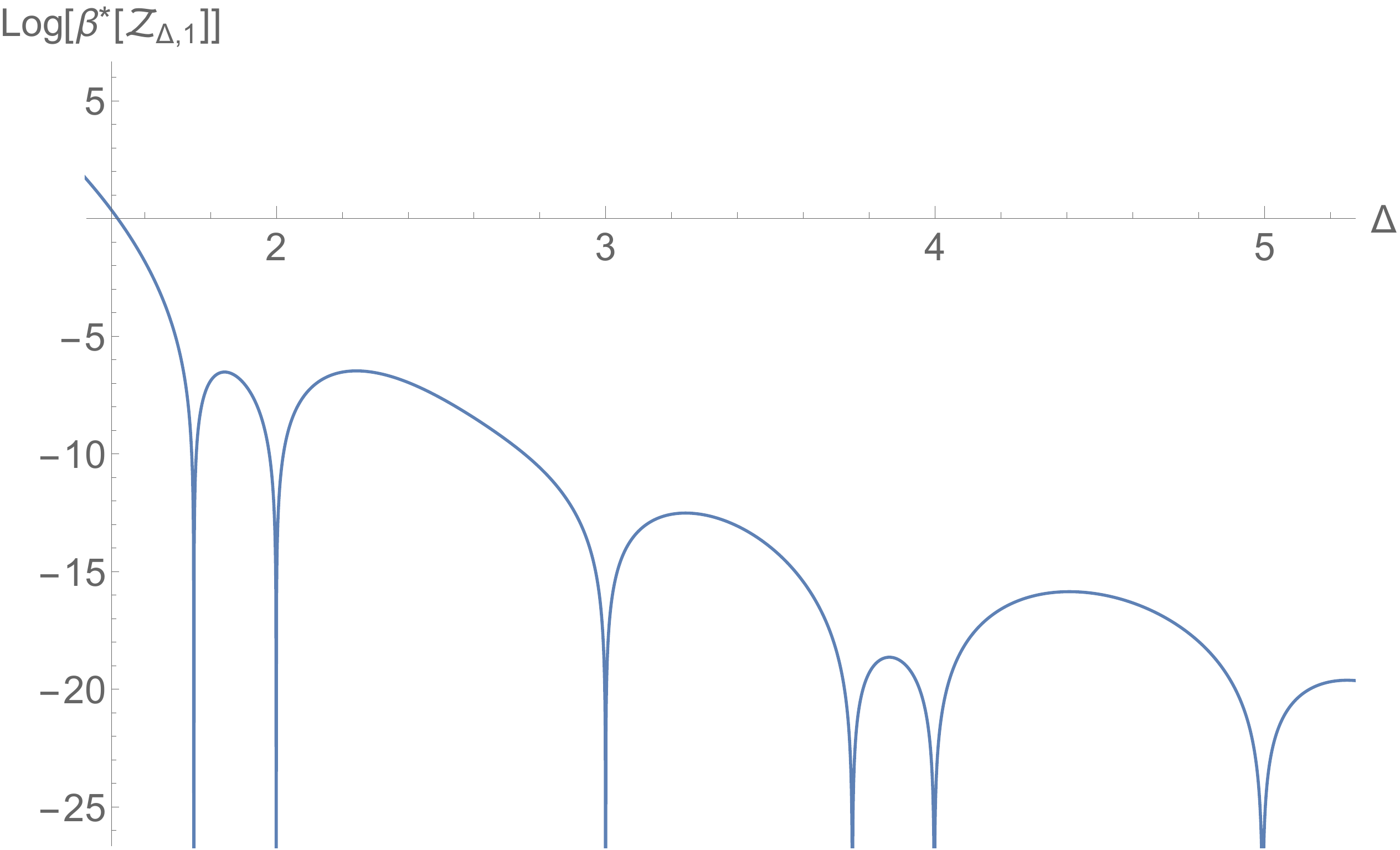}
\end{center}
\caption{Extremal Functional Method applied to $c=3$ CFT with $\mathcal{W}(2,3,4)$-algebra. In this plot, we set $N_\text{max}=51$ and $j_{\textrm{max}}=30$.}
\label{EFM_A3}
\end{figure}

\begin{table}[ht]
\centering
{
\begin{tabular}{|c |c || c| c || c | c ||}
\hline
 \rule{0in}{3ex}     $(h,\bar{h})$         & Max. Deg & $(h,\bar{h})$ &  Max. Deg  & $(h,\bar{h})$ &  Max. Deg \\
\hline
\rule{0in}{3ex} $(\frac{3}{8}, \frac{3}{8})$ &  32.00000 & $(\frac{1}{2}, \frac{1}{2})$ & 36.000000  & $(1, 1)$ &  225.25714 \\
\hline
\rule{0in}{3ex} $(\frac{3}{8}, \frac{11}{8})$ &  96.00000 & $(\frac{1}{2}, \frac{3}{2})$ & 48.00000  & $(1, 2)$ &  75.00020  \\
\hline
\rule{0in}{3ex} $(\frac{11}{8}, \frac{11}{8})$ &  288.01585 & $(\frac{3}{2}, \frac{3}{2})$ & 64.11818   & $(2, 2)$ &  25.00500 \\
 \hline
\end{tabular}
\caption{Here $N_\text{max}=51$, $j_\text{max}=30$. We used the $\mathcal{W}(2,3,4)$ character with $c=3$.}
\label{A3deg}
}
\end{table}

Let us discuss the point ($c=3, \D_t = \frac{3}{4}$) placed at the numerical boundary on $\D_t$ for the $\CW(A_3) = \mathcal{W}(2,3,4)$-algebra. The extremal spectrum at this point and their maximal degeneracies are illustrated in Figure \ref{EFM_A3} and Table \ref{A3deg}. From these results, the conformal dimensions of spin-0 and spin-1 primaries can be obtained as $\Delta_{j=0} = \{\frac{3}{4} +2n, 1 +n \}$ and $\Delta_{j=1}  = \{\frac{7}{4} +2n, 2 +n \}$, for $n \in \mathbb{Z}_{n \ge 0}$.

We can recast the partition function of a CFT containing the primaries in Table \ref{A3deg} as the following form
\begin{equation} \label{A3 partition}
Z_{A_3}^{\mathcal{W}(2,3,4)}(q, \bar{q}) = |f_{\text{vac}}^{c=3} (q)|^2 + |f_{\frac{1}{2}}^{c=3} (q)|^2 +2|f_{\frac{3}{8}}^{c=3} (q)|^2 \ ,
\end{equation}
where $f_{\text{vac}}^{c=3} (q)$, $f_{\frac{1}{2}}^{c=3}(q)$ and $f_{\frac{3}{8}}^{c=3} (q)$ are $(\widehat{A}_3)_1$ characters given by,
\begin{align}
f_{\text{vac}}^{c=3} (q) &= \chi_{[1;0,0,0]} (q)= q^{-\frac{1}{8}} \left( 1 + 15 q+ 51 q^2 + 172 q^3 + 453 q^4 +  \mathcal{O}(q^5) \right), \nonumber \\
f_{\frac{1}{2}}^{c=3} (q) &=\chi_{[0;0,1,0]} (q)= q^{\frac{1}{2}-\frac{1}{8}} \left(6 + 26q + 102q^2 +276q^3 + 728q^4 +\mathcal{O}(q^5) \right),  \\
f_{\frac{3}{8}}^{c=3} (q) &=\chi_{[0;1,0,0]} (q)=\chi_{[0;0,0,1]} (q)= q^{\frac{3}{8}-\frac{1}{8}} \left( 4 + 24q + 84 q^2 + 248 q^3 + 648 q^4 + \mathcal{O}(q^5) \right). \nonumber
\end{align}
\eqref{A3 partition} is nothing but the partition function of the $(\widehat{A}_3)_1$ WZW model. We thus identify a putative CFT of our interest with $c=3$ as the $(\widehat{A}_3)_1$ WZW model.

\end{itemize}

\subsection{Accumulation of the Spectrum}
\begin{figure}[h!]
\subfigure[$c=4$, Virasoro algebra]{\includegraphics[width = 3in]{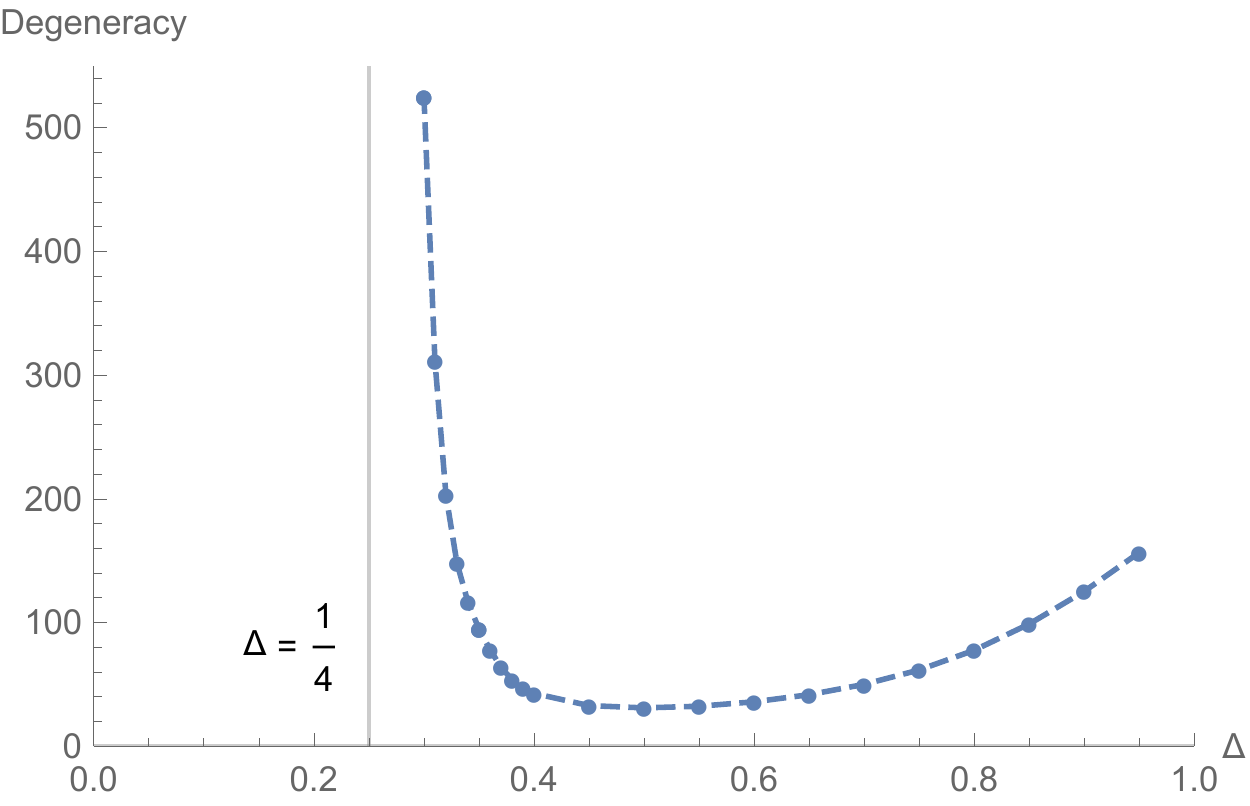}} \
\subfigure[$c=6$, Virasoro algebra]{\includegraphics[width = 3in]{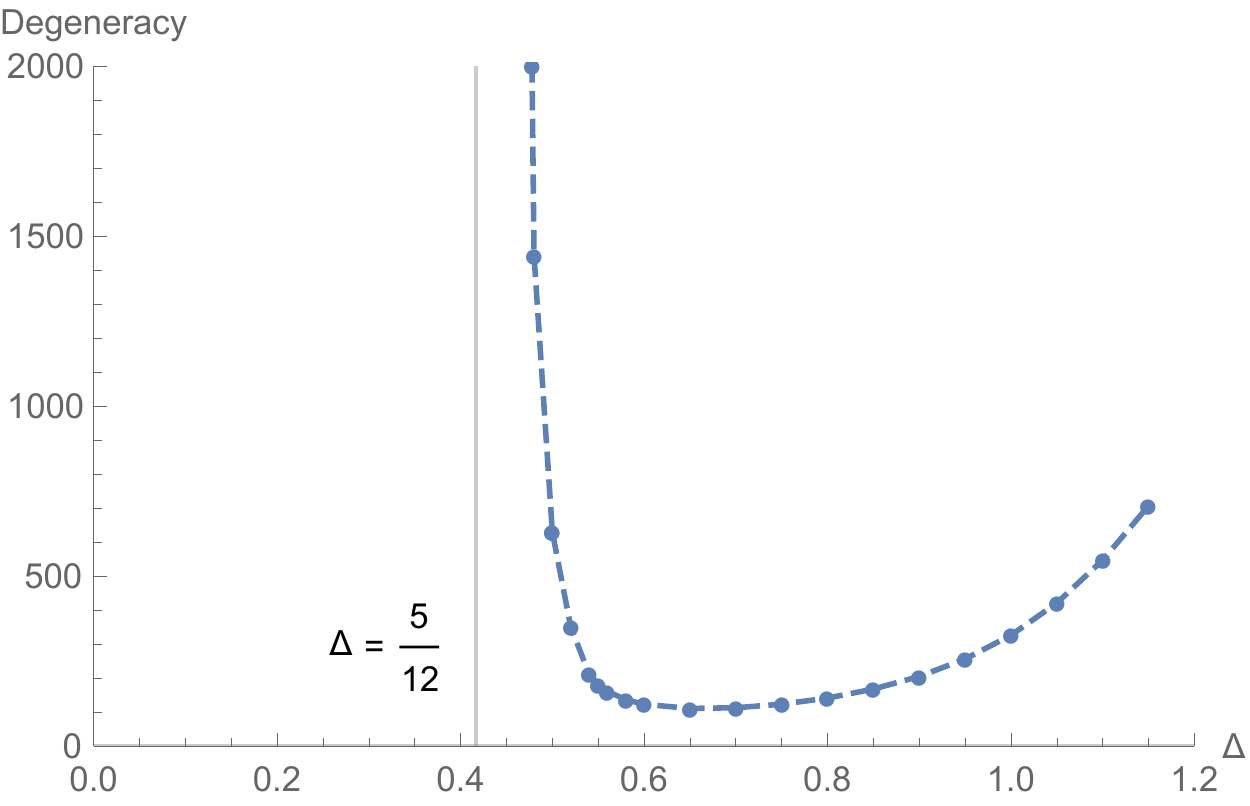}} \vspace*{-0.4cm} \newline
\subfigure[$c=\frac{14}{5}$, $\mathcal{W}(2,3)$-algebra]{\includegraphics[width = 3in]{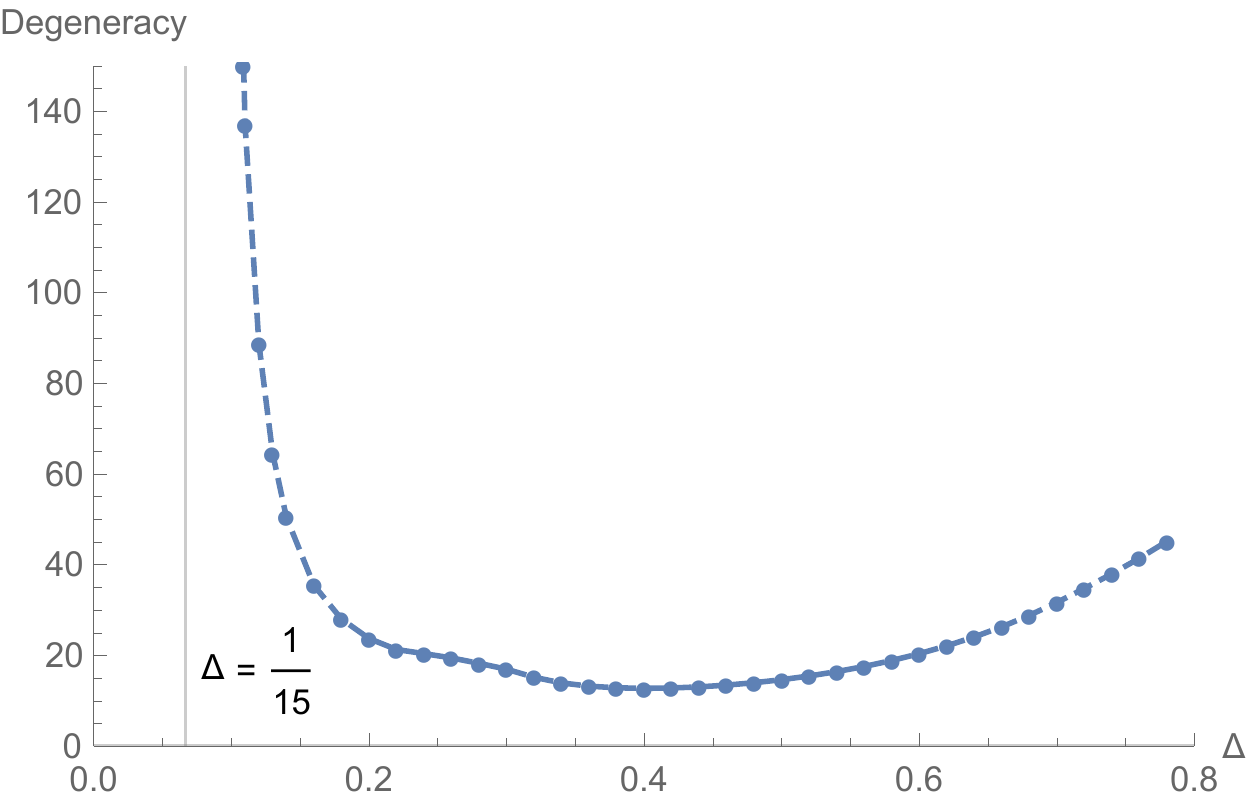}}  \
\subfigure[$c=6$, $\mathcal{W}(2,6,8,12)$-algebra]{\includegraphics[width = 3in]{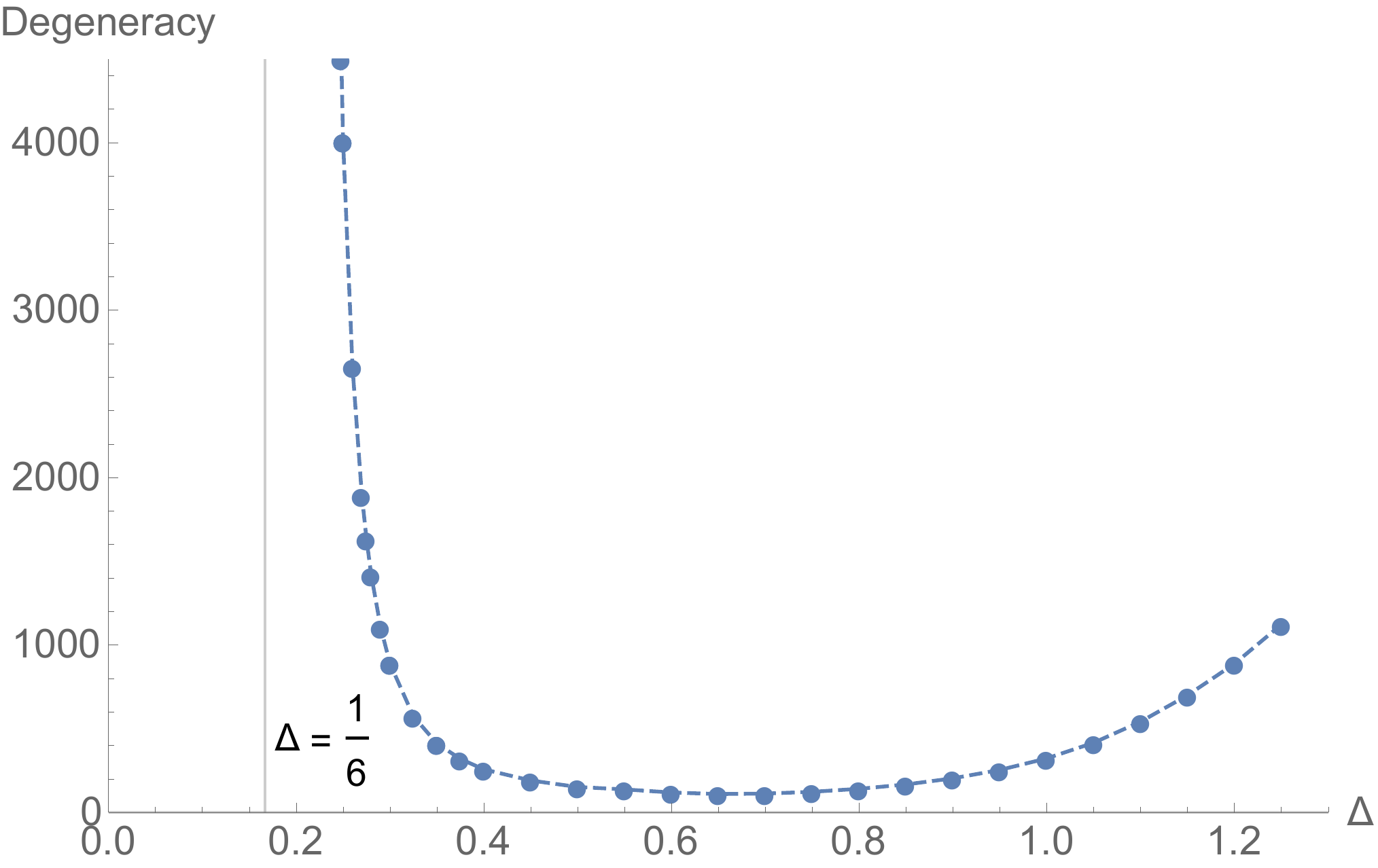}} \vspace*{-0.4cm} \newline
\subfigure[$c=4$, $\mathcal{W}(2,6,8,12)$-algebra]{\includegraphics[width = 3in]{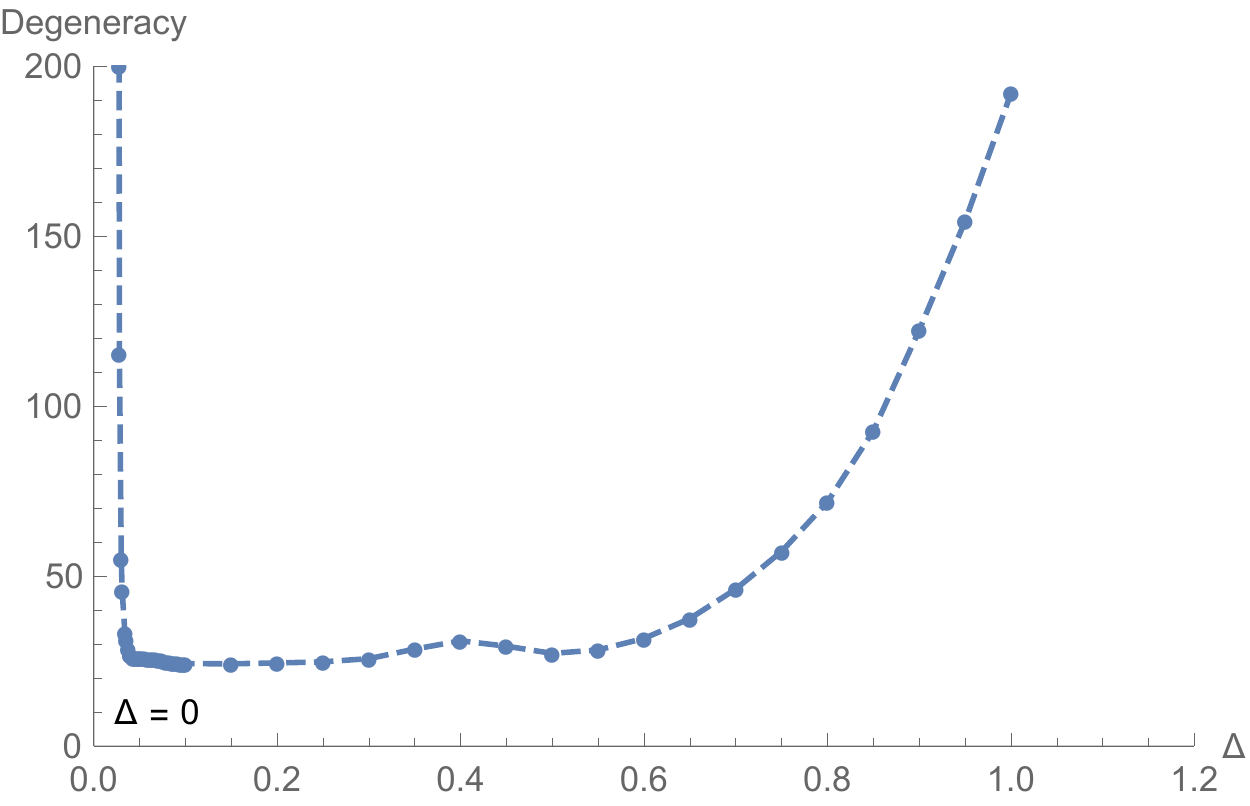}} \
\subfigure[$c=2$, $\mathcal{W}(2,3)$-algebra]{\includegraphics[width = 3in]{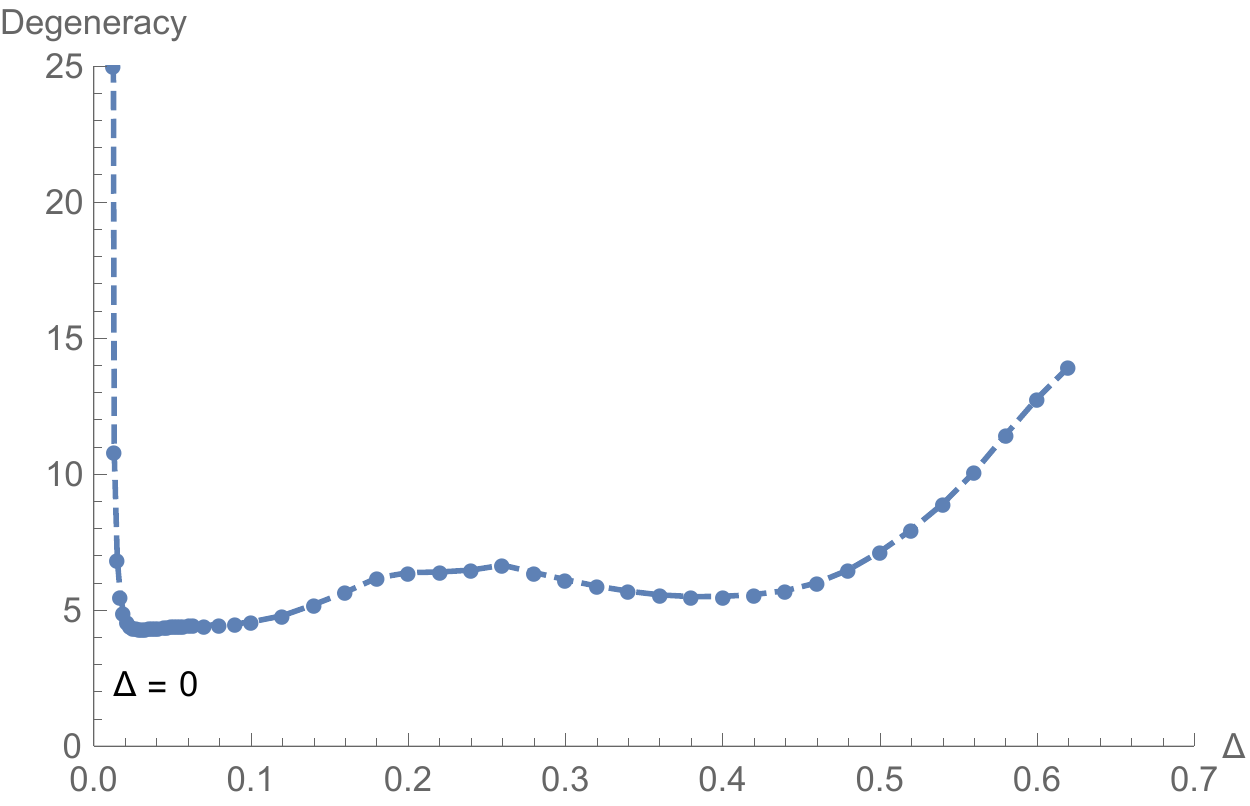}}
\caption{The maximal degeneracies of the lowest primary with various $\mathcal{W}$-algebras and central charges.}
\label{W2_Deg_scalar_fixedC4and6}
\end{figure}
Figure \ref{W2_Deg_scalar_fixedC4and6} shows the plots of the maximal degeneracy versus
the conformal dimension of the lowest primary in a theory with $\mathcal{W}(d_1,d_2,\cdots,d_r)$ symmetry
at various central charges. We notice that there is no upper bound on the degeneracy as the conformal dimension $\Delta_t$
approaches to $\frac{c-r}{12}$. We also observe that the location at which the degeneracy diverge is independent of
the presence of holomorphic/anti-holomorphic currents other than $\CW$-symmetry.
The divergence can be explained
from the fact that infinitely many primaries get accumulated at $h = \frac{c-r}{24}$ \cite{Collier:2016cls, Afkhami-Jeddi:2017idc}.
For the self-containment, we briefly present the derivation of \cite{Collier:2016cls, Afkhami-Jeddi:2017idc} below.

In order to understand the origin of the divergence, it is sufficient
to consider the theories without conserved currents. In the limit $\bar{\tau} \rightarrow i \infty$,
the vacuum character of $\CW$-algebra dominates the partition function:
\begin{align}
\lim_{\bar \t \to - i \infty} \left[ \frac{Z(\tau, \bar{\tau})}{\bar \chi_0(\bar \t)} \right] = \chi_0(\tau)
\end{align}
We further consider the limit $\tau \rightarrow i 0^+$ where the vacuum character behaves as
\begin{align}
\lim_{\t \to i 0^+} \chi_0(\tau) = \lim_{\t \to i 0^+}  (q')^{-\frac{r}{24}} \ ,
\end{align}
where $q' \equiv e^{-\frac{2 \pi i}{\tau}}$. As a consequence, the modular invariance of the partition function
$Z(\t,\bar \t)=Z(-1/\t, - 1/\bar \t)$ in the above special limits requires
\begin{align}
\begin{split}
\lim_{\t \to i 0^+}  e^{\frac{ 2 \pi i }{24 \tau} r } = \lim_{\t \to i 0^+} \lim_{\bar \t \to - i \infty}
\left[ \chi_0(-\frac{1}{\t}) \cdot \frac{\bar \chi_0 (- \frac{1}{\bar \t})}{\bar \chi_0 (\bar \t)} + \sum_{h, \bar{h}} d_{h,\bar h}~
\chi_h(-\frac{1}{\t}) \cdot \frac{\bar \chi_{\bar h} (- \frac{1}{\bar \t})}{\bar \chi_0 (\bar \t)} \right].
\label{MBESL}
\end{split}
\end{align}
Note that $\lim_{\bar \t \to -i\infty} \frac{\bar \chi_{\bar h} (- \frac{1}{\bar \t})}{\bar \chi_0 (\bar \t)} = \lim_{\bar \t \to -i\infty} (\bar q)^{(c-r)/24} \rightarrow 0$ for $c>r$.
If the limit and the summation in the RHS of (\ref{MBESL}) were to commute, we would therefore say that the RHS of (\ref{MBESL}) vanish, which is inconsistent with
the LHS of (\ref{MBESL}). One can resolve the above contradiction only when there exist infinitely many primaries of weight $h$ accumulating to $\frac{c-r}{24}$.
The infinite degeneracy at $h=\frac{c-r}{24}$ and $\bar h=\frac{c-r}{24}$ then
explains the divergence behavior in Figure \ref{W2_Deg_scalar_fixedC4and6}.

\section*{Acknowledgments}
We would like to thank Jeffrey Harvey, Sunil Mukhi, Kimyeong Lee and Soo-Jong Rey for useful discussions.
We thank KIAS Center for Advanced Computation for providing computing resources.
The research of S.L. is supported in part by the National Research Foundation of Korea (NRF) Grant NRF-$2017$R$1$C$1$B$1011440$.
S.L. and J.S. thank the organizers of the Pollica Summer Workshop $2017$ for the hospitality, and was partly supported by the ERC STG grant $306260$ during the Pollica Summer Workshop.
We also thank the organizers of the workshop ``Geometry of String and Gauge Theories'' at CERN, and 
also CERN-Korea Theory Collaboration funded by National Research Foundation (Korea) for the hospitality and support.
J.S. also thanks the organizers of the Summer Simons Workshops in Mathematics and Physics 2017 for the hospitality.


\bibliographystyle{jhep}
\bibliography{refs}

\def\cprime{$'$}
\providecommand{\href}[2]{#2}\begingroup\raggedright\begin{thebibliography}{10}

\bibitem{Belavin:1984vu}
A.~A. Belavin, A.~M. Polyakov, and A.~B. Zamolodchikov, {\it {Infinite
  Conformal Symmetry in Two-Dimensional Quantum Field Theory}},  {\em Nucl.
  Phys.} {\bf B241} (1984) 333--380.

\bibitem{Moore:1988qv}
G.~W. Moore and N.~Seiberg, {\it {Classical and Quantum Conformal Field
  Theory}},  {\em Commun. Math. Phys.} {\bf 123} (1989) 177.

\bibitem{Rattazzi:2008pe}
R.~Rattazzi, V.~S. Rychkov, E.~Tonni, and A.~Vichi, {\it {Bounding scalar
  operator dimensions in 4D CFT}},  {\em JHEP} {\bf 12} (2008) 031,
  [\href{http://arxiv.org/abs/0807.0004}{{\tt arXiv:0807.0004}}].

\bibitem{Poland:2011ey}
D.~Poland, D.~Simmons-Duffin, and A.~Vichi, {\it {Carving Out the Space of 4D
  CFTs}},  {\em JHEP} {\bf 05} (2012) 110,
  [\href{http://arxiv.org/abs/1109.5176}{{\tt arXiv:1109.5176}}].

\bibitem{Chang:2015qfa}
C.-M. Chang and Y.-H. Lin, {\it {Bootstrapping 2D CFTs in the Semiclassical
  Limit}},  {\em JHEP} {\bf 08} (2016) 056,
  [\href{http://arxiv.org/abs/1510.02464}{{\tt arXiv:1510.02464}}].

\bibitem{Chang:2016ftb}
C.-M. Chang and Y.-H. Lin, {\it {Bootstrap, universality and horizons}},  {\em
  JHEP} {\bf 10} (2016) 068, [\href{http://arxiv.org/abs/1604.01774}{{\tt
  arXiv:1604.01774}}].

\bibitem{Lin:2015wcg}
Y.-H. Lin, S.-H. Shao, D.~Simmons-Duffin, Y.~Wang, and X.~Yin, {\it {$
  \mathcal{N} $ = 4 superconformal bootstrap of the K3 CFT}},  {\em JHEP} {\bf
  05} (2017) 126, [\href{http://arxiv.org/abs/1511.04065}{{\tt
  arXiv:1511.04065}}].

\bibitem{Lin:2016gcl}
Y.-H. Lin, S.-H. Shao, Y.~Wang, and X.~Yin, {\it {(2, 2) superconformal
  bootstrap in two dimensions}},  {\em JHEP} {\bf 05} (2017) 112,
  [\href{http://arxiv.org/abs/1610.05371}{{\tt arXiv:1610.05371}}].

\bibitem{Hellerman:2009bu}
S.~Hellerman, {\it {A Universal Inequality for CFT and Quantum Gravity}},  {\em
  JHEP} {\bf 08} (2011) 130, [\href{http://arxiv.org/abs/0902.2790}{{\tt
  arXiv:0902.2790}}].

\bibitem{Friedan:2013cba}
D.~Friedan and C.~A. Keller, {\it {Constraints on 2d CFT partition functions}},
   {\em JHEP} {\bf 10} (2013) 180, [\href{http://arxiv.org/abs/1307.6562}{{\tt
  arXiv:1307.6562}}].

\bibitem{Hartman:2014oaa}
T.~Hartman, C.~A. Keller, and B.~Stoica, {\it {Universal Spectrum of 2d
  Conformal Field Theory in the Large c Limit}},  {\em JHEP} {\bf 09} (2014)
  118, [\href{http://arxiv.org/abs/1405.5137}{{\tt arXiv:1405.5137}}].

\bibitem{Collier:2016cls}
S.~Collier, Y.-H. Lin, and X.~Yin, {\it {Modular Bootstrap Revisited}},
  \href{http://arxiv.org/abs/1608.06241}{{\tt arXiv:1608.06241}}.

\bibitem{Hellerman:2010qd}
S.~Hellerman and C.~Schmidt-Colinet, {\it {Bounds for State Degeneracies in 2D
  Conformal Field Theory}},  {\em JHEP} {\bf 08} (2011) 127,
  [\href{http://arxiv.org/abs/1007.0756}{{\tt arXiv:1007.0756}}].

\bibitem{Keller:2012mr}
C.~A. Keller and H.~Ooguri, {\it {Modular Constraints on Calabi-Yau
  Compactifications}},  {\em Commun. Math. Phys.} {\bf 324} (2013) 107--127,
  [\href{http://arxiv.org/abs/1209.4649}{{\tt arXiv:1209.4649}}].

\bibitem{Qualls:2013eha}
J.~D. Qualls and A.~D. Shapere, {\it {Bounds on Operator Dimensions in 2D
  Conformal Field Theories}},  {\em JHEP} {\bf 05} (2014) 091,
  [\href{http://arxiv.org/abs/1312.0038}{{\tt arXiv:1312.0038}}].

\bibitem{Qualls:2014oea}
J.~D. Qualls, {\it {Universal Bounds in Even-Spin CFTs}},  {\em JHEP} {\bf 12}
  (2015) 001, [\href{http://arxiv.org/abs/1412.0383}{{\tt arXiv:1412.0383}}].

\bibitem{Qualls:2015bta}
J.~D. Qualls, {\it {Universal Bounds on Operator Dimensions in General 2D
  Conformal Field Theories}},  \href{http://arxiv.org/abs/1508.00548}{{\tt
  arXiv:1508.00548}}.

\bibitem{Cardy:2017qhl}
J.~Cardy, A.~Maloney, and H.~Maxfield, {\it {A new handle on three-point
  coefficients: OPE asymptotics from genus two modular invariance}},
  \href{http://arxiv.org/abs/1705.05855}{{\tt arXiv:1705.05855}}.

\bibitem{Cho:2017fzo}
M.~Cho, S.~Collier, and X.~Yin, {\it {Genus Two Modular Bootstrap}},
  \href{http://arxiv.org/abs/1705.05865}{{\tt arXiv:1705.05865}}.

\bibitem{Keller:2017iql}
C.~A. Keller, G.~Mathys, and I.~G. Zadeh, {\it {Bootstrapping Chiral CFTs at
  Genus Two}},  \href{http://arxiv.org/abs/1705.05862}{{\tt arXiv:1705.05862}}.

\bibitem{Afkhami-Jeddi:2017idc}
N.~Afkhami-Jeddi, K.~Colville, T.~Hartman, A.~Maloney, and E.~Perlmutter, {\it
  {Constraints on Higher Spin CFT$_2$}},
  \href{http://arxiv.org/abs/1707.07717}{{\tt arXiv:1707.07717}}.

\bibitem{Apolo:2017xip}
L.~Apolo, {\it {Bounds on CFTs with $W_3$ algebras and AdS$_3$ higher spin
  theories}},  \href{http://arxiv.org/abs/1705.10402}{{\tt arXiv:1705.10402}}.

\bibitem{Brown:1986nw}
J.~D. Brown and M.~Henneaux, {\it {Central Charges in the Canonical Realization
  of Asymptotic Symmetries: An Example from Three-Dimensional Gravity}},  {\em
  Commun. Math. Phys.} {\bf 104} (1986) 207--226.

\bibitem{Simmons-Duffin:2015qma}
D.~Simmons-Duffin, {\it {A Semidefinite Program Solver for the Conformal
  Bootstrap}},  {\em JHEP} {\bf 06} (2015) 174,
  [\href{http://arxiv.org/abs/1502.02033}{{\tt arXiv:1502.02033}}].

\bibitem{Benjamin:2016fhe}
N.~Benjamin, E.~Dyer, A.~L. Fitzpatrick, and S.~Kachru, {\it {Universal Bounds
  on Charged States in 2d CFT and 3d Gravity}},  {\em JHEP} {\bf 08} (2016)
  041, [\href{http://arxiv.org/abs/1603.09745}{{\tt arXiv:1603.09745}}].

\bibitem{Mathur:1988na}
S.~D. Mathur, S.~Mukhi, and A.~Sen, {\it {On the Classification of Rational
  Conformal Field Theories}},  {\em Phys. Lett.} {\bf B213} (1988) 303--308.

\bibitem{Hampapura:2015cea}
H.~R. Hampapura and S.~Mukhi, {\it {On 2d Conformal Field Theories with Two
  Characters}},  {\em JHEP} {\bf 01} (2016) 005,
  [\href{http://arxiv.org/abs/1510.04478}{{\tt arXiv:1510.04478}}].

\bibitem{frenkel1984natural}
I.~B. Frenkel, J.~Lepowsky, and A.~Meurman, {\it A natural representation of
  the fischer-griess monster with the modular function j as character},  {\em
  Proceedings of the National Academy of Sciences} {\bf 81} (1984), no.~10
  3256--3260.

\bibitem{Hoehn:2007aa}
G.~Hoehn, {\it Selbstduale vertexoperatorsuperalgebren und das babymonster
  (self-dual vertex operator super algebras and the baby monster)},
  \href{http://arxiv.org/abs/0706.0236}{{\tt arXiv:0706.0236}}.

\bibitem{Henneaux:2010xg}
M.~Henneaux and S.-J. Rey, {\it {Nonlinear $W_{infinity}$ as Asymptotic
  Symmetry of Three-Dimensional Higher Spin Anti-de Sitter Gravity}},  {\em
  JHEP} {\bf 12} (2010) 007, [\href{http://arxiv.org/abs/1008.4579}{{\tt
  arXiv:1008.4579}}].

\bibitem{Mathur:1988gt}
S.~D. Mathur, S.~Mukhi, and A.~Sen, {\it Reconstruction of conformal field
  theories from modular geometry on the torus},  {\em Nucl. Phys.} {\bf B318}
  (1989) 483--540.

\bibitem{Gannon:1992nq}
T.~Gannon, {\it {WZW commutants, lattices, and level 1 partition functions}},
  {\em Nucl. Phys.} {\bf B396} (1993) 708--736,
  [\href{http://arxiv.org/abs/hep-th/9209043}{{\tt hep-th/9209043}}].

\bibitem{Tuite:2008pt}
M.~P. Tuite, {\it {Exceptional Vertex Operator Algebras and the Virasoro
  Algebra}},  {\em Contemp. Math.} {\bf 497} (2009) 213--225,
  [\href{http://arxiv.org/abs/0811.4523}{{\tt arXiv:0811.4523}}].

\bibitem{Hampapura:2016mmz}
H.~R. Hampapura and S.~Mukhi, {\it {Two-dimensional RCFT's without Kac-Moody
  symmetry}},  {\em JHEP} {\bf 07} (2016) 138,
  [\href{http://arxiv.org/abs/1605.03314}{{\tt arXiv:1605.03314}}].

\bibitem{Frenkel:1988xz}
I.~Frenkel, J.~Lepowsky, and A.~Meurman, {\em {VERTEX OPERATOR ALGEBRAS AND THE
  MONSTER}}.
\newblock 1988.

\bibitem{Bouwknegt:1992wg}
P.~Bouwknegt and K.~Schoutens, {\it {W symmetry in conformal field theory}},
  {\em Phys. Rept.} {\bf 223} (1993) 183--276,
  [\href{http://arxiv.org/abs/hep-th/9210010}{{\tt hep-th/9210010}}].

\bibitem{Zamolodchikov:1985wn}
A.~B. Zamolodchikov, {\it {Infinite Additional Symmetries in Two-Dimensional
  Conformal Quantum Field Theory}},  {\em Theor. Math. Phys.} {\bf 65} (1985)
  1205--1213. [Teor. Mat. Fiz.65,347(1985)].

\bibitem{Iles:2013jha}
N.~J. Iles and G.~M.~T. Watts, {\it {Characters of the $W_3$ algebra}},  {\em
  JHEP} {\bf 02} (2014) 009, [\href{http://arxiv.org/abs/1307.3771}{{\tt
  arXiv:1307.3771}}].

\bibitem{Iles:2014gra}
N.~J. Iles and G.~M.~T. Watts, {\it {Modular properties of characters of the
  W$_{3}$ algebra}},  {\em JHEP} {\bf 01} (2016) 089,
  [\href{http://arxiv.org/abs/1411.4039}{{\tt arXiv:1411.4039}}].

\bibitem{Blumenhagen:1990jv}
R.~Blumenhagen, M.~Flohr, A.~Kliem, W.~Nahm, A.~Recknagel, and R.~Varnhagen,
  {\it {W algebras with two and three generators}},  {\em Nucl. Phys.} {\bf
  B361} (1991) 255--289.

\bibitem{Drinfeld:1984qv}
V.~G. Drinfeld and V.~V. Sokolov, {\it {Lie algebras and equations of
  Korteweg-de Vries type}},  {\em J. Sov. Math.} {\bf 30} (1984) 1975--2036.

\bibitem{Bershadsky:1989mf}
M.~Bershadsky and H.~Ooguri, {\it {Hidden SL(n) Symmetry in Conformal Field
  Theories}},  {\em Commun. Math. Phys.} {\bf 126} (1989) 49.

\bibitem{Feigin:1990pn}
B.~Feigin and E.~Frenkel, {\it {Quantization of the Drinfeld-Sokolov
  reduction}},  {\em Phys. Lett.} {\bf B246} (1990) 75--81.

\bibitem{ElShowk:2012hu}
S.~El-Showk and M.~F. Paulos, {\it {Bootstrapping Conformal Field Theories with
  the Extremal Functional Method}},  {\em Phys. Rev. Lett.} {\bf 111} (2013),
  no.~24 241601, [\href{http://arxiv.org/abs/1211.2810}{{\tt
  arXiv:1211.2810}}].

\bibitem{landsberg2006sextonions}
J.~M. Landsberg and L.~Manivel, {\it {The sextonions and $E_{7\frac{1}{2}}$}},
  {\em Advances in Mathematics} {\bf 201} (2006), no.~1 143--179.

\bibitem{GAP4}
The GAP~Group, {\em {GAP -- Groups, Algorithms, and Programming, Version
  4.8.7}}, 2017.

\end{thebibliography}\endgroup

\end{document}